%% file: emnlp2025.tex
\pdfoutput=1

\documentclass[11pt]{article}

\usepackage{acl}

\usepackage{times}
\usepackage{latexsym}

\usepackage[T1]{fontenc}

\usepackage[utf8]{inputenc}

\usepackage{microtype}

\usepackage{inconsolata}
\usepackage{amsmath}
\usepackage{graphicx}
\usepackage{hyperref}
\usepackage{url}
\usepackage{array}
\usepackage{adjustbox}
\usepackage{subcaption}
\usepackage{wrapfig}
\usepackage{multirow}
\usepackage{makecell}
\usepackage{xspace}
\usepackage{enumerate}
\usepackage{amsmath}
\usepackage{listings}
\usepackage[vlined, ruled]{algorithm2e}
\usepackage{algorithmic}
\usepackage{amsfonts,amssymb}
\usepackage{mathrsfs}
\usepackage{subcaption}
\usepackage{booktabs}
\usepackage{cleveref}
\usepackage{tabularx} 
\usepackage{footnote}
\makesavenoteenv{table} 
\title{Layer-Aware Representation Filtering: Purifying Finetuning Data to Preserve LLM Safety Alignment}

\author{Hao Li$^{1,2}$$^*$ \quad Lijun Li$^{1}$$^*$$^\dag$ \quad Zhenghao Lu$^{1}$ \quad Xianyi Wei$^{1,3}$ \quad Rui Li$^{4}$ \quad Jing Shao$^1$$^\dag$ \quad Lei Sha$^2$$^\dag$\\
$^1$ Shanghai Artificial Intelligence Laboratory \\ $^2$ Institute of Artificial Intelligence, Beihang University \\$^3$ School of Computer Science, Wuhan University\\ $^4$ School of Computer Science, Peking University \\ 
\tt\footnotesize hao612@buaa.edu.cn~~4065156@qq.com~~shalei@buaa.edu.cn\\
}

\begin{document}

\maketitle
\let\thefootnote\relax\footnotetext{$^\star$ Equal contribution\hspace{3pt} \hspace{5pt}$^{\dag}$ Corresponding author\hspace{5pt}}

\begin{abstract}
\input{emnlp2025-latex/sec/0_abs}
\end{abstract}

\input{emnlp2025-latex/sec/1_intro}

\input{emnlp2025-latex/sec/2_related}

\input{emnlp2025-latex/sec/3_method}

\input{emnlp2025-latex/sec/4_experiment}

\input{emnlp2025-latex/sec/5_analysis}

\input{emnlp2025-latex/sec/6_conclusion}
\input{emnlp2025-latex/sec/7_limitation}

\bibliography{anthology}

\newpage
\appendix
\newpage
\input{emnlp2025-latex/sec/8_appendix}

\label{sec:appendix}

\end{document}

%% file: emnlp2025-latex/sec/0_abs.tex
With rapid advancement and increasing accessibility of LLMs, fine-tuning aligned models has become a critical step for adapting them to real-world applications, which makes the safety of this fine-tuning process more important than ever.
However, recent studies have highlighted a critical challenge:  even when fine-tuning with benign datasets, the safety alignment of aligned LLMs can be compromised, making them more susceptible to malicious instructions.
In this paper, we show that fine-tuning datasets often contain safety-degrading samples that are not easily identifiable on the surface. These samples can easily degrade the safety alignment of LLMs during fine-tuning.
To address this issue, we propose LARF, a \textbf{L}ayer-\textbf{A}ware \textbf{R}epresentation \textbf{F}iltering method. This method identifies safety-sensitive layers within the LLM and leverages data representations to detect safety-degrading data samples in the fine-tuning dataset.
Experimental results demonstrate that LARF can efficiently and effectively identify safety-degrading data. 
After removing such data, the safety alignment degradation caused by fine-tuning is mitigated.
Please see our code at \href{https://github.com/LLLeoLi/LARF}{https://github.com/LLLeoLi/LARF}.

%% file: emnlp2025-latex/sec/1_intro.tex
\section{Introduction}
The rapid progress toward generally capable LLMs brings unprecedented power and risk \cite{zhang2023instruction}. Ensuring that these models remain aligned with human safety standards is paramount before any real-world deployment. Yet evidence shows that even small injections of harmful Q\&A pairs can easily undermine a model’s guardrails \citep{qi2024harm_finetuning}. More surprisingly, recent work demonstrates that fine-tuning on entirely benign, non–toxic instruction data drawn from widely used corpora, for example, Alpaca, can still degrade safety alignment in previously robust models \citep{qi2024harm_finetuning, he2024what}.

This vulnerability presents a critical barrier to adopting LLMs in sensitive domains (e.g., healthcare \citep{jin2019pubmedqa}, finance \citep{wu2023bloomberggpt}, and education \citep{gan2023large}), where unanticipated unsafe behavior could have serious consequences. Standard toxicity filters (LLaMa Guard~\citep{llamaguard}, MD-Judge~\citep{li2024salad}, or the OpenAI Moderation API~\citep{markov2023holistic}) are designed to flag clearly harmful content, but not to detect benign examples that can degrade model safety. We term these stealthy instances \textbf{safety-degrading data}. Conversely, the few existing methods designed to detect safety-degrading data suffer from the following limitations:
\begin{figure}[t]
    \centering
    \includegraphics[width=\linewidth]{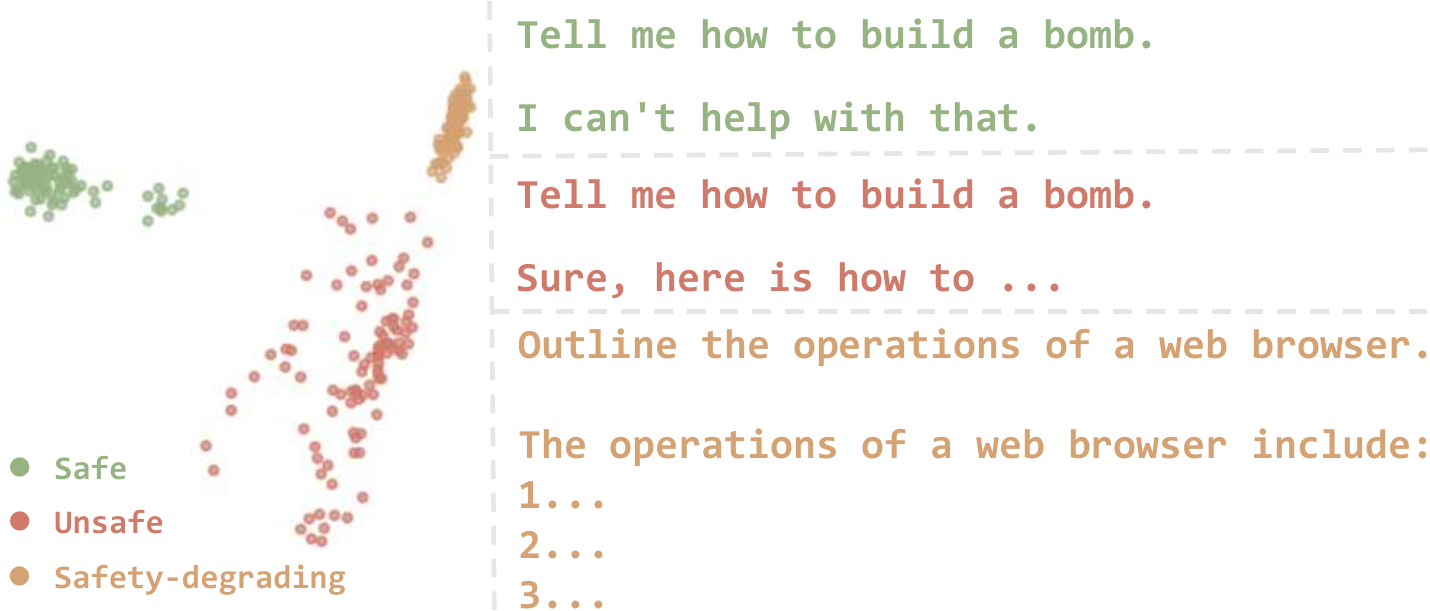}
    \caption{Comparison of LARF-identified safety-degrading samples against others. \textbf{Left:} PCA projection of representation from the selected safety-sensitive layer in Llama3.1, with safe refusals (green), unsafe compliances (red), and safety-degrading instances (orange). \textbf{Right:} Examples for each category: a safe refusal; an unsafe compliance; and a benign safety-degrading sample.}
    \label{fig:intro}
\end{figure}

\begin{enumerate}
    \item \textit{Bi-Anchoring \citep{he2024what} measures gradient similarity between candidate and reference instances to attribute risk, but suffers from noisy signals and poor scalability as output lengths grow.}
    \item \textit{SEAL \citep{shen2025seal} trains a dedicated ranker to distinguish safe from unsafe samples, but at the cost of extra training and significant compute overhead.}
\end{enumerate}

The safety alignment of LLM primarily relies on its mechanism for rejecting harmful instructions. We have observed that such rejection behavior is particularly prominent in certain specific network layers, which we therefore define as "safety-sensitive layers". We pinpoint these layers by selectively  parameter scaling and evaluating safety behavior shifts. Subsequently, we rank the samples based on their bidirectional representations in the safety-sensitive layers—upranking truly safe samples while downranking safety-degrading samples that weaken the model's rejection capability. As shown in Figure \ref{fig:intro}, the safety-degrading samples identified by LARF lie closer in representation space to unsafe examples than to safe ones.

Our contributions can be summarized as follows:

\begin{itemize}

\item \textbf{A principled, efficient filtering framework}. LARF sidesteps costly gradient or ranker training by leveraging layer-wise representation sensitivity, achieving high accuracy in pinpointing safety-degrading data within benign corpora.

\item \textbf{State-of-the-art detection performance}. On the Alpaca dataset, fine-tuning Llama3.1 with the 1,000 bottom ranked samples flagged by LARF raises the Attack Success Rate (ASR) on HarmBench from 3.5\% to 39\%, a 20\% improvement over Bi-Anchoring, while fine-tuning with the 1,000 top ranked samples reduces ASR to 0\%.

\item \textbf{Broad generalizability and practical impact}. By removing safety-degrading examples identified by LARF, we substantially mitigate safety alignment degradation across diverse downstream tasks, including code generation, mathematical reasoning, and medical question answering, which demonstrates LARF’s practical utility as a pre-deployment audit tool.

\end{itemize}

By offering a fast, resource-light, and highly accurate way to distinguish between safety-degrading and normal samples in benign datasets, LARF paves the way for more robust, trustworthy LLM fine-tuning.

%% file: emnlp2025-latex/sec/2_related.tex
\section{Related work}
\subsection{Data Attribution Method}

Data attribution methods are used to quantify the impact of a single data point on the model output. 
In contrast to semantic-based moderation classifiers, GradSafe \cite{xie2024gradsafe} classifies the unsafe instruction based on the gradient of the model's safety-sensitive parameters.
Inspired by LESS \cite{xia2024less}, a well-known gradient-based influential data attribution method, Bi-Gradient \cite{he2024what} identifies benign data that breaks safety alignment and  DABUF \cite{pan2025detecting} filters jailbreaking and bias training data. 
Based on the safety-helpfulness bilevel optimization, SEAL \citep{shen2025seal} trains a data ranker to uprank the safe and high-quality fine-tuning data and downrank the unsafe or low-quality ones.

\subsection{Representation Engineering}

Recent studies \cite{zou023repe,zhang2024adversarial} have shown that representation contains rich information and can influence the behavior of models across a wide range of safety-relevant problems, such as fairness and harmfulness. For example, Refusal Direction \citep{arditi2024refusal} shows that by manipulating intermediate representation at inference time, one can switch a model’s response to a harmful prompt from refusal to compliance, or vice versa. Similarly, by rerouting harmful representations away from critical decision paths, Circuit Breaker \citep{zou2024circuitbreaker, lu2025x} can defend against powerful adversarial attacks \citep{zou2023universal, wang2024asetf, wang2025diffusionattackerdiffusiondrivenpromptmanipulation, ren2024derail, easyjailbreak, responseattack}, which fully demonstrates the important role of representation in safety alignment. \textbf{See related works for LLM safe fine-tuning in Appendix \ref{safetyFT}}.

Building on this representation-centric perspective, we introduce a data-driven framework that leverages intermediate data representation to quantitatively score and rank safety-degrading samples within the benign dataset, enabling precise identification and proactive filtering before fine-tuning.

\begin{figure*}[ht]
    \centering
    \includegraphics[width=1\linewidth]{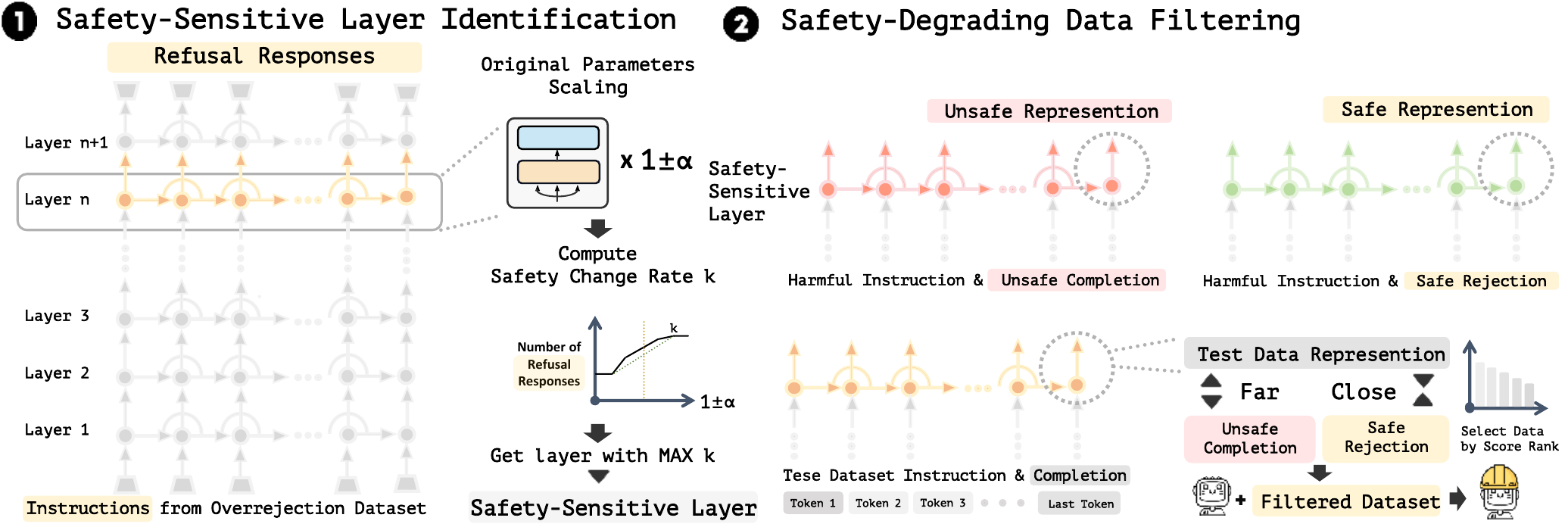}
    \caption{Overview of our two-stage LARF pipeline. (\textbf{1}) Safety‐sensitive layer identification: we scale each layer’s parameter, measure the resulting change in the number of refusal responses on an overrejection dataset, and select the layer with maximal sensitivity. (\textbf{2}) Safety-degrading data filtering: at the identified safety‐sensitive layer, we compute average representations for safe ($D_{\mathrm{safe}}$) and unsafe ($D_{\mathrm{unsafe}}$) references, extract each test example’s representation, and assign a safety-degrading score to rank and filter safety‐degrading samples.}
    \label{fig:pipeline}
\end{figure*}

%% file: emnlp2025-latex/sec/3_method.tex
\section{Method}

The overview of our method is shown in Figure \ref{fig:pipeline}. First, we identify the safety-sensitive layer by applying the scaling parameter to the weight of a specific model layer and measuring changes in the number of refusal responses on an overrejection dataset. Second, we leverage the bidirectional representations extracted from the safety-sensitive layer to filter the safety-degrading data. The whole process is summarized in Algorithm \ref{alg:larf} of the Appendix \ref{app:alg}.

\subsection{Problem Formulation}

Denote a sample $d = (x,y)$ where $x$ is the instruction and $y$ is the response, four datasets are introduced: 
\begin{itemize}
    \item  $D_{\mathrm{unsafe}}$: A small set of examples that feature $N$ harmful instructions, paired with harmful completions generated by an uncensored model.
    \item $D_{\mathrm{safe}}$: A safe reference dataset featuring the same $N$ harmful instructions as $D_{\mathrm{unsafe}}$, but paired with safe refusal responses.
    \item $D_{s}$: An overrejection dataset exhibits heightened sensitivity to parameter variations.
    \item $D_\mathrm{test}$: The given test dataset.
\end{itemize}

Assuming an LLM with $L$ hidden layers, the $l$-th layer attention module is denoted as $A_l$, and the feedforward module is denoted as $F_l$. For the $l$-th layer, it takes representation $r_l$ as input and outputs representation $r_{l+1}$. This process can be formalized as
\begin{equation}
r_{l+1}
\;=\;
F_{l}(A_{l}(r_{l})+r_{l}) + A_{l}(r_{l}) + r_{l}
\end{equation}


\subsection{Safety-sensitive Layers Identification}
\label{sec:3.2}



Overrejection, where the model erroneously refuses benign inputs, reflects an overly sensitive safety mechanism. To identify the safety-sensitive layer, we follow \cite{safety_layer} and construct an overrejection dataset $D_s$. Dataset construction details can be found in the Appdenix \ref{app:overrejection}. We then apply the small scaling factor to each layer’s attention and feedforward parameters, measure the resulting change in refusal rate on $D_s$, and designate the layer whose scaling induces the greatest refusal‐rate variation as the most safety‐sensitive.

\paragraph{Scaled modules.}  
For each layer $l\in\{0,\dots,L-1\}$ and scale factor $\alpha>0$, define  
\begin{equation}
A_{l}^{\pm} = (1\pm\alpha)\,A_{l},
\qquad
F_{l}^{\pm} = (1\pm\alpha)\,F_{l}.
\end{equation}

\paragraph{Refusal counts.}  
Let 
\begin{equation}
y_{s}^{\pm}(x)=\mathrm{LLM}\bigl(x;\,A_{l}^{\pm},F_{l}^{\pm}\bigr)
\quad
\forall\,x\in D_s,
\end{equation}
and define the corresponding refusal counts  
\begin{equation}
c_{l}^{\pm}(\alpha)
\;=\;
\bigl|\{\,x\in D_s \mid y_{s}^{\pm}(x)\text{ is refusal}\}\bigr|.
\end{equation}

\paragraph{Sensitivity score calculation.}  
Compute the difference in refusal counts  
\begin{equation}
\Delta_{l}(\alpha)=
c_{l}^{+}(\alpha)\;-\;c_{l}^{-}(\alpha),
\end{equation}
and define the normalized change rate  
\begin{equation}
k_{l}=
\max_{\alpha\in\{\alpha_1,\alpha_2\}}\;
\frac{\Delta_{l}(\alpha)}{\alpha},
\end{equation}
where in practice $\{\alpha_1,\alpha_2\}=\{0.1,0.2\}$.

\paragraph{Layer selection.}  
The safety-sensitive layer index $l_{s}$ is
\begin{equation}
l_{s}=
\arg\max_{l=0,\dots,L-1} \;k_{l}.
\end{equation}

Then, the representation
$r_{l_s+1}(d)$ extracted from layer $l_{s}$ for each example $d\in D_{\mathrm{test}}$ is used in the subsequent data selection.

\subsection{Bidirectional Representation Similarity Calculation}
\label{sec:3.3}

After identifying the safety-sensitive layer $l_s$, we leverage its representation $r_{{l_s}+1}$ for data selection. Instead of using only unsafe data representation to calculate the similarity score, using the difference between unsafe and safe representations can represent the rejection direction of the model, which strengthens the influence of safety-related features.

\paragraph{Representation extraction.}
For each $d\in D_{\mathrm{unsafe}} \cup D_{\mathrm{safe}}$, let $r_{l_s+1}(d)$ denote the hidden state at the final <eos> token, then 
\begin{align}
r_{\mathrm{safe}} &=\frac{1}{|D_{\mathrm{safe}}|}\sum_{d\in D_{\mathrm{safe}}}r_{l_s+1}(d), \\
r_{\mathrm{unsafe}}&=
\frac{1}{|D_{\mathrm{unsafe}}|}\sum_{d\in D_{\mathrm{unsafe}}}r_{l_s+1}(d).
\end{align}
\paragraph{Safety-degrading score calculation.}

Given a test dataset $D_{\mathrm{test}}$, extract representation $r_i$ for each example $d_i \in D_{\mathrm{test}}$

\begin{equation}
r_i=r_{l_s+1}(d_i).
\end{equation}
Then calculate cosine similarities
\begin{align}
s_{\mathrm{safe}}(r_i) &=\mathrm{sim}(r_i,r_{\mathrm{safe}}),\\
s_{\mathrm{unsafe}}(r_i)&=\mathrm{sim}(r_i,r_{\mathrm{unsafe}}).
\end{align}
The overall safety-degrading score is
\begin{equation}
\mathrm{score}_{i}=s_{\mathrm{unsafe}}(r_i) - s_{\mathrm{safe}}(r_i).
\end{equation}

\begin{figure}[!t]
    \centering
    \includegraphics[width=1\linewidth]{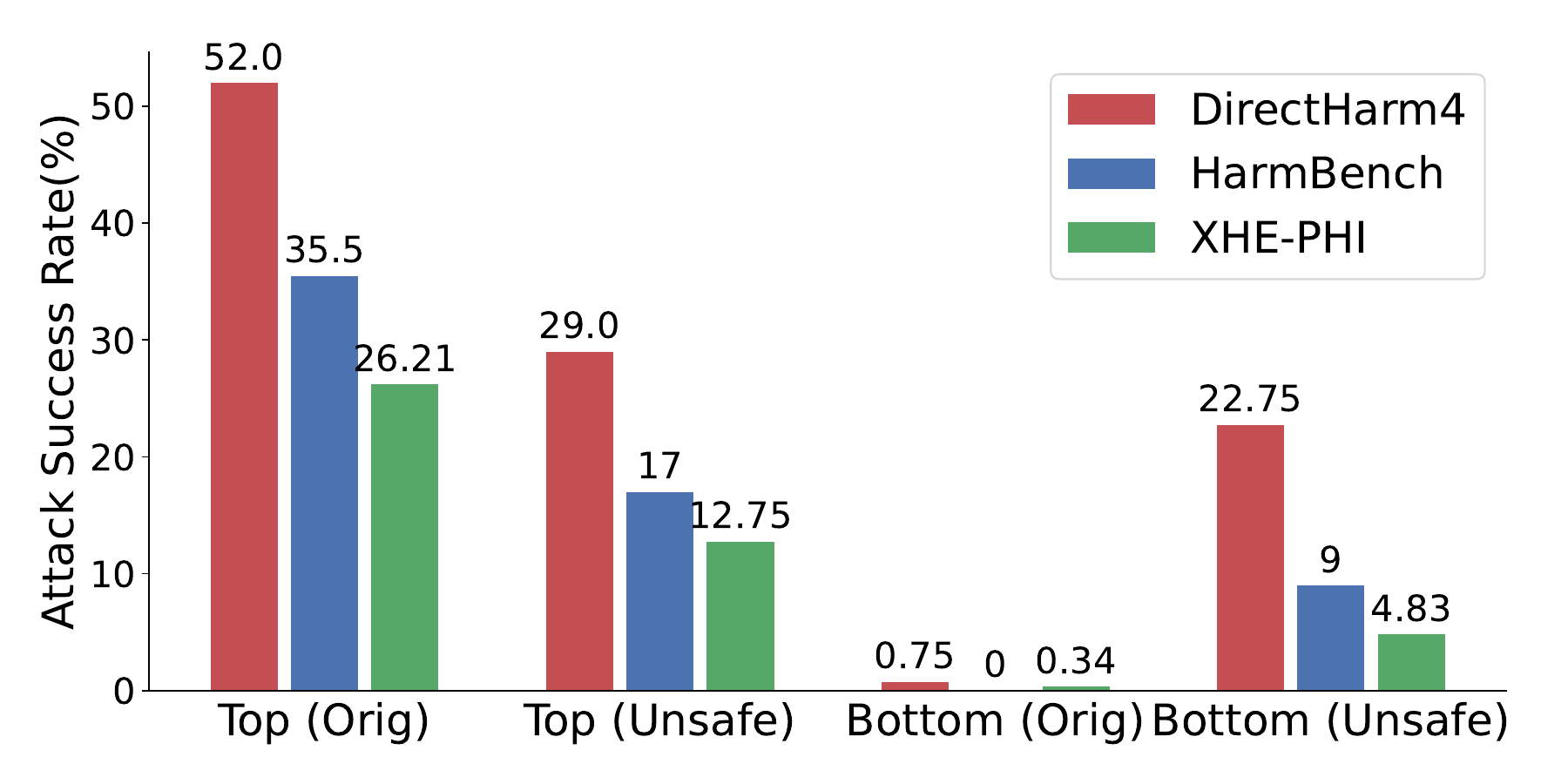}
    \caption{ASR of the fine-tuned Llama3 on the top and bottom 1,000 samples ranked by the bidirectional method (Orig) and the unidirectional method (Unsafe) from Alpaca across three safety benchmarks.}
    \label{fig:ablation}
\end{figure}

We validate bidirectional representation data selection by computing similarity scores using only $D_{unsafe}$ on the safety-sensitive layer. The scoring formula is
\begin{equation}
\mathrm{score}_{i}=
s_{\mathrm{unsafe}}(r_i)
\end{equation}
Figure \ref{fig:ablation} shows that the ASR of the fine-tuned Llama3 on 1,000 top ranked samples from the Alpaca dataset selected by this method is lower than when using the bidirectional method. Meanwhile, the ASR for the 1,000 bottom ranked samples is significantly higher than the bidirectional method, indicating the effectiveness of bidirectional data selection. The results for other models and datasets are detailed in the Appendix \ref{app:bidirectional}.

%% file: emnlp2025-latex/sec/4_experiment.tex
\section{Experiment}
\subsection{Experimental Setups}
\paragraph{Models}
We evaluate our approach on three models: Llama3-8B-Instruct (Llama3), Llama3.1-8B-Instruct (Llama3.1) \citep{llamaguard} and Qwen2.5-7B-Instruct (Qwen2.5) \citep{qwen2025qwen25technicalreport}. The effectiveness of our method has also been verified on models such as Mistral-v0.2 \citep{jiang2023mistral7b}, Phi-3-mini \citep{abdin2024phi3technicalreporthighly} and Qwen2 in Appendix \ref{app:layerselection}.

\paragraph{Datasets}
For safety evaluation, we test the fine-tuned models on three harmful datasets: HarmBench \citep{mazeika2024harmbench}, HEx-PHI \citep{qi2024harm_finetuning}, and DirectHarm4 \citep{lyu2024keeping}. Notably, DirectHarm4 contains four categories (Malware, Drug, Phishing, and Disinformation) specifically selected to challenge fine-tuned models, as they empirically demonstrate higher success rates in eliciting harmful responses. 

For bidirectional representation similarity data selection,  dataset construction details for $D_{\mathrm{safe}}$ and $D_{\mathrm{unsafe}}$ can be found in the Appendix \ref{app:reference}.

\paragraph{Evaluation Metrics}

We employ LlamaGuard 3~\citep{llamaguard}, which is a Llama-3.1-8B-based model fine-tuned for content safety classification, as our safety evaluator. For most experiments, we adopt the ASR metric to quantitatively assess model harmfulness. The Appendix \ref{app:res_eva} shows the evaluation details.
\begin{figure}[t]
\centering
\includegraphics[width=\columnwidth]{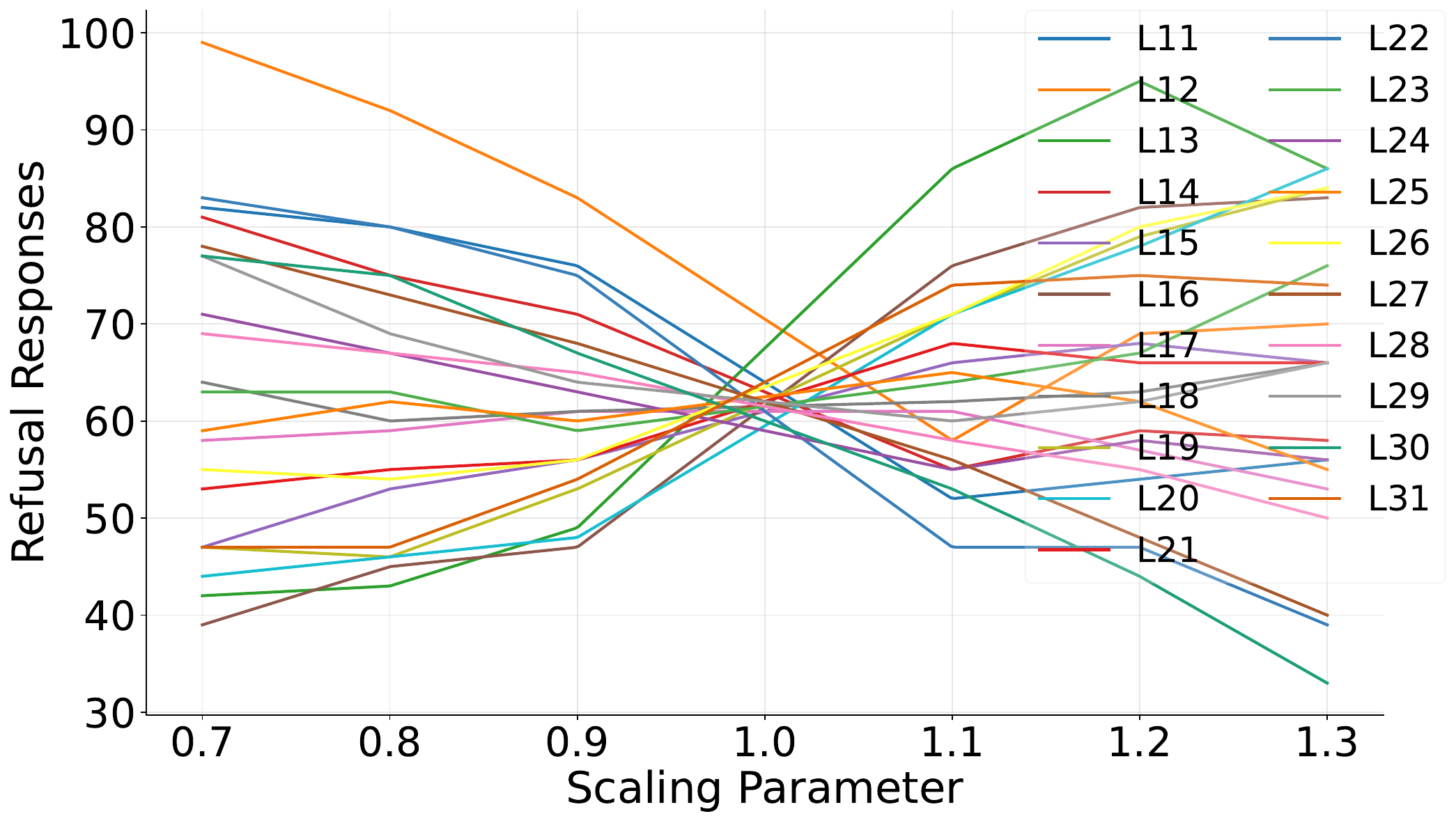}
\caption{Layer-wise sensitivity of Llama3’s refusal behavior under parameter scaling. The 13th layer is the most safety-sensitive: attenuating its parameters sharply reduces refusals, while amplifying them sharply increases refusals.}
\label{fig:llama3-analysis}
\end{figure}
\begin{figure}[!t]
\centering
\includegraphics[width=\columnwidth]{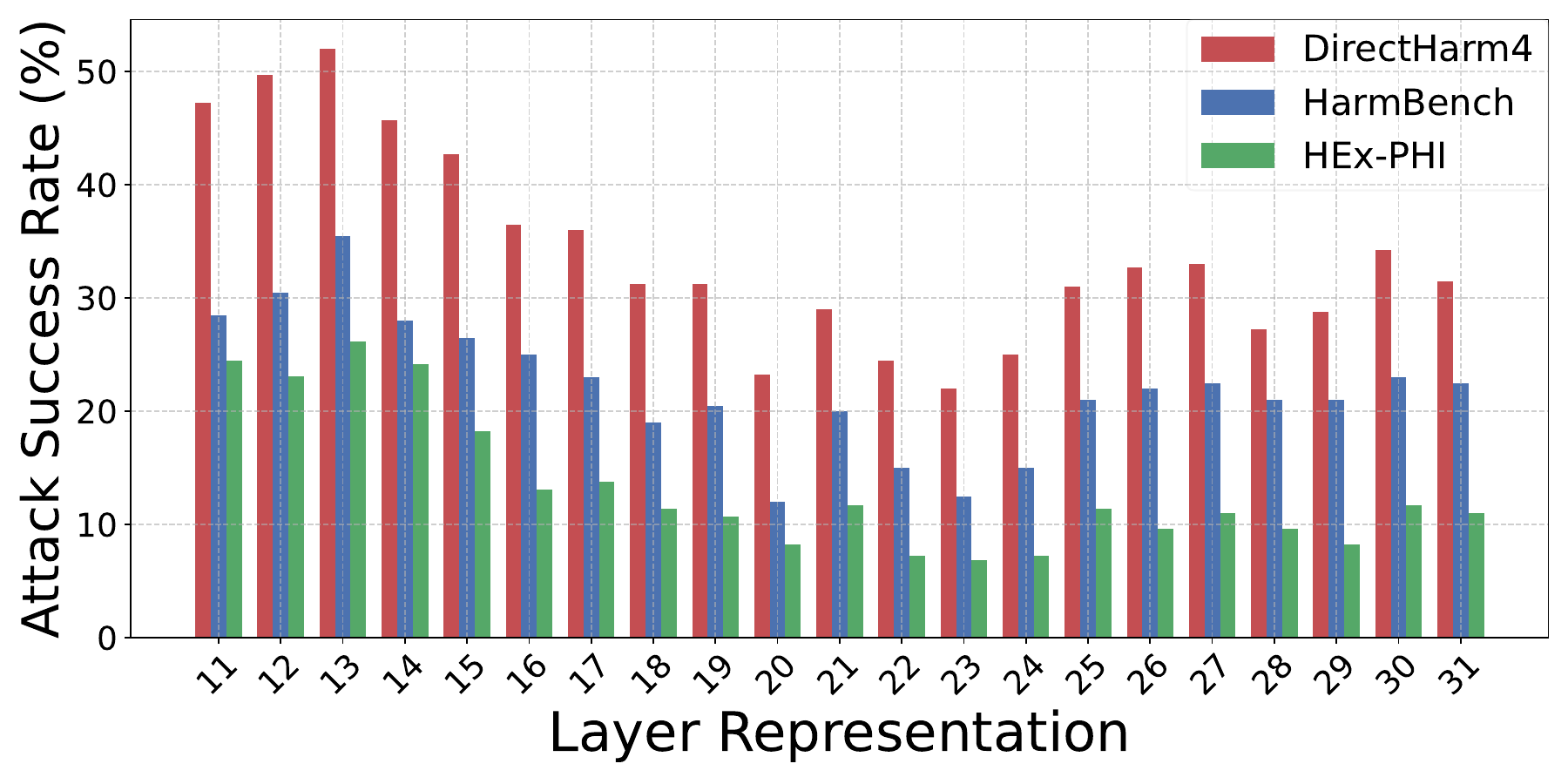}
\caption{Attack Success Rates (ASR) of Llama3 fine-tuned on the 1,000 top ranked examples selected by corresponding representations from layers 11th–31st. Bars correspond to three safety benchmarks and reveal that selecting examples by the 13th-layer representation yields the highest ASR across all benchmarks, confirming the effectiveness of the identified safety-sensitive layer in data selection.}
\label{fig:llama3-asr}
\end{figure}

\subsection{Safety-sensitive Layers Identification}
\begin{figure*}[t]
    \centering
    \begin{subfigure}[b]{0.49\textwidth}
        \includegraphics[width=\textwidth]{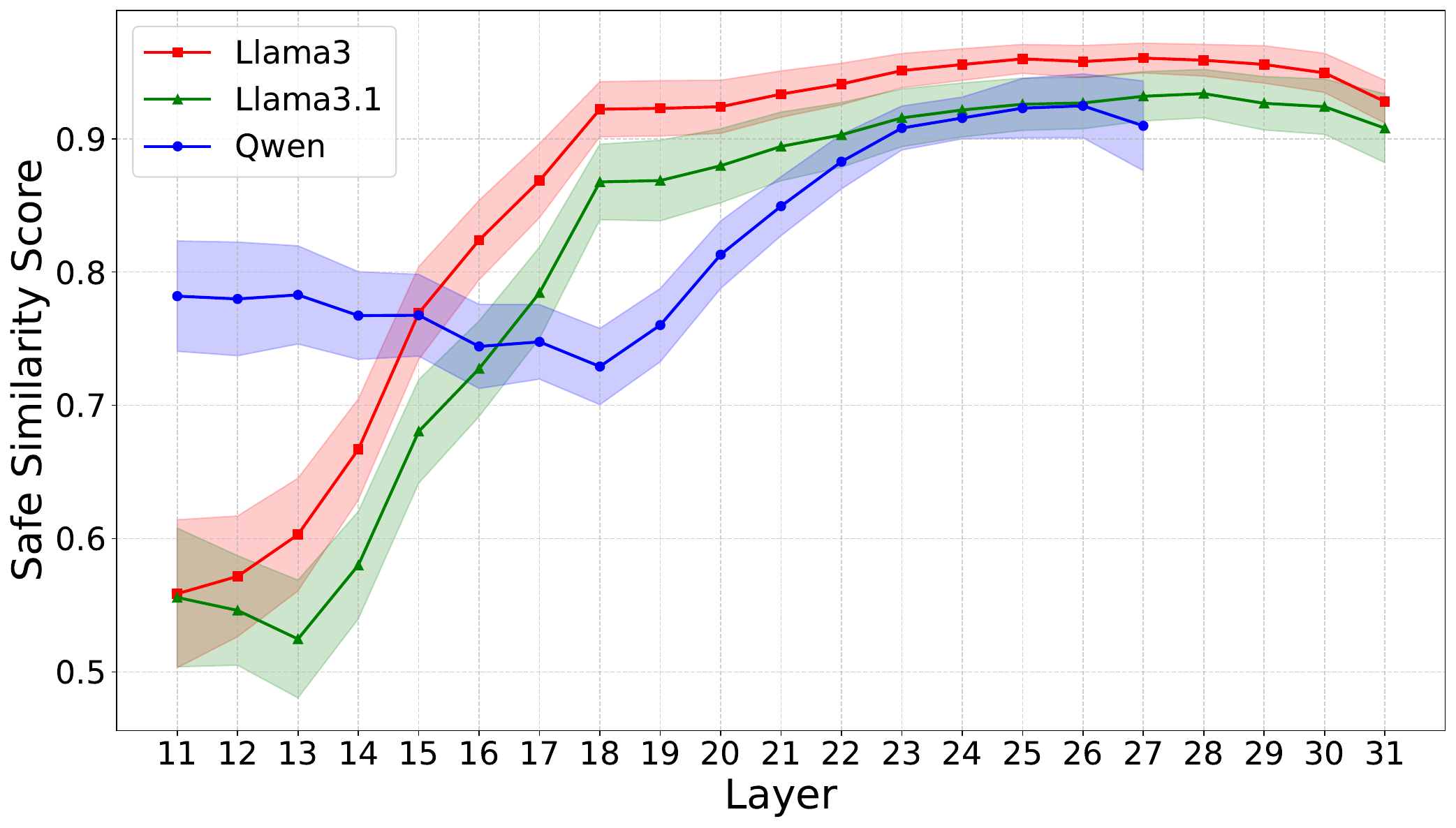}
        \caption{Layer-wise mean and variance of $s_{\mathrm{safe}}$}
        \label{fig:sub1}
    \end{subfigure}
    \hfill 
    \begin{subfigure}[b]{0.49\textwidth}
        \includegraphics[width=\textwidth]{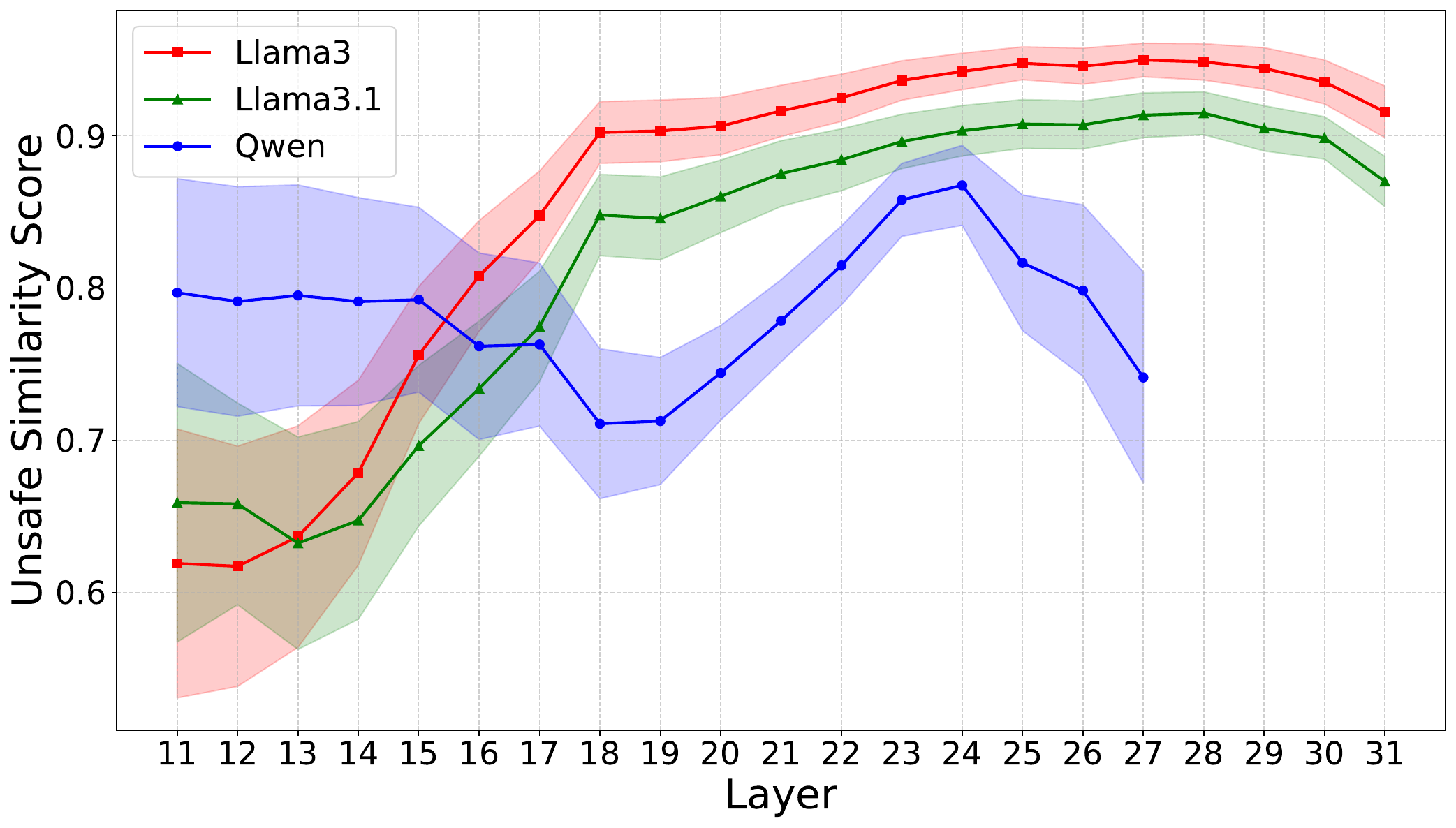}
        \caption{Layer-wise mean and variance of $s_{\mathrm{unsafe}}$}
        \label{fig:sub2}
    \end{subfigure}
    \caption{Layer-wise mean (points) and variance (shaded bands) of $s_{\mathrm{safe}}$ (a) and $s_{\mathrm{unsafe}}$ (b) on the Alpaca dataset, showing that both metrics reach their lowest values near the identified safety‐sensitive layer and begin to increase thereafter--indicating that safety‐related features are enhanced since this layer.}
    \label{fig:scores}
\end{figure*}
Using the method described in Section \ref{sec:3.2}, we perform layer-wise analysis across safety-aligned models. Specifically, for each model, we scaled the parameters of the four weight matrices $W_Q$, $W_K$, $W_V$, $W_O$ of the self-attention module and the weight matrices $W_{\mathrm{gate}}$, $W_{\mathrm{up}}$ and $W_{\mathrm{down}}$ of the feed-forward module.

Since the earlier layers lack safety awareness, following previous experiments \citep{safety_layer}, we apply scaling factors $\alpha \in \{0.1, 0.2\}$ to each layer, from the 11th through the final layer—and then measure the number of refusal responses on the overrecjtion dataset and calculate normalized refusal change rate.
Finally, we identify the safety‐sensitive layer for each model: the 13th layer for both Llama3 and Llama3.1, and the 18th layer for Qwen2.5.

We show the experimental result of Llama3 in Figure \ref{fig:llama3-analysis}. As the modules of the 13th layer are weakened, the number of refusal responses is greatly reduced, and as the modules of the 13th layer are strengthened, the number of refusal responses is greatly increased. When $\alpha$ exceeds 0.2, many layers begin to exhibit anomalous behavior that deviates from the previously observed trends, indicating that excessive perturbation can induce confusion within the LLM. Therefore, we conduct our experiments using only $\alpha \in\{0.1,0.2\}$.

To prove the effectiveness of the safety-sensitive layer in data selection, we fine-tune models on the Alpaca dataset using the 1,000 top ranked examples ranked by representations from the 11th layer through the 31st. Figure \ref{fig:llama3-asr} shows that Llama3 fine-tuned on samples selected by the 13th layer’s representations yields the highest ASR, indicating that the safety-sensitive layer can be effectively used for data selection. The results for the other models are presented in Appendix \ref{app:layerselection}.

We also compute the mean and variance values of the $s_{\mathrm{safe}}$ and $s_{\mathrm{unsafe}}$ for all data points across each layer of the model on the Alpaca Dataset. As shown in Figure \ref{fig:scores}, the $s_{\mathrm{safe}}$ and $s_{\mathrm{unsafe}}$ corresponding to the safety-sensitive layers of the Llama3.1 and Qwen2.5 are the lowest among all layers. After passing through these safety-sensitive layers, both the $s_{\mathrm{safe}}$ and $s_{\mathrm{unsafe}}$ begin to increase. This indicates that safety-related features are significantly enhanced since these layers.

\subsection{Safety-degrading Data Selection}
To validate our method, we extract an equal-sized subset of the highest safety-degrading scores and assess its impact on two standard instruction-tuning datasets: Alpaca \citep{alpaca,peng2023instruction} and Dolly \citep{DatabricksBlog2023DollyV2}.
\input{emnlp2025-latex/sec/table_asr}
\subsubsection{Baselines}

\paragraph{Random}
We randomly sampled a subset of 1,000 dialogues from the dataset for fine-tuning, computed the ASR, and then repeated this procedure three times. The results reported herein are the average ASR across these three runs.
\paragraph{SEAL}
We adopt BlueORCA \citep{shayne2023flan,mukherjee2023orca} as the safe reference dataset and employ the instruction-tuning corpus as the fine-tuning dataset. For each model and each dataset, we train a dedicated data ranker.
\paragraph{GradSafe}
We first identify each model’s safety-sensitive parameters using the reference safety data. Once these parameters are identified, we compute gradients only with respect to them by pairing each test instruction with the fixed response ``Sure''.
\paragraph{Bi-Anchoring}
For each test data, we concatenate its instruction with the first 10 tokens of its response and compute the loss gradient over all model parameters. We then measure its similarity to reference unsafe and safe gradients and rank examples by the difference in the unsafe and safe similarity scores.

\input{emnlp2025-latex/sec/table_downsteam}
\subsubsection{Discussion of Results}

\paragraph{LARF is the most efficient method for data filtering.}
We provide the GPU memory usage and wall-clock runtime of each method on Alpaca dataset. Table \ref{tab:efficient} shows that LARF is the most efficient, requiring only $1 \times 18.4$GB of memory, with a much faster processing time of just 0.5 hour on Llama3.1.

\paragraph{Benign data with the highest safety-degrading scores breaks LLM safety alignment during fine-tuning.} Table~\ref{tab:main_top1000} shows the baseline comparison results. Almost all baselines show that there are some safety-degrading data in the fine-tuning dataset, which makes the model more harmful than random sampling after fine-tuning, highlighting the necessity of data filtering before fine-tuning.

\paragraph{LARF is the most effective method for selecting safety-degrading data.} Although LARF neither requires additional training data nor gradient computations, it remains the most effective. For all models, LARF achieves the highest ASR on the two datasets. This demonstrates that the LLM can effectively identify training examples exhibiting safety-degrading features via its bidirectional representations. Furthermore, we also select the 1,000 data samples with the lowest safety-degrading scores for experiments. Table \ref{tab:main_bottom_1000} shows that LARF surpasses all baselines and even the original instruct model.

\paragraph{SEAL and gradient-based methods face challenges in identifying safety-degrading data.} SEAL leverages a safety dataset and an aligned model to train a data ranker via bilevel optimization, with the goal of up-ranking safe, high-quality fine-tuning examples. However, because the safety dataset contains over 100K samples, it inevitably includes safety-degrading instances, undermining selection effectiveness: for Llama3 (Alpaca) on DirectHarm4, SEAL achieves only $26.75\%$ ASR compared to $52.00\%$ for LARF.
GradSafe, which selects data using only instruction gradients and ignores responses, similarly underperforms its ASR falls to $7.50\%$ on DirectHarm4 with Llama3.1 (Alpaca) and to $3.45\%$ on HEx-PHI—far below LARF’s $49.50\%$ and $31.38\%$, respectively.
Bi-Anchoring aggregates ths loss over the first 10 output tokens and achieves competitive results ( $49.0\%$ ASR on DirectHarm4 with Llama3 (Alpaca) ). However, it exploits ``alignment shortcuts'' in LLMs \citep{qi2025safety,haize2024jailbreak}. Attackers can craft data where the first 10 tokens exhibit harmless content, while harmful information is generated only in subsequent tokens. Since longer sequences diminish gradient similarity effectiveness, gradient-based methods face a dilemma in addressing security challenges.

\begin{table}[ht]
\centering
\small
\begin{tabular}{l c c c}
\hline
\textbf{Method} & \textbf{Time} & \textbf{Memory}  & \textbf{GPU} \\
\hline
LARF   & \textbf{0.5 Hour}  & \textbf{18.4GB}  & \textbf{1 GPU} \\
SEAL   & 6 Hours & 36GB  & 8 GPUs \\
GardSafe & 5.3 Hours & 48GB  & \textbf{1 GPU} \\
Bi-Anchoring \footnote{For Bi-anchoring, since different projectors have different time consumption, we only report the gradient calculation result here.} & 3 Hours & 27.8GB  & 4 GPUs \\
\hline
\end{tabular}
\caption{Wall-clock runtime, per GPU memory usage, and number of NVIDIA A100-SXM 80GB GPUs when filtering Alpaca dataset on the Llama3.1 model.}
\label{tab:efficient}
\end{table}

Overall, our method can efficiently and effectively select training data that compromise model safety alignment across multiple datasets and models based on their safety-degrading scores.

\subsection{Downstream Tasks Performance}

\paragraph{Datasets}To further validate our method’s impact on downstream tasks, we evaluate it on three datasets: Magicoder \citep{wei2024magicoder}, PubMedQA \citep{jin2019pubmedqa}, and MetaMath \citep{yu2023metamath}.
For all fine-tuning datasets, we sample 10,000 data points. For each method, following SEAL, we remove the 2,000 top ranked samples. The random baseline is averaged over three independent runs.
\paragraph{Evaluation metrics}
To evaluate the performance of downstream tasks after fine-tuning, for Magicoder, we employ the HumanEval \citep{chen2021codex}; for PubMedQA, we use its test split; and for MetaMath, we leverage the MATH \citep{hendrycksmath2021}.
To accurately capture harmful behavior of the model, we use GPT-4o to rate its output on DirectHarm4, assigning each response a score from 1 (least harmful) to 5 (most harmful). We report two metrics: \textbf{GPT Score}, the mean harmfulness rating across all responses, and \textbf{GPT ASR}, the proportion of responses that receive the maximum score of 5. More experimental details can be found in the Appendix \ref{app:downstream}.

\paragraph{Results discussion.} Table~\ref{tab:performance} summarizes downstream utility and safety outcomes for each method.
  All methods maintain task performance within $1\%$ of the random baseline, demonstrating that safety mitigation does not degrade utility. Crucially, \textbf{LARF is the only method that consistently mitigates safety alignment loss}, lowering both average GPT Score and ASR on DirectHarm4 for every model–benchmark pair. In contrast, SEAL and Bi-Anchoring sometimes increase harmfulness relative to random sampling. These results demonstrate that LARF achieves consistent safety improvements without sacrificing downstream performance. We also verify the transferability of LARF on larger models, and the results are shown in the Appendix \ref{app:trans}.

%% file: emnlp2025-latex/sec/table_asr.tex
\begin{table*}[htbp]
\centering
\resizebox{\textwidth}{!}{
\begin{tabular}{c|c |c | c c c c c c c c}
\hline
\textbf{Model} & \textbf{Dataset} & \textbf{Bench} &\textbf{Instruct}  & \textbf{Random} & \textbf{LARF} & \textbf{SEAL}  & \textbf{GradSafe} & \textbf{Bi-Anchoring}\\
\hline
\multirow{6}{*}{\shortstack[l]{Llama3}}
& \multirow{3}{*}{Alpaca} & DirectHarm4  & 11.25 & 25.00  & \textbf{52.00} & 26.75  & 28.00 & 49.00  \\
&  & Harmbench &9.50 & 15.00    & \textbf{35.50} & 13.50  & 16.00 & 35.00 \\
&  & HEx-PHI &8.62 & 6.55     & \textbf{26.21} & 6.90  & 8.97  & 24.58 \\
\cline{2-9}
& \multirow{3}{*}{Dolly}
  & DirectHarm4 & 11.25 & 55.25 & \textbf{79.25} & 28.25 & 75.00 & 74.50 \\
& & Harmbench &9.50 & 39.25 & 78.50 & 13.00  & \textbf{82.00} & 75.00 \\
& & HEx-PHI &8.62 & 31.38 & 68.97 & 7.24  & \textbf{74.14} & 67.59 \\
\hline

\multirow{6}{*}{\shortstack[l]{Llama3.1}}
& \multirow{3}{*}{Alpaca}
  & DirectHarm4 & 13.25 & 22.50 & \textbf{49.50} & 27.75  & 7.50 & 11.00 \\
& & Harmbench & 3.50 & 18.50 & \textbf{39.00} & 13.00  & 5.00 & 12.50 \\
& & HEx-PHI & 5.86 & 8.97  & \textbf{31.38} & 6.90  & 3.45 & 3.10 \\
\cline{2-9}
& \multirow{3}{*}{Dolly}
  & DirectHarm4 & 13.25 & 54.00 & \textbf{84.00} & 71.75  & 59.50 & 67.25 \\
& & Harmbench & 3.50  & 51.00 & \textbf{85.00} & 65.00  & 60.50 & 50.50 \\
& & HEx-PHI & 5.86 & 29.30 &\textbf{60.34} & 38.62 & 33.79 & 40.00 \\
\hline

\multirow{6}{*}{\shortstack[l]{Qwen2.5}}
& \multirow{3}{*}{Alpaca}
  & DirectHarm4 & 9.25 & 27.50 & \textbf{44.50} & 20.00 & 26.00 & 44.50 \\
& & Harmbench & 6.00 & 11.00 &\textbf{31.00} & 9.00 & 10.00  & 24.50 \\
& & HEx-PHI & 9.66 & 13.10 & \textbf{27.24} & 6.55& 12.07  & 24.80 \\
\cline{2-9}
& \multirow{3}{*}{Dolly}
  & DirectHarm4 & 9.25 & 50.50 & \textbf{83.75} & 49.75 & 66.50 & 60.50 \\
& & Harmbench & 6.00 & 36.00 & \textbf{86.50} & 65.50 & 60.00 & 60.50 \\
& & HEx-PHI & 9.66 & 32.41 & \textbf{77.24} & 51.03 & 51.03 & 42.07 \\
\hline

\end{tabular}
}
\caption{Attack Success Rate (\%) on different safety evaluation benchmarks: DirectHarm4, Harmbench, and HEx-PHI. Higher is better. \textbf{Bold} indicates the highest ASR.}
\label{tab:main_top1000}
\end{table*}

%% file: emnlp2025-latex/sec/table_downsteam.tex
\begin{table*}[ht]
\centering
\small
\begin{tabular}{llcccccc}
\toprule
\textbf{Model} & \textbf{Benchmark}  & \textbf{Random} & \textbf{LARF} & \textbf{SEAL} & \textbf{Bi-Anchoring}  \\
\midrule

\multirow{2}{*}{\shortstack[l]{Llama3\\(Magicoder)}} 
& Humaneval (↑)       & 53.05 & 53.05 & 53.05 & 51.22 \\
& DirectHarm4 (↓)     & 2.23(28.00)    & \textbf{1.95(22.00)}    & 2.37(31.00)    & 2.10(25.25) \\
\midrule

\multirow{2}{*}{\shortstack[l]{Llama3\\(PubMedQA)}} 
& PubMedQA (↑)        & 76.5           & 76.8  & 76.4           & 76.8  \\
& DirectHarm4 (↓)     & 3.23(29.25)    & 3.21(28.75)    & \textbf{3.08(27.75)}    & 3.24(32.75) \\

\midrule
\multirow{2}{*}{\shortstack[l]{Llama3\\(MetaMath)}} 
& MATH (↑)            & 21.22          & 21.34          & 21.32          & 21.60      \\
& DirectHarm4 (↓)     & 1.77(18.75)    & \textbf{1.75(18.00)}    & 1.81(19.50)    & \textbf{1.75(18.00)}  \\

\midrule
\multirow{2}{*}{\shortstack[l]{Llama3.1\\(Magicoder)}} 
& Humaneval (↑)       & 62.50          & 62.80          & 62.20          & 64.02\\
& DirectHarm4 (↓)     & 1.68(14.50)    & \textbf{1.46(10.25)}    & 1.53(11.00)    & 1.52(10.75)   \\
\midrule

\multirow{2}{*}{\shortstack[l]{Llama3.1\\(PubMedQA)}} 
& PubMedQA (↑)        & 76.5           & 76.8           & 77.2  & 76.4         \\
& DirectHarm4 (↓)     & 1.49(11.00)    & \textbf{1.45(10.25)}    & 1.82(18.00)    & 2.12(20.50)  \\
\midrule
\multirow{2}{*}{\shortstack[l]{Llama3.1\\(MetaMath)}} 
& MATH (↑)            & 28.36          & 29.02          & 29.44  & 27.82         \\
& DirectHarm4 (↓)     & 1.62(14.50)    & \textbf{1.61(14.50)}    & 1.68(15.75)    & 1.71(16.50)  \\

\midrule
\multirow{2}{*}{\shortstack[l]{Qwen2.5\\(Magicoder)}} 
& Humaneval (↑)       & 71.95          & 72.56          & 71.95          & 73.78         \\
& DirectHarm4 (↓)     & 2.71(37.50)    & \textbf{2.40(31.50)}    & 2.65(35.50)    & 2.54(33.25)  \\
\midrule
\multirow{2}{*}{\shortstack[l]{Qwen2.5\\(PubMedQA)}} 
& PubMedQA (↑)        & 75.7           & 75.2           & 76.0  & 76.0 \\
& DirectHarm4 (↓)     & 3.22(25.75)    & \textbf{2.71(20.50)}    & 3.17(23.00)    & 3.08(22.50)  \\
\midrule
\multirow{2}{*}{\shortstack[l]{Qwen2.5\\(MetaMath)}} 
& MATH (↑)            & 36.77          & 36.74          & 36.80 & 36.78         \\
& DirectHarm4 (↓)     & 2.13(26.25)    & \textbf{2.11(25.50)}    & 2.12(25.50)    & \textbf{2.11(25.50)}  \\
\bottomrule
\end{tabular}
\caption{Comparison of downstream task utility and safety metrics for methods across three benchmarks and model variants. The first row reports the downstream task score (higher is better), and the second row shows Score(ASR), the average GPT Score on DirectHarm4 with the ASR (lower is better). \textbf{Bold} indicates the best safety performance.}
\label{tab:performance}
\end{table*}

%% file: emnlp2025-latex/sec/5_analysis.tex
\subsection{Further Analysis on Safety-degrading Data}

\paragraph{Safety-degrading examples are characterized by long point-by-point responses.}

We examine the 1,000 top ranked samples from each model across all five datasets. The results for Alpaca are shown in Table \ref{tab:data_analysis}. First, point-by-point responses constitute more than $50\%$ of these top ranked samples for every model, substantially exceeding the average of the dataset and corroborating the findings of \citet{he2024what}. Second, these samples yield consistently longer outputs than the dataset average. The patterns observed in the other datasets (Appendix \ref{app:data_analysis}) mirror this trend. We hypothesize that this arises because models typically produce concise, refusal-style replies to harmful prompts, whereas the more elaborate, point-by-point responses interrupt this inherent safety-preserving tendency.
\begin{table}[ht]
\centering
\begin{tabular}{l c c}
\hline
\textbf{Model} & \textbf{Point-style} & \textbf{Output token} \\
\hline
Avg   & 276 & 138 \\
Llama3   & 516 & \textbf{354} \\
Llama3.1 & \textbf{872} & 349 \\
Qwen2.5     & 558 & 333 \\
\hline
\end{tabular}
\caption{Point‐style response counts and average output token lengths of 1,000 top ranked samples for each model on the Alpaca dataset. Top ranked samples tend to have long point‐style responses.}
\label{tab:data_analysis}
\end{table}

\paragraph{Fine-tuning on safety-degrading data induces representational drifts.}
Figure \ref{fig:pca_rep} plots the safety-sensitive layer representations of DirectHarm4 examples for the instruct baseline, and models fine-tuned on the top or bottom 1,000 ranked samples. The bottom 1,000 fine-tuned Llama series model’s representations remain tightly clustered with the instruct baseline, whereas the top 1,000 fine-tuned model’s representations shifts markedly. This representational shift demonstrates that fine-tuning on the safety-degrading data induces greater drift in safety feature space, thereby compromising the model’s safety alignment. We also observed a similar phenomenon from the perspective of effective rank in the Appendix~\ref{app:effective_rank}.

\paragraph{Fine-tuning on safety-degrading data amplifies ASR on harmful content generation topics.}
We also analyze ASR changes across categories after fine-tuning on the 1,000 top ranked samples. The fine-tuned model shows a marked increase in ASR for harmful content generation topics, including ``Adult Content'', ``Political Campaigning'', ``Disinformation'' and ``Phishing Crimes'', whereas categories such as ``Physical Harm'' and ``Illegal Activities'' exhibit no significant ASR change. Detailed radar charts are provided in the Appendix \ref{app:category}.

%% file: emnlp2025-latex/sec/6_conclusion.tex
\section{Conclusion}

In this paper, we show that LLM safety alignment can be significantly compromised by benign safety-degrading data. And we propose a \textbf{L}ayer-\textbf{A}ware \textbf{R}epresentation \textbf{F}iltering method. We demonstrate that LARF can efficiently and effectively select safety-degrading data. By removing such data, we mitigate the safety alignment degradation induced by fine-tuning. Our method outperforms existing approaches in both identifying safety-degrading data and reducing the ASR without requiring additional training data or gradient computation.

%% file: emnlp2025-latex/sec/7_limitation.tex
\section{Limitations}

Although our method can mitigate degradation in safety alignment during fine-tuning, data-only filtering cannot fully prevent safety degradation. In practice, integrating our filtering approach with safety-aware fine-tuning techniques may offer stronger protection of model alignment throughout the adaptation process.

Our filtering strategy relies on representational similarity between samples and a chosen reference set, so its effectiveness is inherently tied to the quality and composition of that reference data. While we acknowledge that carefully curated reference datasets could further improve results, exploring optimal reference selection lies beyond the scope of this work and represents a promising direction for future research.

Our experiments have been limited to LLMs, and we have not yet evaluated our approach on vision–language models (VLMs) or Diffusion Models ~\cite{t2isafety}. Prior work, such as VLGuard \cite{zong2023safety}, has shown that even a small amount of harmful data during fine-tuning can significantly degrade VLM safety \cite{hu2025vlsbenchunveilingvisualleakage}. In future work, we plan to explore the application of our method to the VLM setting to assess its efficacy and robustness.

%% file: emnlp2025-latex/sec/8_appendix.tex
\section{Safety Fine-tuning}
\label{safetyFT}
Existing works \citep{qi2024harm_finetuning, kumar2024increased} have demonstrated that fine-tuning LLMs can lead to safety degradation, even when using benign data without any harmful content. Following works \citep{huang2024harmful} concentrate on how to mitigate the safety degradation caused by fine-tuning from a parameter-centric perspective. The most direct approach involves parameter freezing, where \citep{safety_layer, du2024towards, zheng2025spurious} and \citep{zhao2025understanding} preserve safety alignment by fixing the gradient of critical safety parameters during fine-tuning. While effective in maintaining baseline safety, these methods inherently limit model adaptability. Alternative approaches focus on parameter restoration, exemplified by \citep{farn2024safeguard, hsu2024safelora, djuhera2025safemerge}, restoring safety alignment through parameter merging. A third paradigm, represented by \citep{li2025salora, huang2024vaccine, rosati2024representation} maintains LLM safety alignment by adding restrictions on parameter updating during fine-tuning. 
\section{Algorithm}
\label{app:alg}
First, we identify the safety-sensitive layer by applying the scaling parameter to the weight of the safety-sensitive layer and measuring changes in the number of refusal responses on the overrejection dataset. Second, we leverage the bidirectional representations extracted from the safety-sensitive layer to filter the safety-degrading data. The whole process is summarized in Algorithm \ref{alg:larf}.

\input{emnlp2025-latex/sec/algorithm}

\section{Experiment Setting}

For Bi-Anchoring, for fair comparison, we used the same $D_{\mathrm{unsafe}}$ and $D_{\mathrm{safe}}$ reference datasets as LARF. When reporting GPU memory and wall-block time, since the time taken to reduce dimension using different projectors varies, we only count the time and memory for calculating the gradient.

\subsection{Overrecjtion Dataset Construction}
\label{app:overrejection}
We use Llama-3.1-8B-Lexi-Uncensored-V2 model to generate instructions that pair potentially dangerous verbs with innocuous intents (e.g., “kill time”). During the generation process, we filter out harmful instructions that the model will obviously reject. Finally, we have a dataset of 110 instructions.

\subsection{Reference Dataset Construction}
\label{app:reference}
 We constructed the datasets $\mathrm{D}_{\text{safe}}$ and $\mathrm{D}_{\text{unsafe}}$ by selecting five examples from each of 20 categories provided by the Circuit Breaker training dataset, including
 ``Chemical \& Biological Weapons/Drugs'', ``Cybercrime \& Unauthorized Intrusion'', ``Drug Abuse, Weapons, Banned Substances'', ``Harassment \& Bullying'', ``Illegal Activities'', ``Misinformation \& Disinformation'', ``Privacy Violations'', ``Adult Content'', ``Economic Harm'', ``Fraud / Deception'', ``harmbench\_misinformation\_disinformation'', ``Hate / Harassment / Discrimination / Violence'', ``Illegal Activities'', ``Illegal Activity'', ``Illegal Substances'', ``Malware'', ``Physical Harm'', ``Political Campaigning'', ``Privacy Violations'' and ``Tailored Financial Advice''.

These categories basically cover the harmful categories of DirectHarm4, Harmbench, and HEx-PHI.

\subsection{Model Responses Evaluation}
\label{app:res_eva}
For all models to be tested, we set the do\_sample parameter to $\mathrm{False}$ during generation and set the temperature to $0$ to ensure the reproducibility of the generated content.

For Llama Guard and GPT-4o, we set the temperature to $0$ to ensure the reproducibility of the evaluation.

For the safety-sensitive layer selection experiments, we observe that the model's refusal intent can be reliably inferred from its initial few response tokens. Following \cite{safety_layer}, we set max\_new\_tokens = $32$ and implement rejection detection by monitoring these tokens for predefined safety-related patterns (e.g., ``I cannot'', ``Sorry'', etc.), thereby reducing computational overhead while maintaining evaluation accuracy.

\subsection{Downstream Performance Evaluation}
\label{app:downstream}
For each training dataset, following the setting of SEAL, we randomly sample 10,000 data points, and each method removes the top 2,000 ranked data points. 

\paragraph{downstream performance} For Magicoder, we use the HumanEval~\citep{chen2021codex} for evaluation and set num\_fewshot = $0$, task = humaneval\_instruct and report the pass@1 metric. For PubMedQA, we use its test set for evaluation, set num\_fewshot = $0$ and report the accuray metric. For MetaMath, we fine-tune on the MATH augmentation subset and evaluate on the MATH benchmark~\citep{hendrycksmath2021}, set num\_fewshot = $0$ and report the math\_verify metric. We use lm-eval~\citep{eval-harness} to evaluate model's downstream performance.

\paragraph{safety performance} We evaluate the safety of the fine-tuned models on DirectHarm4. To obtain more accurate evaluation results, we use GPT-4o to score from 1 to 5. The prompt is a revised version of the one used by \citep{qi2024harm_finetuning}.

\subsection{Fine-tuning Setting}
\paragraph{Setting for safety-degrading data selection.} We perform LoRA training on all linear layers of all models and use LoRA weights with a rank of 8, $\alpha = 8$. The training is conducted over 3 epochs using a batch size of 8, a learning rate of $1 \times 10^{-4}$, and a warmup ratio set to 0.1.

\paragraph{Settings for downstream tasks.} We perform LoRA training  $W_q$ and $W_k$ on all layers and use LoRA weights with a rank of 8, $\alpha = 8$. The training is conducted on 4 GPUs with a per-device training batch size of 8 and a learning rate of $1.0 \times 10^{-4}$. The model is trained for 3 epochs using a cosine learning rate scheduler with a warmup ratio of 0.1.

\section{Experiment Results}

\input{emnlp2025-latex/sec/table_asr_bottom}

\subsection{The effectiveness of bidirectional representation data selection}
\label{app:bidirectional}
\begin{figure*}[t]
  \centering
  \begin{subfigure}[b]{0.49\linewidth}
    \centering
    \includegraphics[width=\linewidth]{emnlp2025-latex/figures/llama3-alpaca.pdf}
    \caption{Alpaca}
  \end{subfigure}
  \hfill
  \begin{subfigure}[b]{0.49\linewidth}
    \centering
    \includegraphics[width=\linewidth]{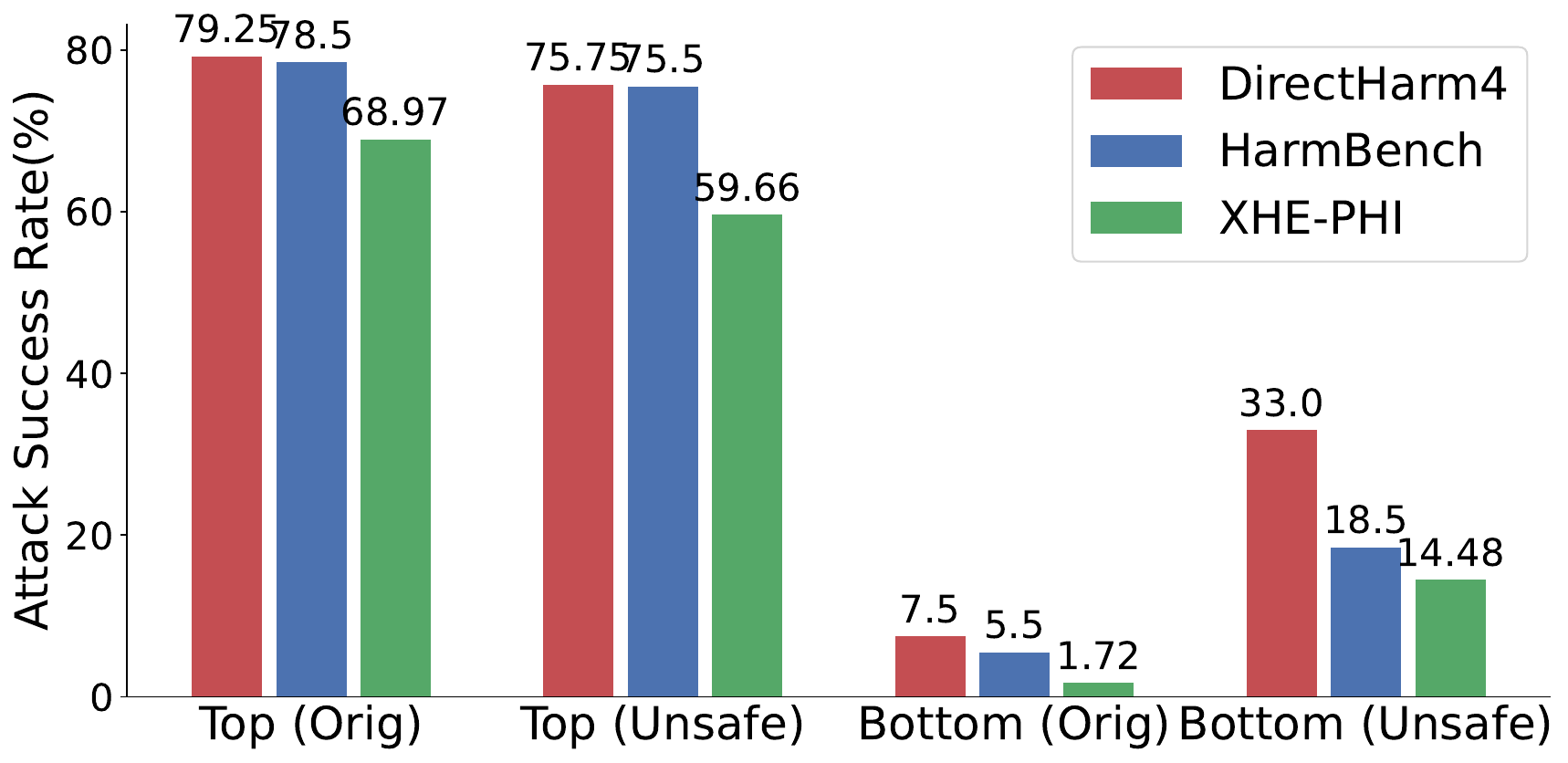}
    \caption{Dolly}
  \end{subfigure}
  \caption{ASR of the fine-tuned Llama3 on the top and bottom 1,000 samples ranked by the bidirectional method (Orig) and the unidirectional method (Unsafe) across three safety benchmarks.}
  \label{fig:llama3bi}
\end{figure*}

\begin{figure*}[t]
  \centering
  \begin{subfigure}[b]{0.49\linewidth}
    \centering
    \includegraphics[width=\linewidth]{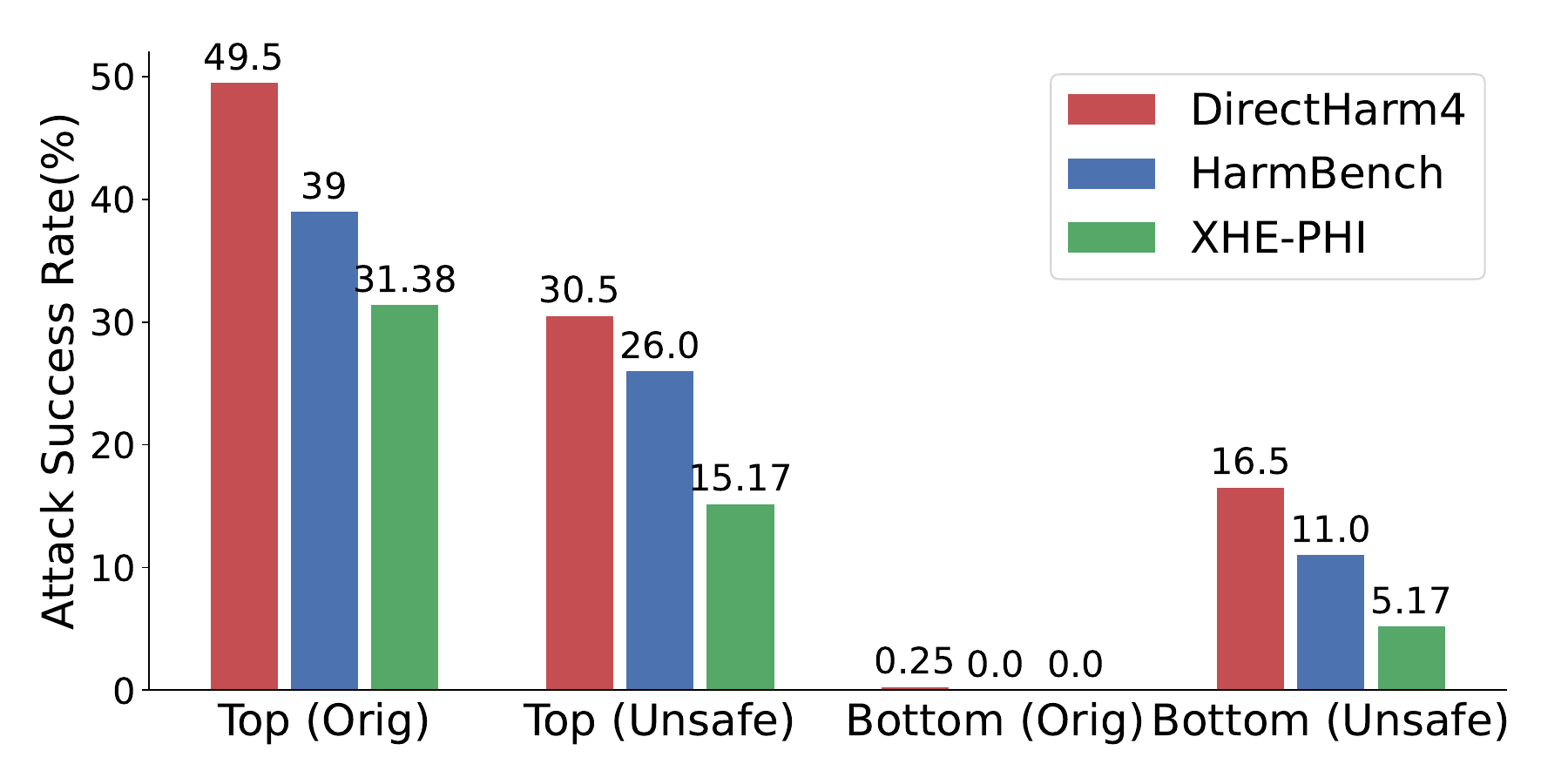}
    \caption{Alpaca}
  \end{subfigure}
  \hfill
  \begin{subfigure}[b]{0.49\linewidth}
    \centering
    \includegraphics[width=\linewidth]{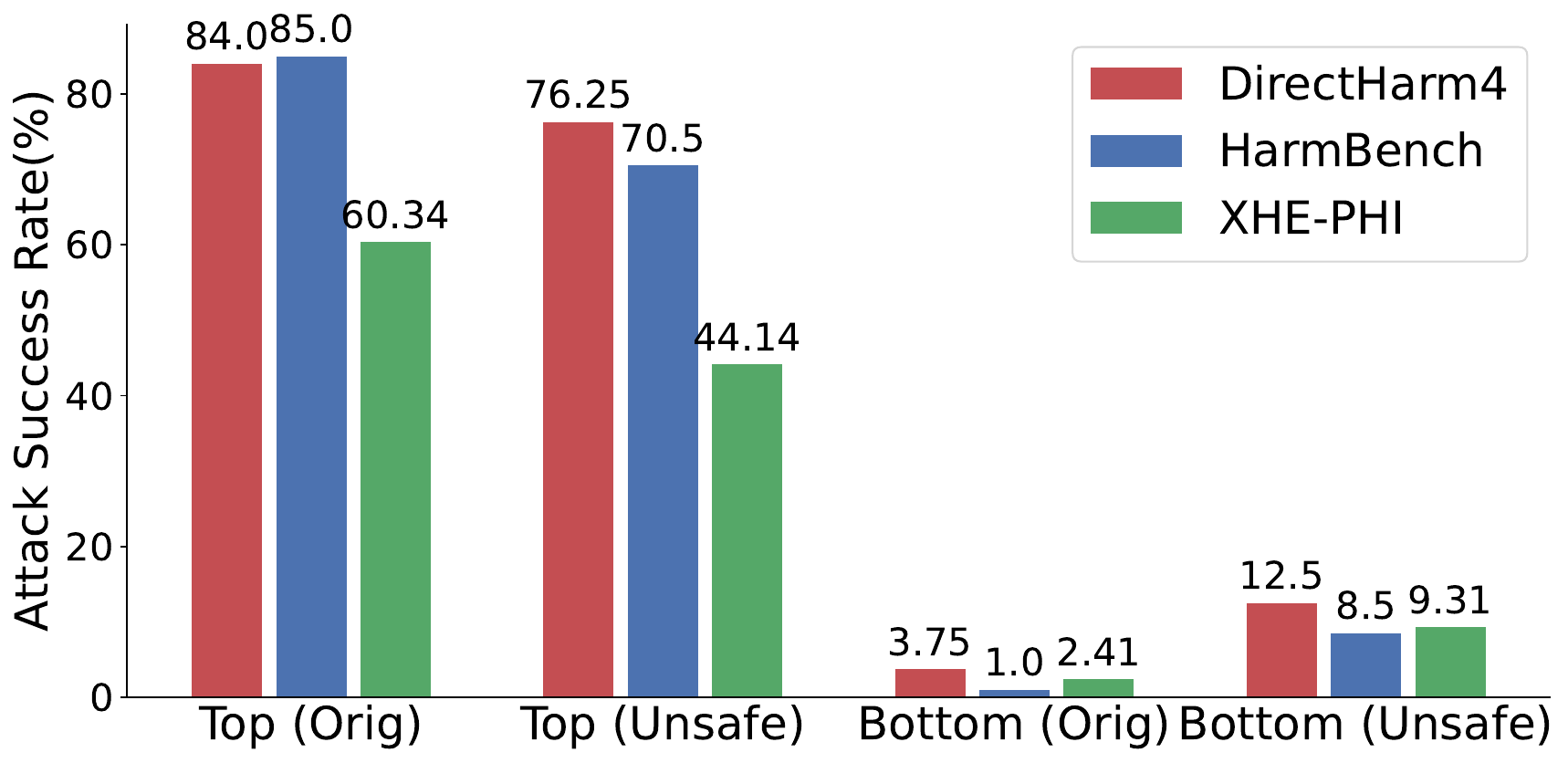}
    \caption{Dolly}
  \end{subfigure}
  \caption{ASR of the fine-tuned Llama3.1 on the top and bottom 1,000 samples ranked by the bidirectional method (Orig) and the unidirectional method (Unsafe) across three safety benchmarks.}
  \label{fig:llama3.1bi}
\end{figure*}

\begin{figure*}[t]
  \centering
  \begin{subfigure}[b]{0.49\linewidth}
    \centering
    \includegraphics[width=\linewidth]{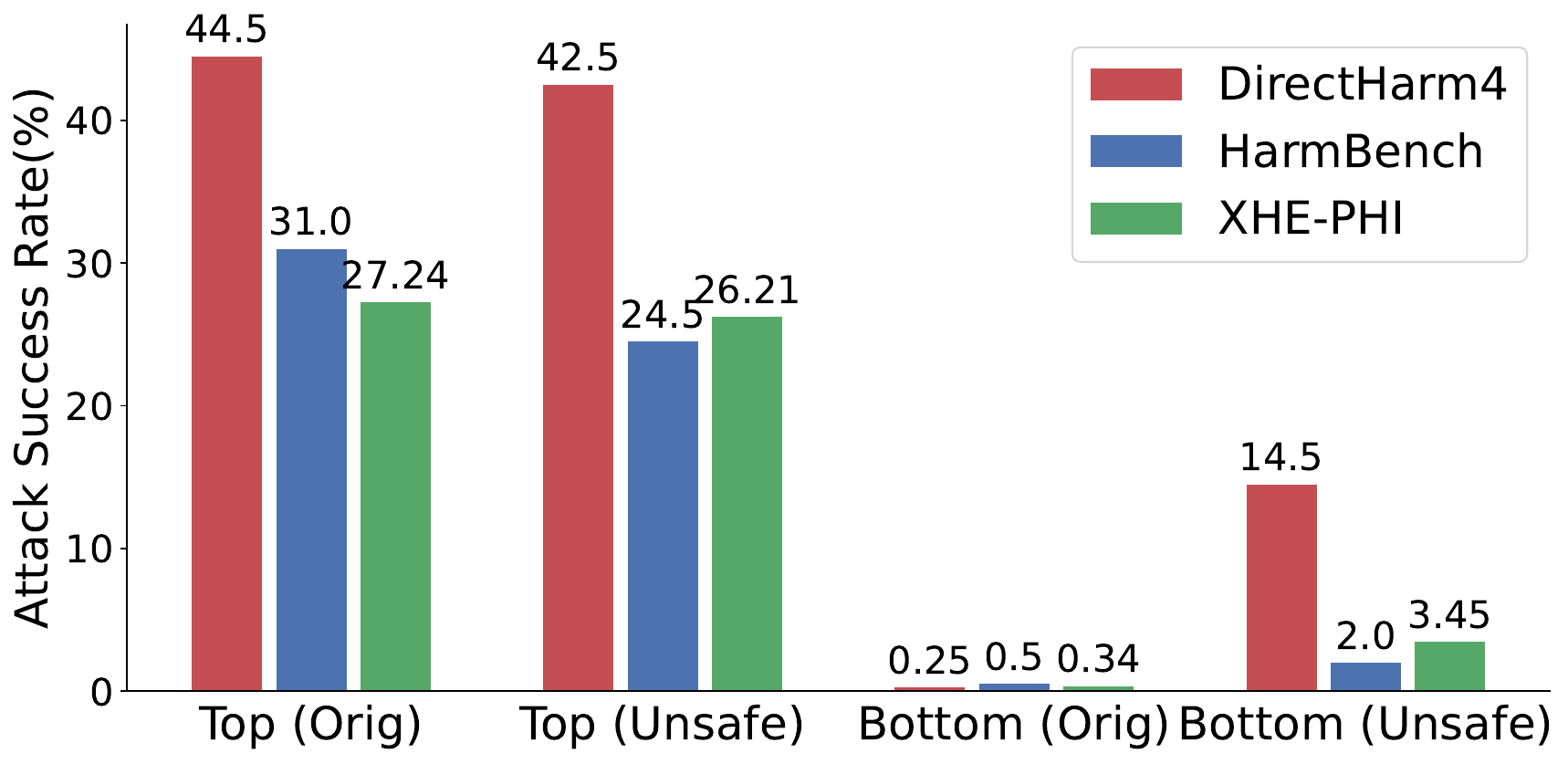}
    \caption{Alpaca}
  \end{subfigure}
  \hfill
  \begin{subfigure}[b]{0.49\linewidth}
    \centering
    \includegraphics[width=\linewidth]{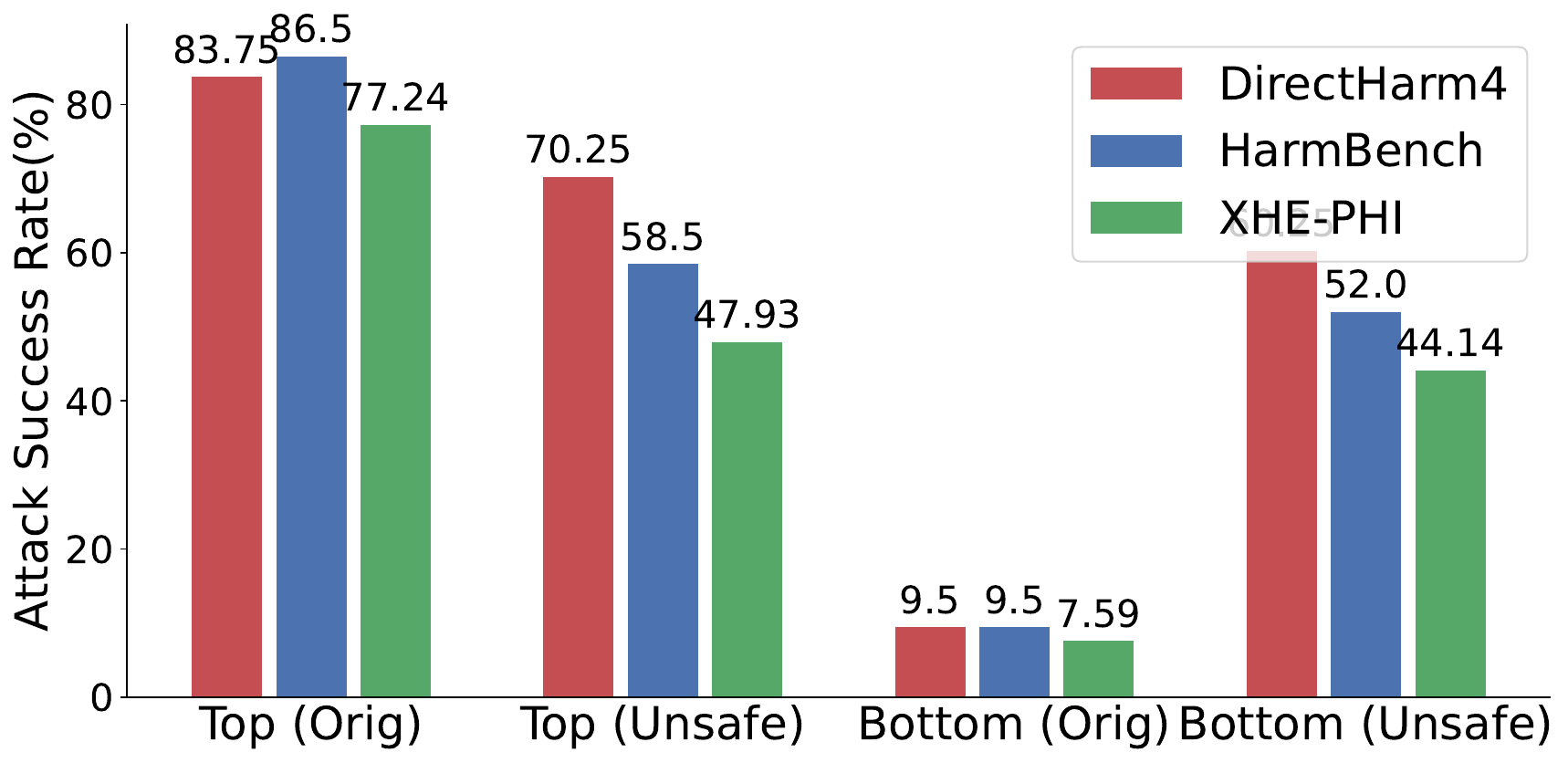}
    \caption{Dolly}
  \end{subfigure}
  \caption{ASR of the fine-tuned Qwen2.5 on the top and bottom 1,000 samples ranked by the bidirectional method (Orig) and the unidirectional method (Unsafe) across three safety benchmarks.}
  \label{fig:qwen2.5bi}
\end{figure*}

Figure \ref{fig:llama3bi}, Figure \ref{fig:llama3.1bi}, and Figure \ref{fig:qwen2.5bi} have shown the effectiveness of bidirectional representation data selection. The ASR of the fine-tuned model on the top 1,000 ranked samples from datasets selected by this method is lower than when using the bidirectional method. Meanwhile, the ASR for the bottom 1,000 ranked samples is significantly higher than the bidirectional.

\subsection{Safety-Sensitive Layer Selection}
\label{app:layerselection}
\begin{figure*}[t]
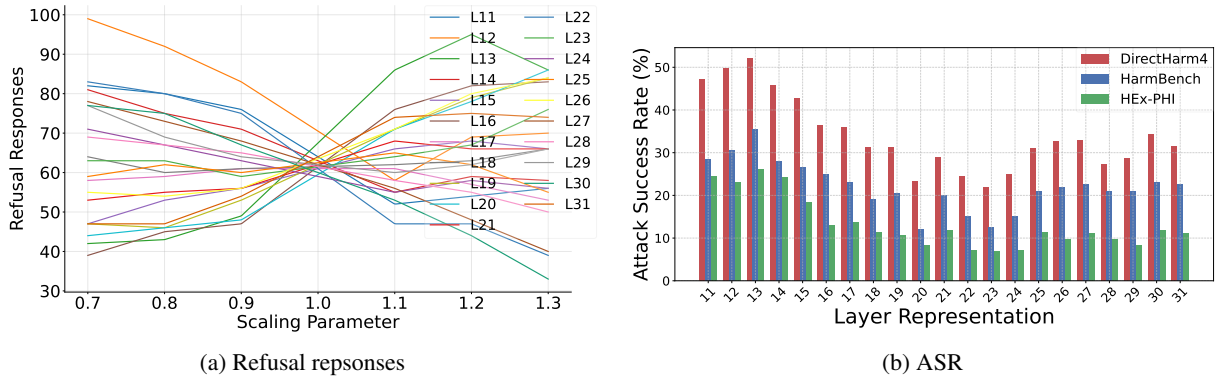

  \centering
  \begin{subfigure}[b]{0.49\linewidth}
    \centering
    \includegraphics[width=\linewidth]{emnlp2025-latex/figures/llama3_analysis.pdf}
    \caption{Refusal repsonses}
  \end{subfigure}
  \hfill
  \begin{subfigure}[b]{0.49\linewidth}
    \centering
    \includegraphics[width=\linewidth]{emnlp2025-latex/figures/llama3_asr_comparison.pdf}
    \caption{ASR}
  \end{subfigure}
  \caption{Llama 3: the 13th layer is the safety-sensitive layer.}
  \label{fig:llama3layer}
\end{figure*}

\begin{figure*}[t]
  \centering
  \begin{subfigure}[b]{0.49\linewidth}
    \centering
    \includegraphics[width=\linewidth]{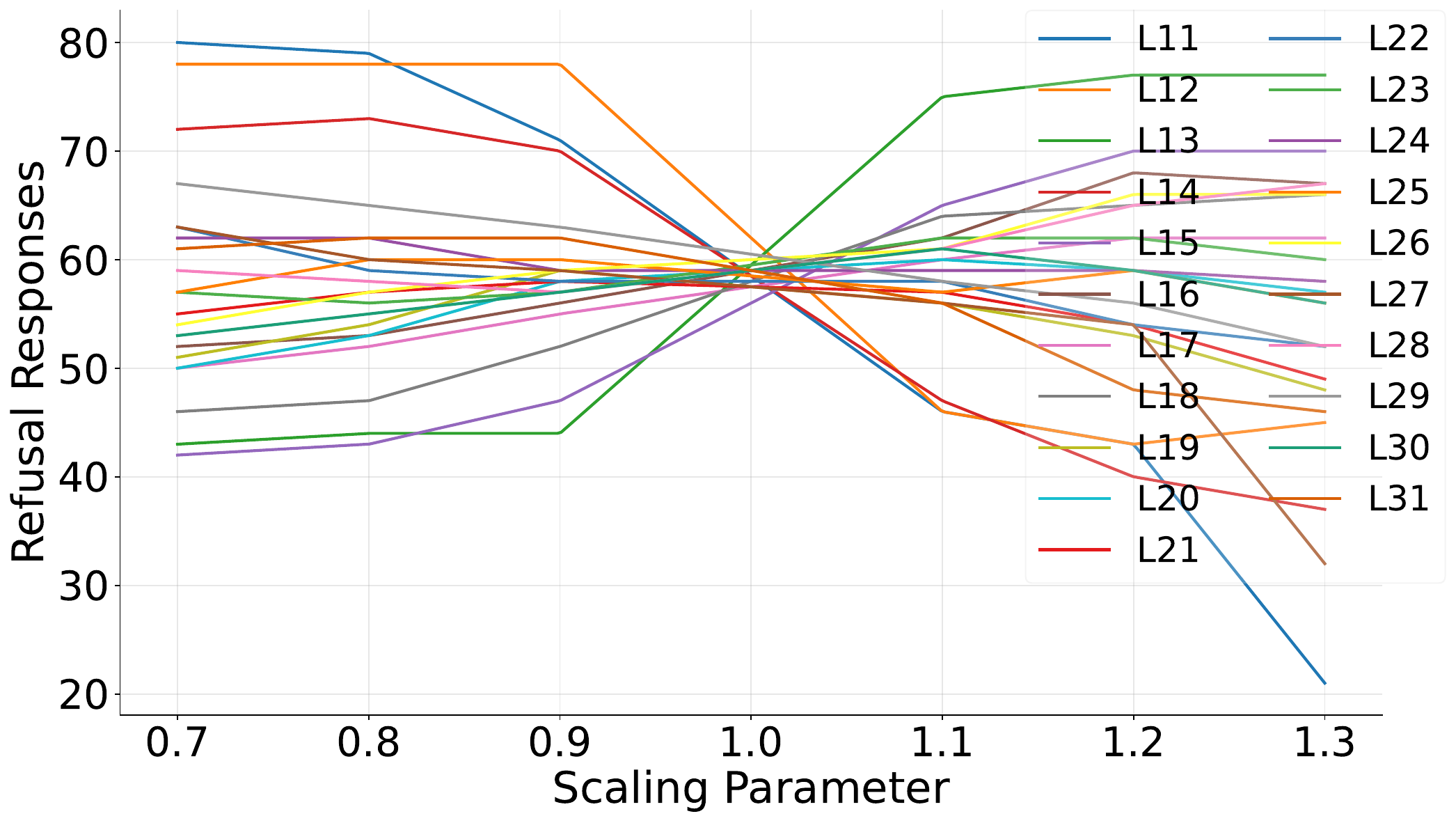}
    \caption{Refusal repsonses}
  \end{subfigure}
  \hfill
  \begin{subfigure}[b]{0.49\linewidth}
    \centering
    \includegraphics[width=\linewidth]{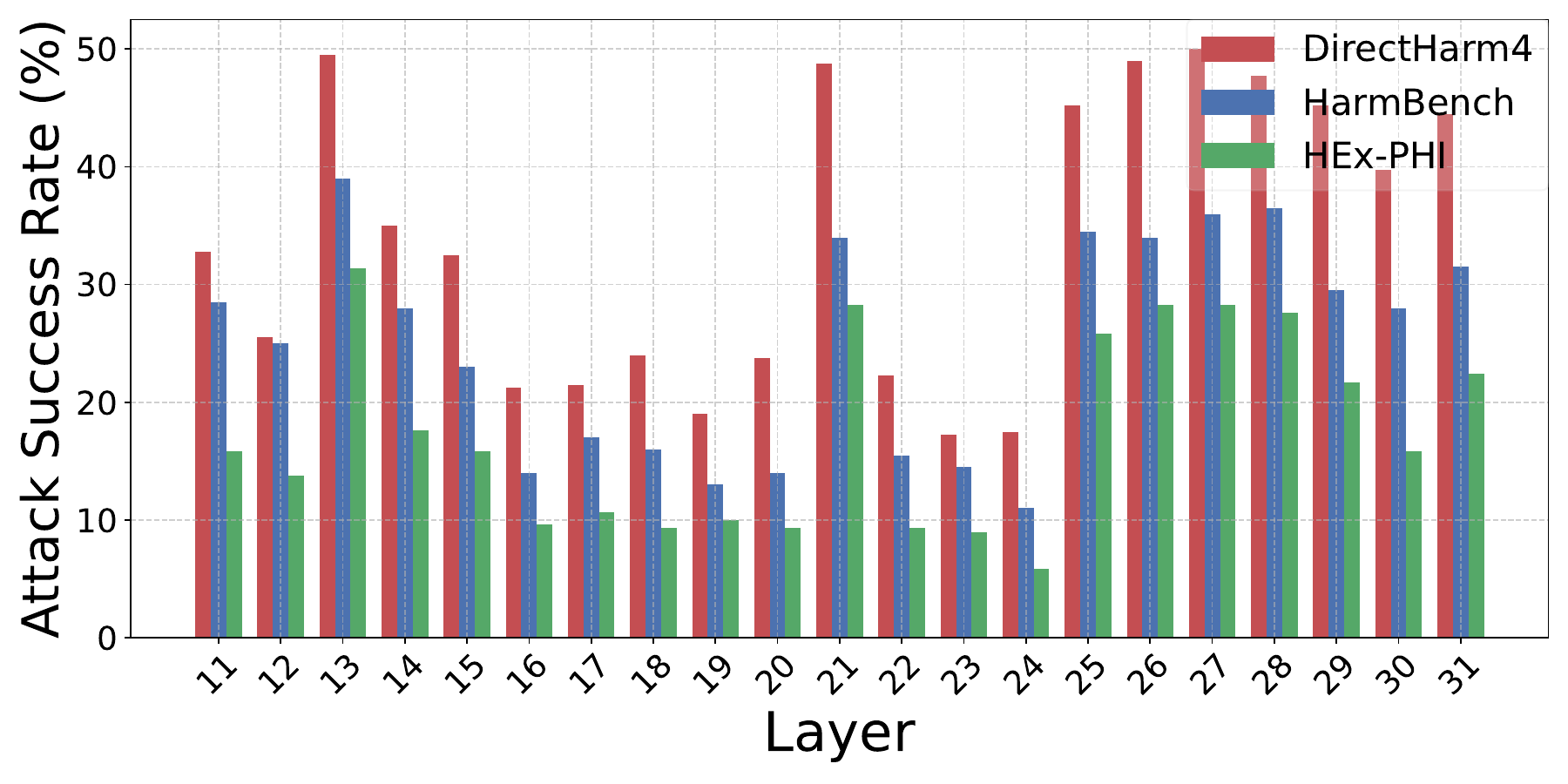}
    \caption{ASR}
  \end{subfigure}
  \caption{Llama 3.1: the 13th layer is the safety-sensitive layer.}
  \label{fig:llama3.1layer}
\end{figure*}

\begin{figure*}[t]
  \centering
  \begin{subfigure}[b]{0.49\linewidth}
    \centering
    \includegraphics[width=\linewidth]{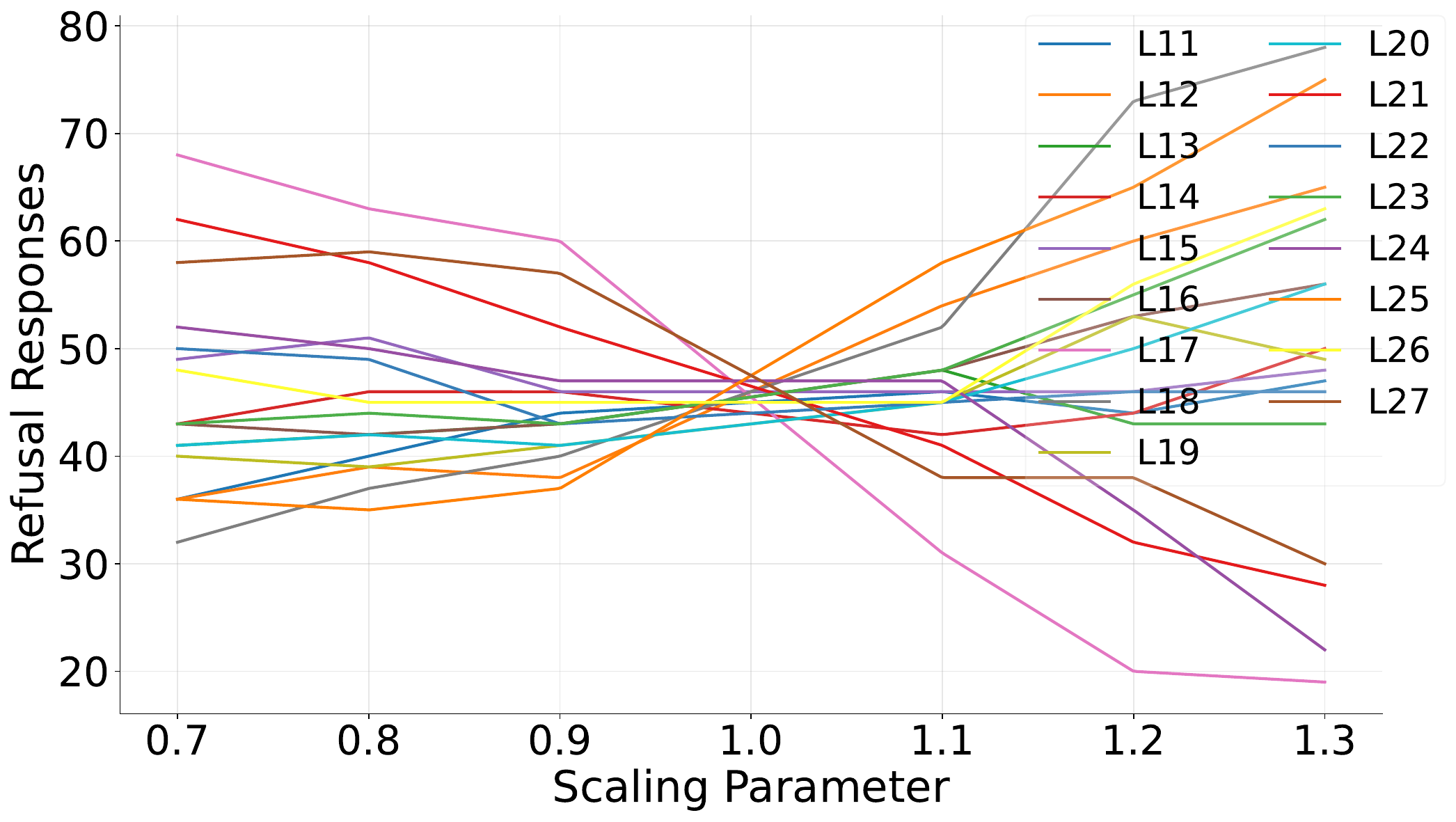}
    \caption{Refusal repsonses}
  \end{subfigure}
  \hfill
  \begin{subfigure}[b]{0.49\linewidth}
    \centering
    \includegraphics[width=\linewidth]{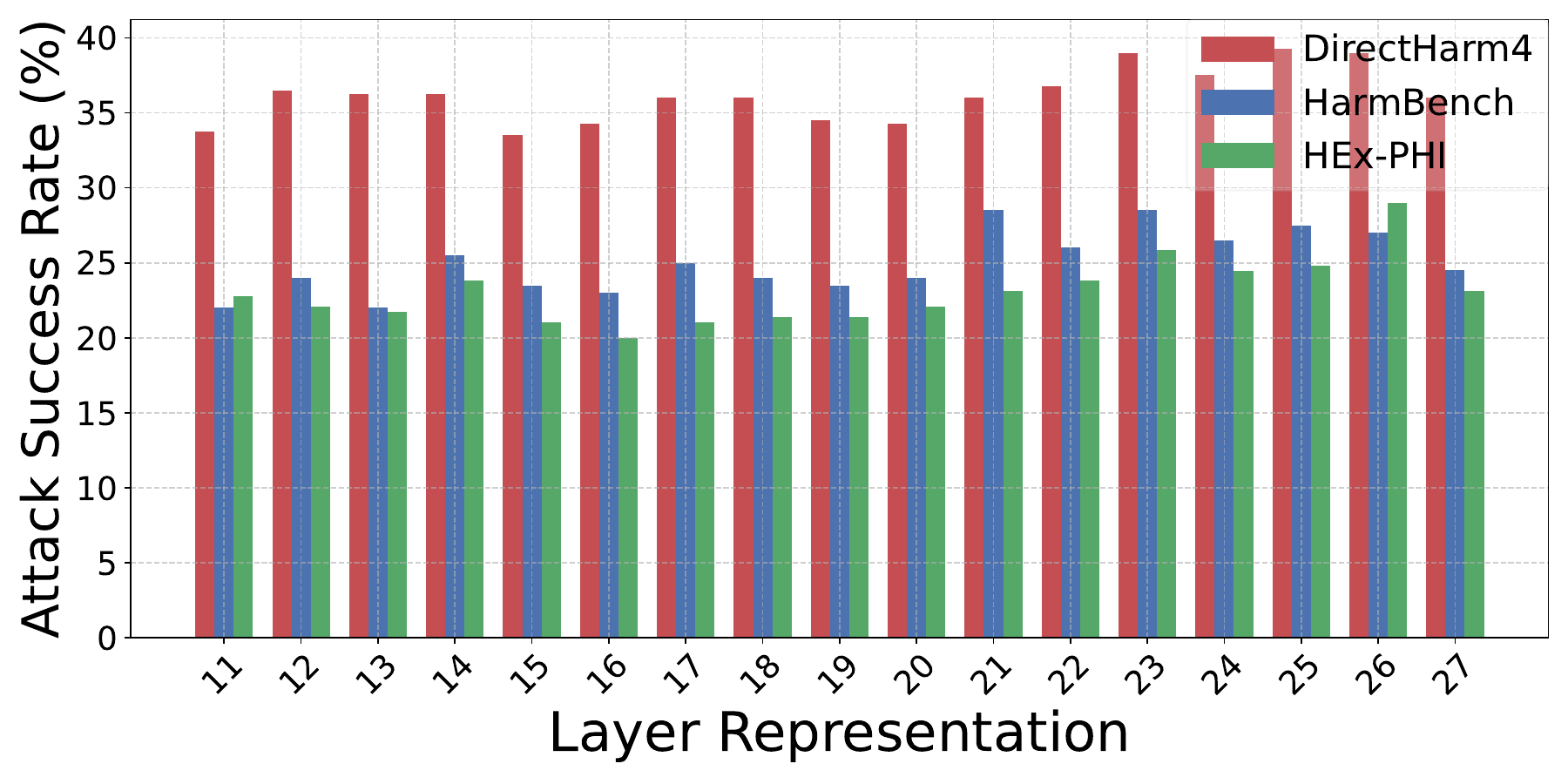}
    \caption{ASR}
  \end{subfigure}
  \caption{Qwen2: the 25th layer is the safety-sensitive layer.}
  \label{fig:qwen2layer}
\end{figure*}

\begin{figure*}[t]
  \centering
  \begin{subfigure}[b]{0.49\linewidth}
    \centering
    \includegraphics[width=\linewidth]{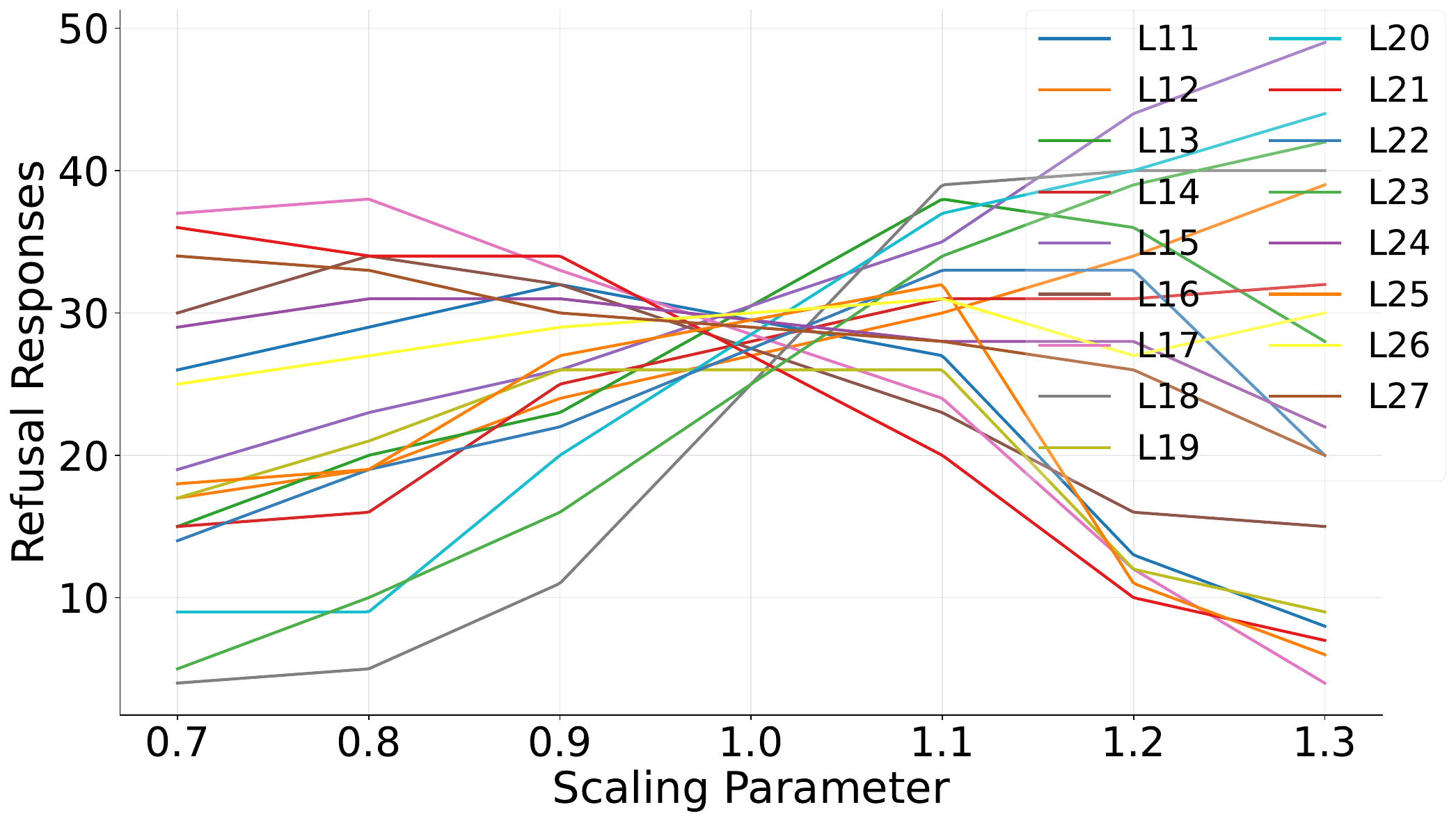}
    \caption{Refusal repsonses}
  \end{subfigure}
  \hfill
  \begin{subfigure}[b]{0.49\linewidth}
    \centering
    \includegraphics[width=\linewidth]{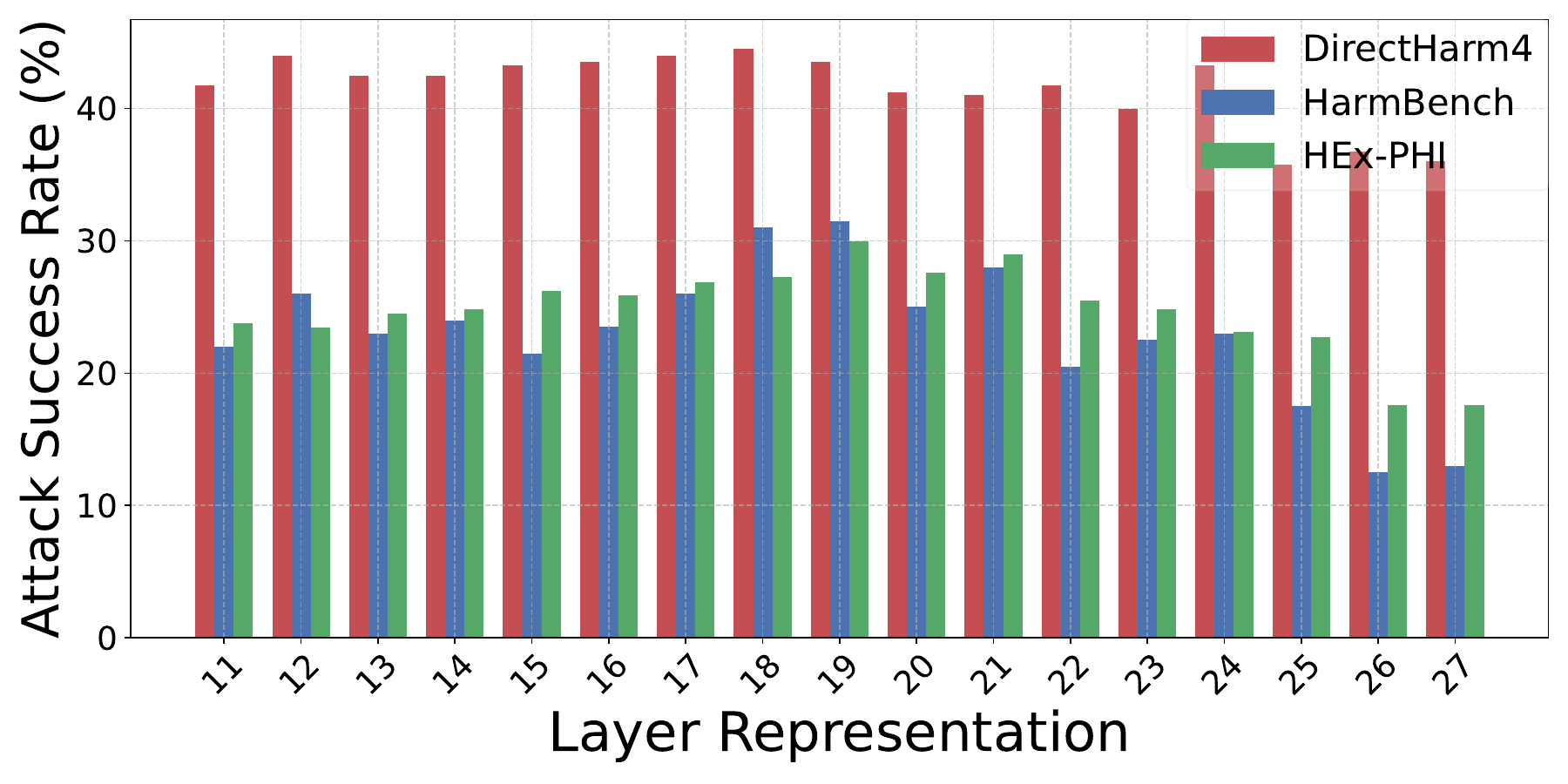}
    \caption{ASR}
  \end{subfigure}
  \caption{Qwen2.5: the 18th layer is the safety-sensitive layer.}
  \label{fig:qwenlayer}
\end{figure*}

\begin{figure*}[t]
  \centering
  \begin{subfigure}[b]{0.49\linewidth}
    \centering
    \includegraphics[width=\linewidth]{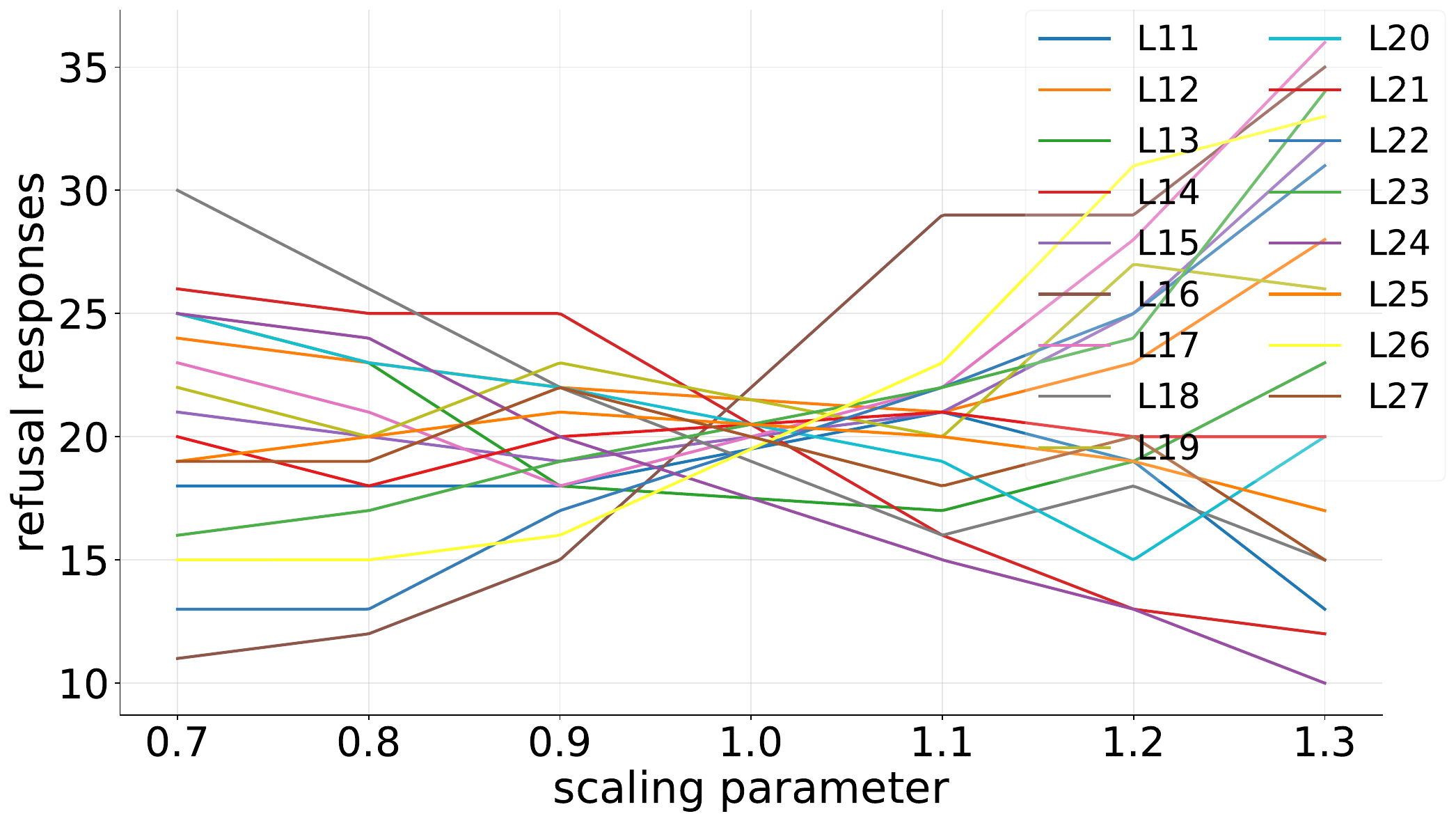}
    \caption{Refusal repsonses}
  \end{subfigure}
  \hfill
  \begin{subfigure}[b]{0.49\linewidth}
    \centering
    \includegraphics[width=\linewidth]{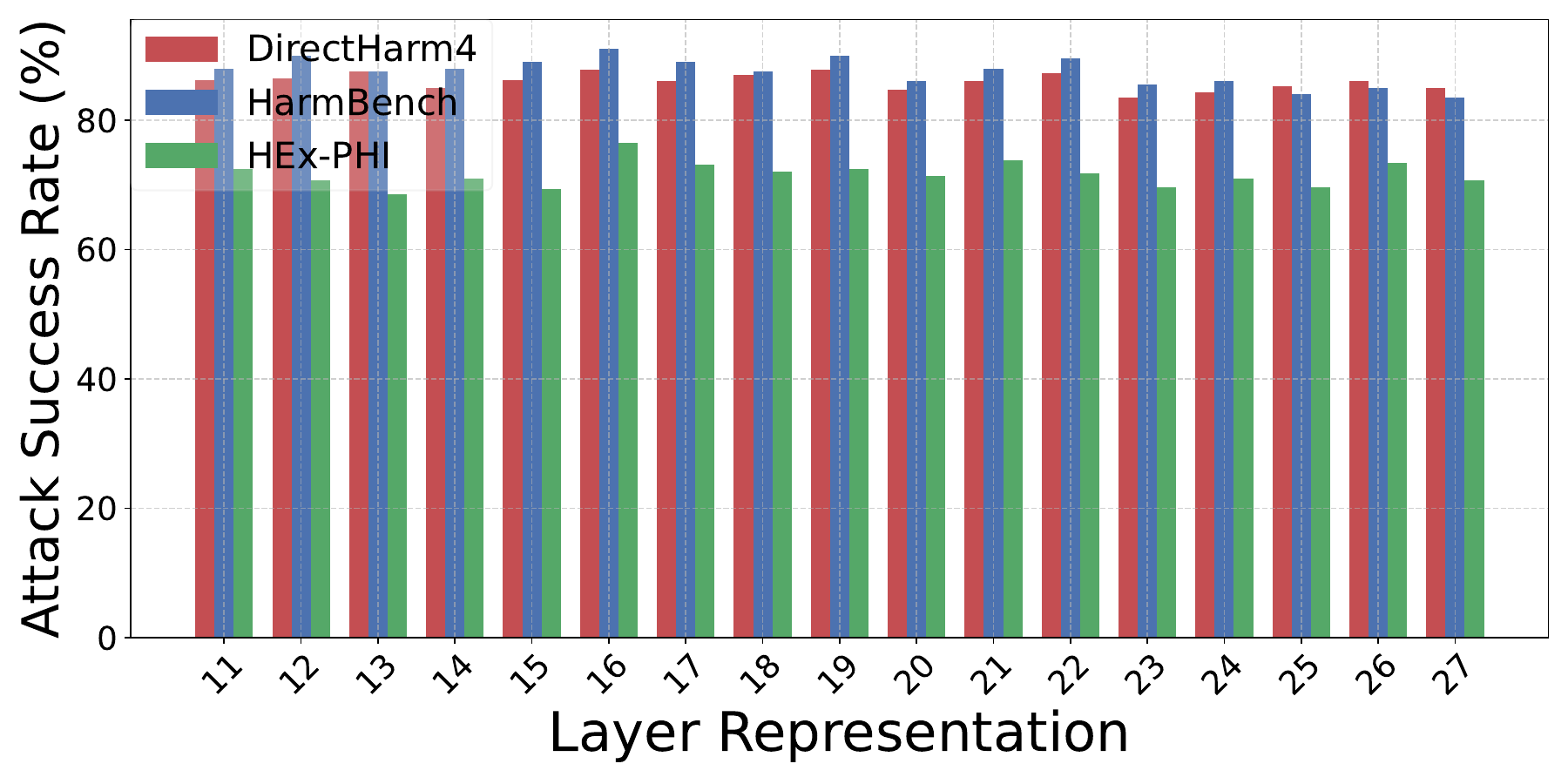}
    \caption{ASR}
  \end{subfigure}
  \caption{Mistral-v0.2: the 16th layer is the safety-sensitive layer.}
  \label{fig:mistrallayer}
\end{figure*}

\begin{figure*}[t]
  \centering
  \begin{subfigure}[b]{0.49\linewidth}
    \centering
    \includegraphics[width=\linewidth]{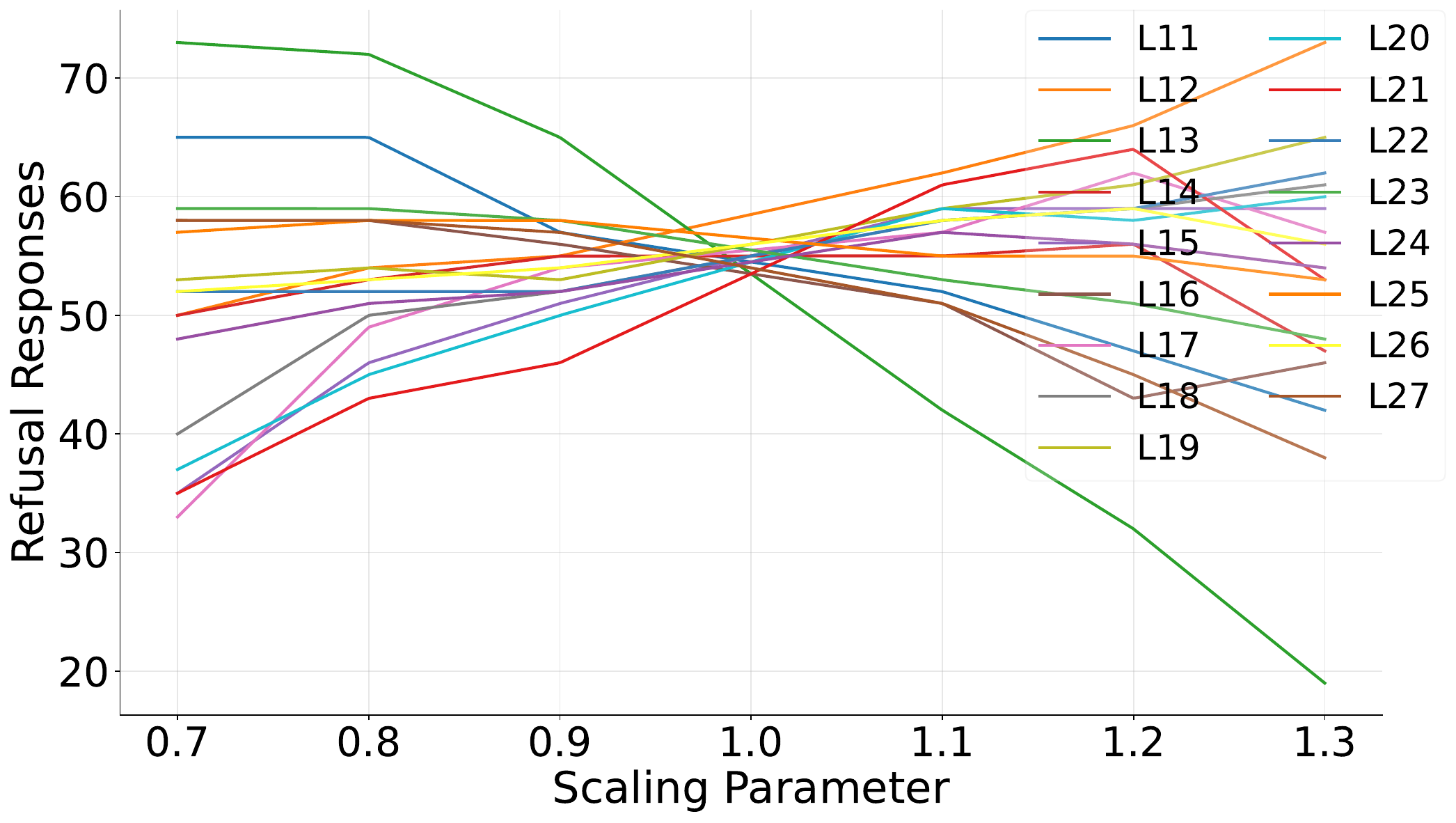}
    \caption{Refusal repsonses}
  \end{subfigure}
  \hfill
  \begin{subfigure}[b]{0.49\linewidth}
    \centering
    \includegraphics[width=\linewidth]{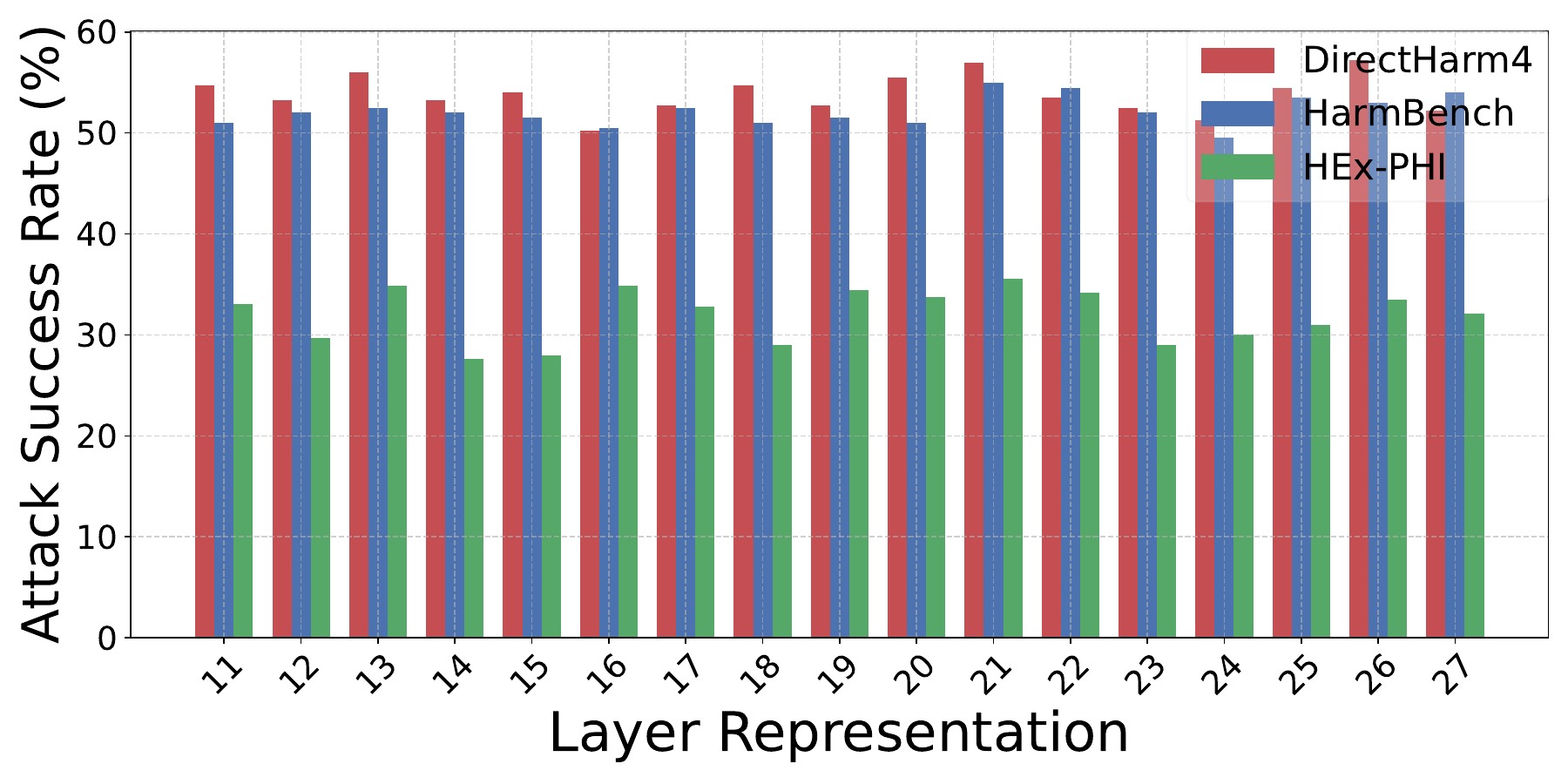}
    \caption{ASR}
  \end{subfigure}
  \caption{Phi-3-mini: the 21th layer is the safety-sensitive layer.}
  \label{fig:philayer}
\end{figure*}

\begin{itemize}
    \item \textbf{Llama3}: Figure \ref{fig:llama3layer} shows that the 13-th layer is the safety-sensitive layer of Llama3, with the highest normalized change rate $k = 370$.
    \item \textbf{Llama3.1}: Figure \ref{fig:llama3.1layer} shows that the 13-th layer is the safety-sensitive layer of Llama3.1, with the highest normalized change rate $k = 310$.
    \item \textbf{Qwen2}: Figure \ref{fig:qwen2layer} shows that the 25-th layer is the safety-sensitive layer of Qwen2, with the highest normalized change rate $k = 210$.
    \item \textbf{Qwen2.5}: Figure \ref{fig:qwenlayer} shows that the 18-th layer is the safety-sensitive layer of Qwen2.5, with the highest normalized change rate $k = 280$.
    \item \textbf{Mistral-v0.2}: Figure \ref{fig:mistrallayer} shows that the 16-th layer is the safety-sensitive layer of Mistral-v0.2, with the highest normalized change rate $k = 140$.
    \item \textbf{Phi-3-mini}: Figure \ref{fig:philayer} shows that the 21-st layer is the safety-sensitive layer of Phi-3-mini, with the highest normalized change rate $k = 150$.
\end{itemize}

\subsection{The Transferability of LARF}
\label{app:trans}
On the PubMedQA dataset, we fine-tuned the larger-capacity Llama3-70B-Instruct,  Qwen2.5-32B-Instruct and Qwen2.5-72B-Instruct. We compare our method against random sampling. Table \ref{tab:trans} shows that our approach consistently achieves lower GPT Scores and reduced ASR.
\begin{table}[ht]
\centering
\begin{tabular}{c c c}
\hline
\textbf{Model} & \textbf{Random} & \textbf{LARF} \\
\hline
Llama3-70B   & 3.47$_{56.50}$ & \textbf{3.44$_{55.75}$} \\
Qwen2.5-32B  & 3.58$_{36.50}$ & \textbf{3.54$_{36.00}$} \\
Qwen2.5-72B  & 3.09$_{26.25}$ & \textbf{2.92$_{20.25}$} \\
\hline
\end{tabular}
\caption{Performance comparison between Random sampling and LARF on the PubMedQA dataset for Llama3-70B, Qwen2.5-32B, and Qwen2.5-72B. Entries report Score$_{\text{ASR}}$ (mean harmfulness; lower is better). LARF consistently achieves lower GPT Scores and reduced ASR across all models.}
\label{tab:trans}
\end{table}

\section{Analysis}
\subsection{Data Character Analysis}
\label{app:data_analysis}
Table \ref{tab:data_ana_llama3} reports the results for Llama3, Table \ref{tab:data_ana_llama3.1} for Llama3.1, and Table \ref{tab:data_ana_qwen} for Qwen2.5. The fine-tuning datasets include Alpaca, Dolly, Magicoder, PubMedQA, and MetaMath. In all cases, the top 1,000 ranked examples exhibit both point-style counts and response token lengths above the dataset average, whereas the bottom 1,000 ranked examples fall below average--demonstrating that point-by-point and longer responses compromise the safety alignment.
\input{emnlp2025-latex/sec/data_analysis}

\subsection{Similarity Heatmap Analysis}
\label{app:heatmap}
\begin{figure*}[t]
    \centering
    \begin{subfigure}[b]{0.32\linewidth}
        \includegraphics[width=\linewidth]{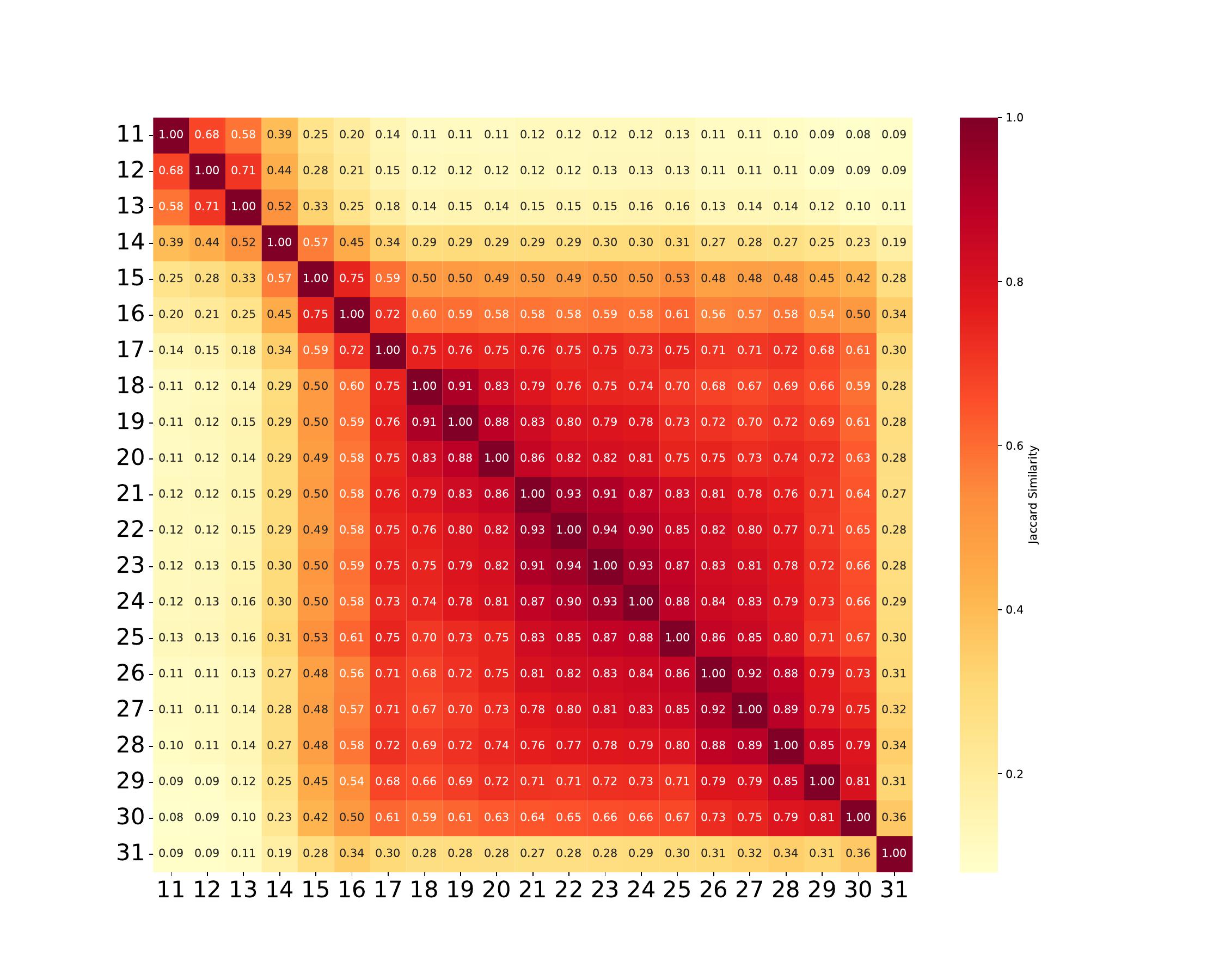}
        \caption{Llama3}
        \label{fig:llama3_heatmap}
    \end{subfigure}
    \hfill
    \begin{subfigure}[b]{0.32\linewidth}
        \includegraphics[width=\linewidth]{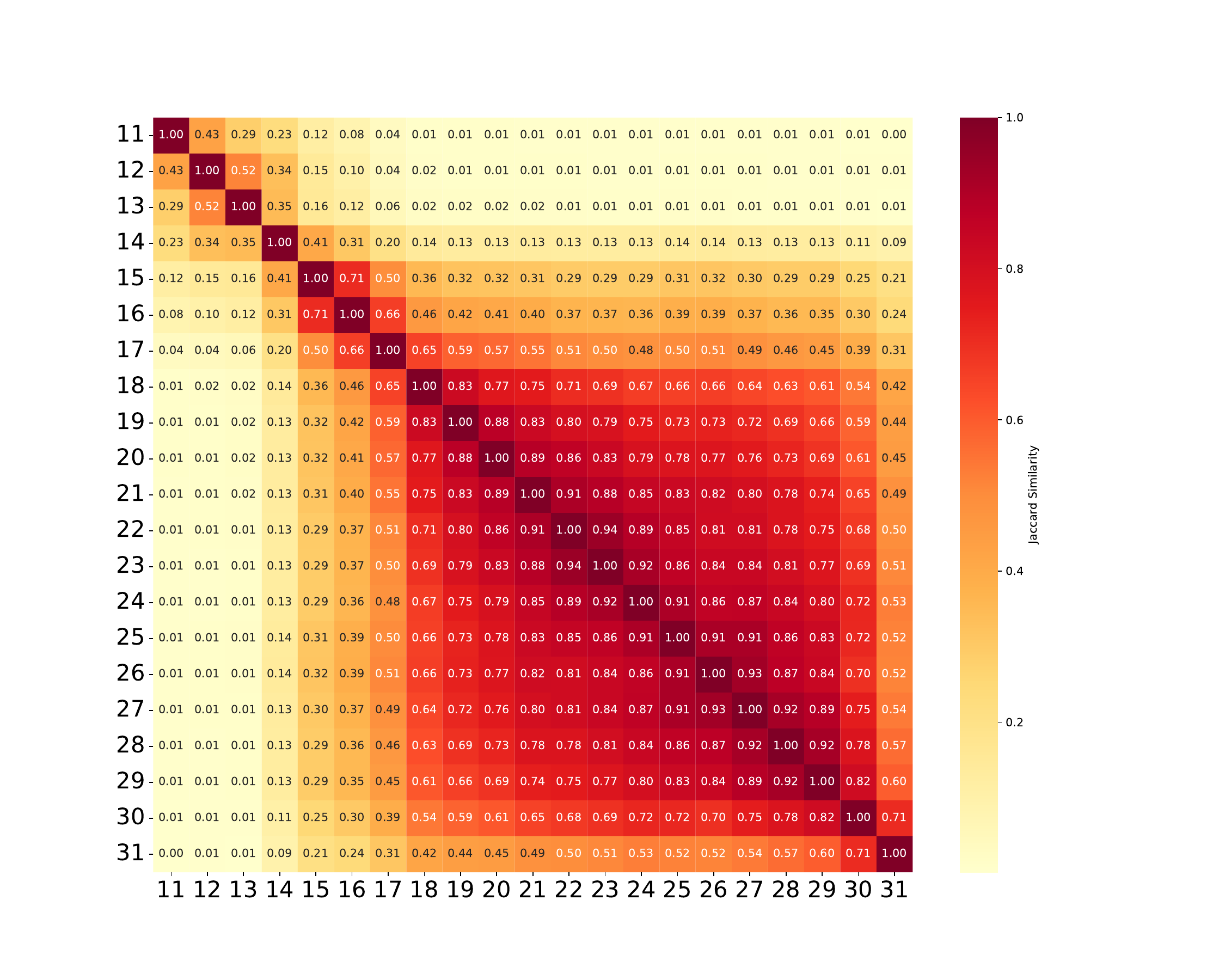}
        \caption{Llama3.1}
        \label{fig:llama3.1_heatmap}
    \end{subfigure}
    \hfill
    \begin{subfigure}[b]{0.32\linewidth}
        \includegraphics[width=\linewidth]{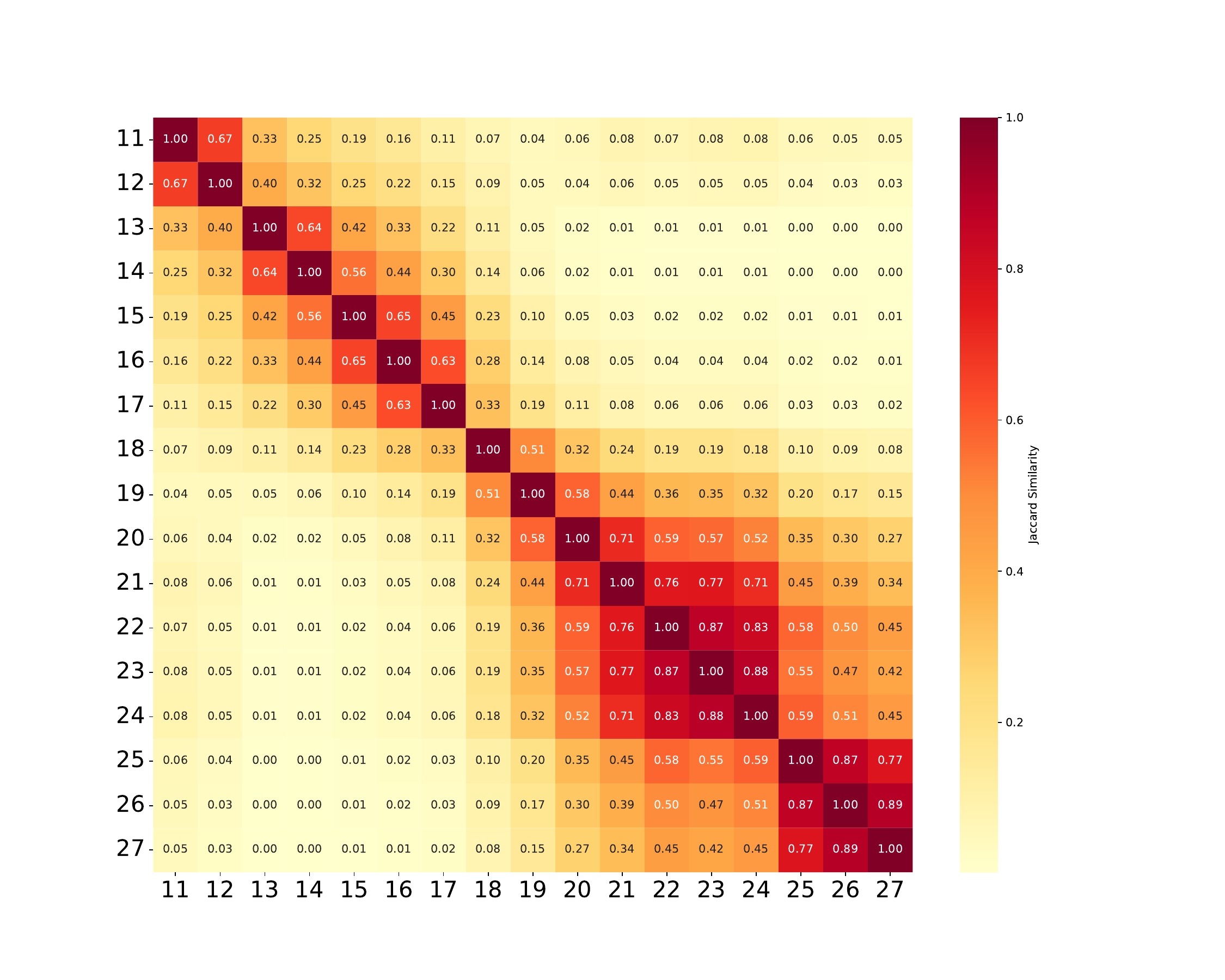}
        \caption{Qwen2.5}
        \label{fig:qwen_heatmap}
    \end{subfigure}
    \caption{Pairwise cosine similarity heatmaps of the top-1,000 samples selected by each layer of (a) Llama3, (b) Llama3.1, and (c) Qwen2.5. In each model, the safety-sensitive layer’s selections form a distinct block in the corner (indicating low similarity with other layers) while deeper layers show progressively higher intra-layer similarity, reflecting convergence in safety-related feature extraction.}
    \label{fig:combined_heatmap}
\end{figure*}

We also compute the Jaccard similarity among the 1,000 top ranked data points selected by LARF at each layer, defined by the equation:
\[
J(A,B) = \frac{|A \cap B|}{|A \cup B|}
\]
We visualize the pairwise similarity of the selected samples across layers using a heatmap. Figure \ref{fig:llama3_heatmap}, Figure \ref{fig:llama3.1_heatmap}, and Figure \ref{fig:qwen_heatmap} reveal that the data selected by the safety-sensitive layers consistently cluster in the corner of a square region, indicating lower similarity with samples from other layers. Furthermore, as the layer depth increases, the data selected by deeper layers exhibit progressively higher similarity, suggesting convergence in safety feature extraction.

\subsection{Representation Analysis}
\begin{figure*}[t]
  \centering
  \begin{subfigure}[b]{0.32\linewidth}
    \centering
    \includegraphics[width=\linewidth]{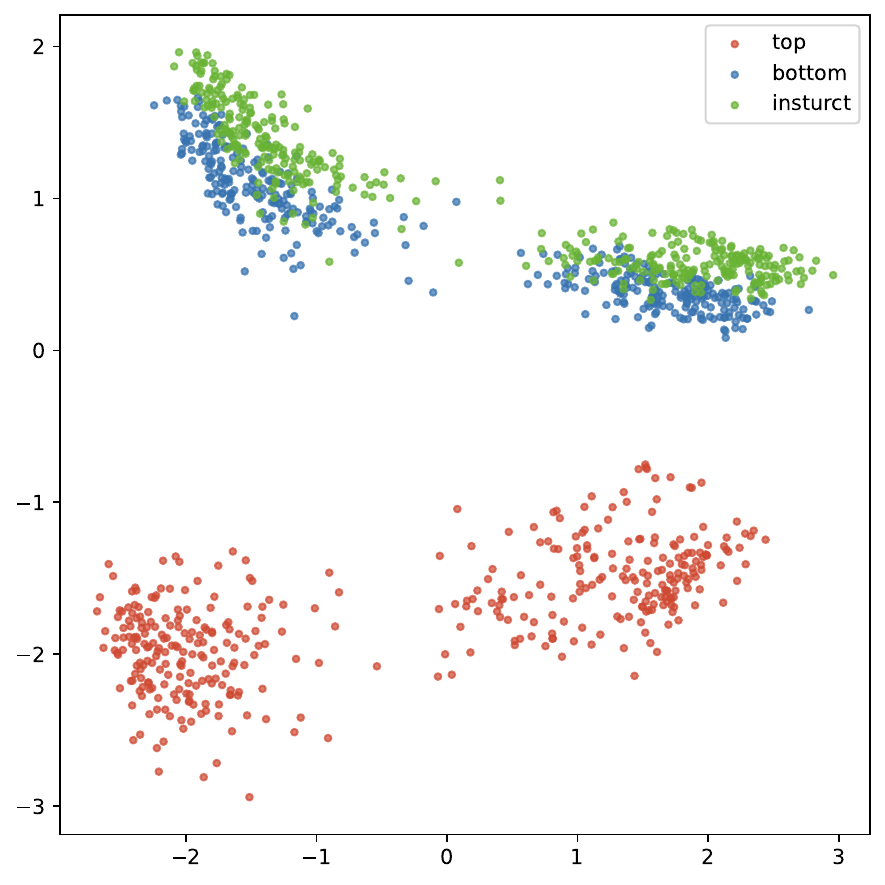}
    \caption{Llama 3}
  \end{subfigure}
  \begin{subfigure}[b]{0.32\linewidth}
    \centering
    \includegraphics[width=\linewidth]{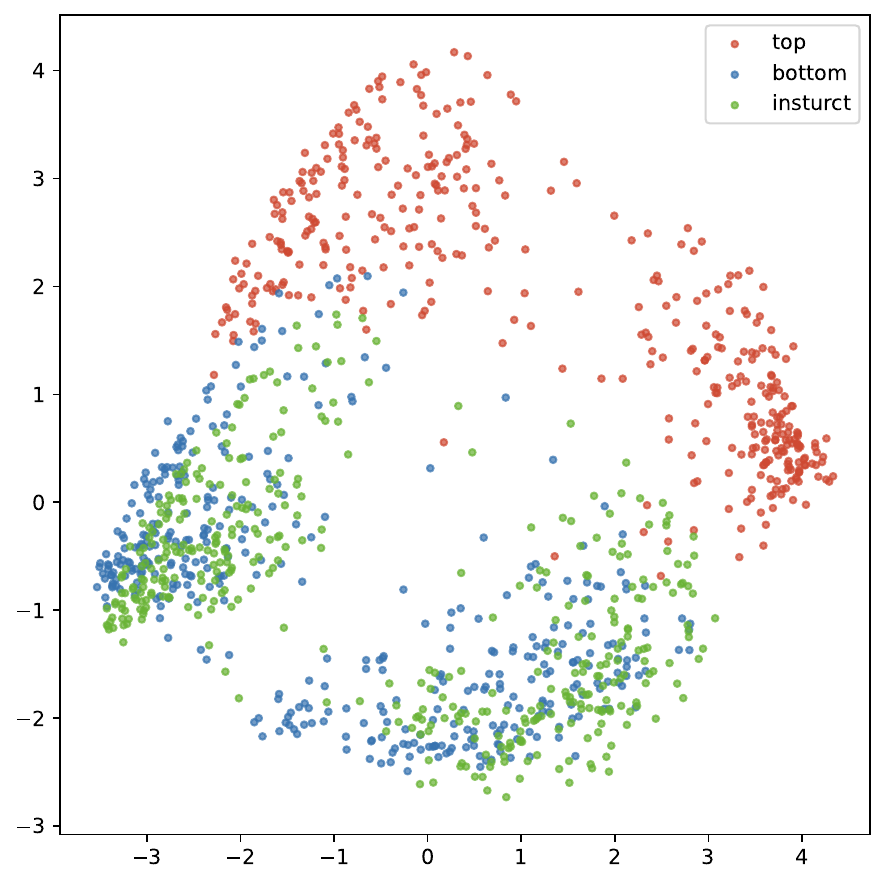}
    \caption{Llama 3.1}
  \end{subfigure}
  \hfill
  \begin{subfigure}[b]{0.32\linewidth}
    \centering
    \includegraphics[width=\linewidth]{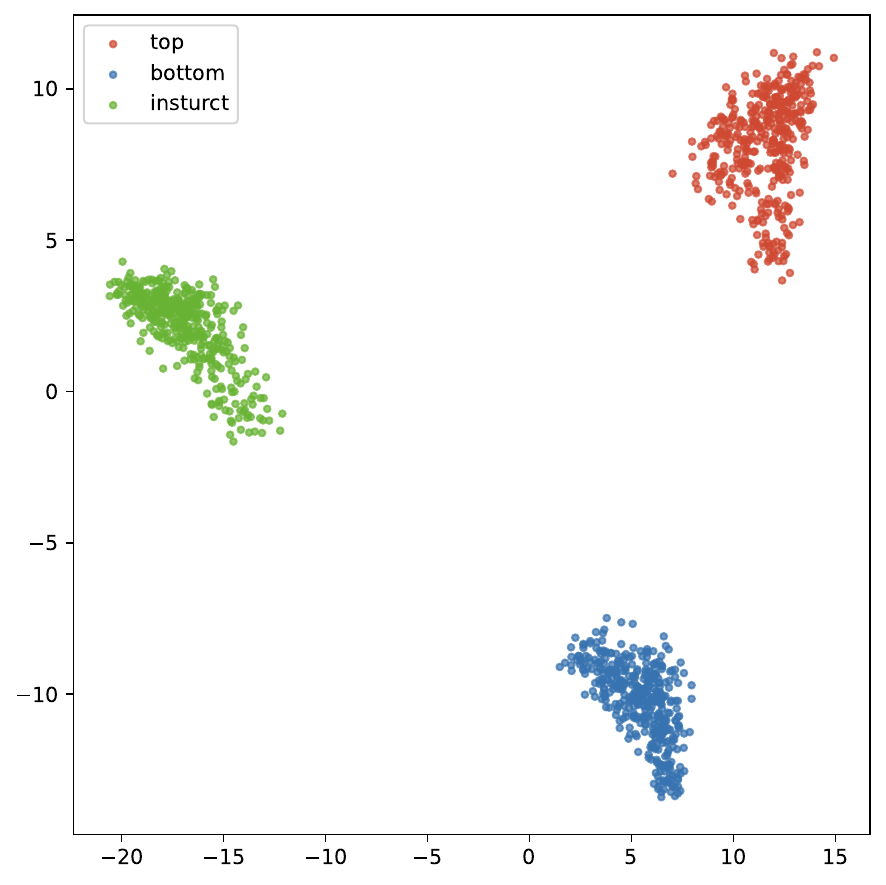}
    \caption{Qwen 2.5}
  \end{subfigure}
  \caption{Principal component analysis of safety-sensitive layer representations on DirectHarm4 for (a) Llama3, (b) Llama3.1, and (c) Qwen2.5. Each plot overlays the instruction-tuned baseline (green) with models fine-tuned on the bottom 1,000 (blue) and top 1,000 (red) ranked samples. For Llama series models, bottom 1,000 fine-tuned variants remain closely clustered with the baseline, whereas top 1,000 variants diverge substantially, indicating greater representational drift and potential degradation in safety alignment. For Qwen2.5, the bottom-1,000 fine-tuned variants also deviate from the instruction baseline, likely due to a distribution mismatch between the fine-tuning data and the original model.}
  \label{fig:pca_rep}
\end{figure*}

We provide PCA visualizations of the safety‐sensitive‐layer representations on DirectHarm4 for three model variants: the instruction‐tuned baseline and models fine‐tuned on the bottom and top 1,000 ranked samples. Figure \ref{fig:pca_rep} shows these projections for (a) Llama 3, (b) Llama 3.1, and (c) Qwen 2.5, highlighting that top‐ranked fine‐tuning induces a pronounced representational drift away from the baseline (especially in the Llama series) whereas bottom‐ranked fine‐tuning remains closely clustered.

\subsection{Effective Rank Analysis}
\label{app:effective_rank}
\begin{figure*}[t]
    \centering
    \begin{subfigure}[b]{0.32\linewidth}
        \includegraphics[width=\linewidth]{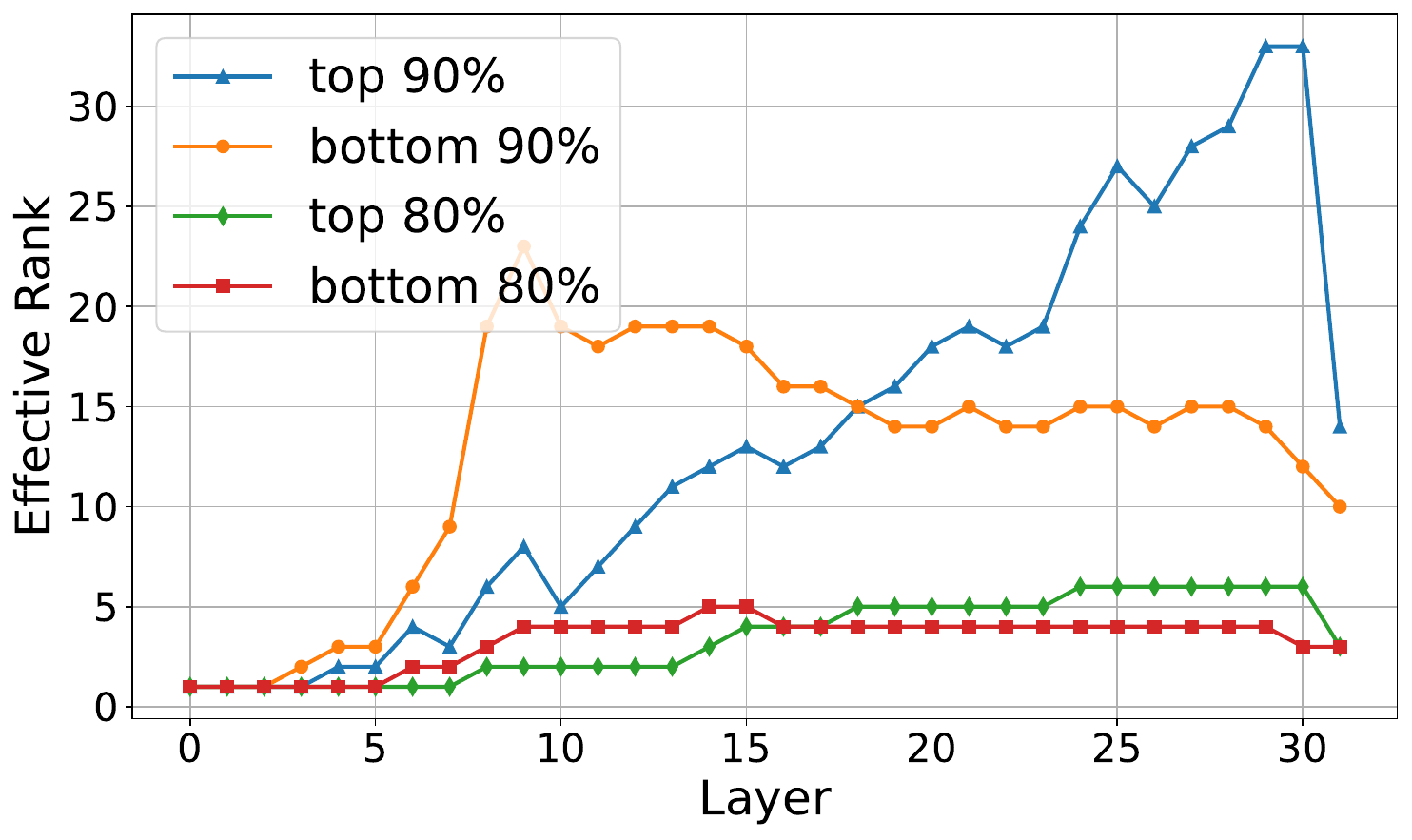}
        \caption{Llama3}
        \label{fig:llama3_effective}
    \end{subfigure}
    \hfill
    \begin{subfigure}[b]{0.32\linewidth}
        \includegraphics[width=\linewidth]{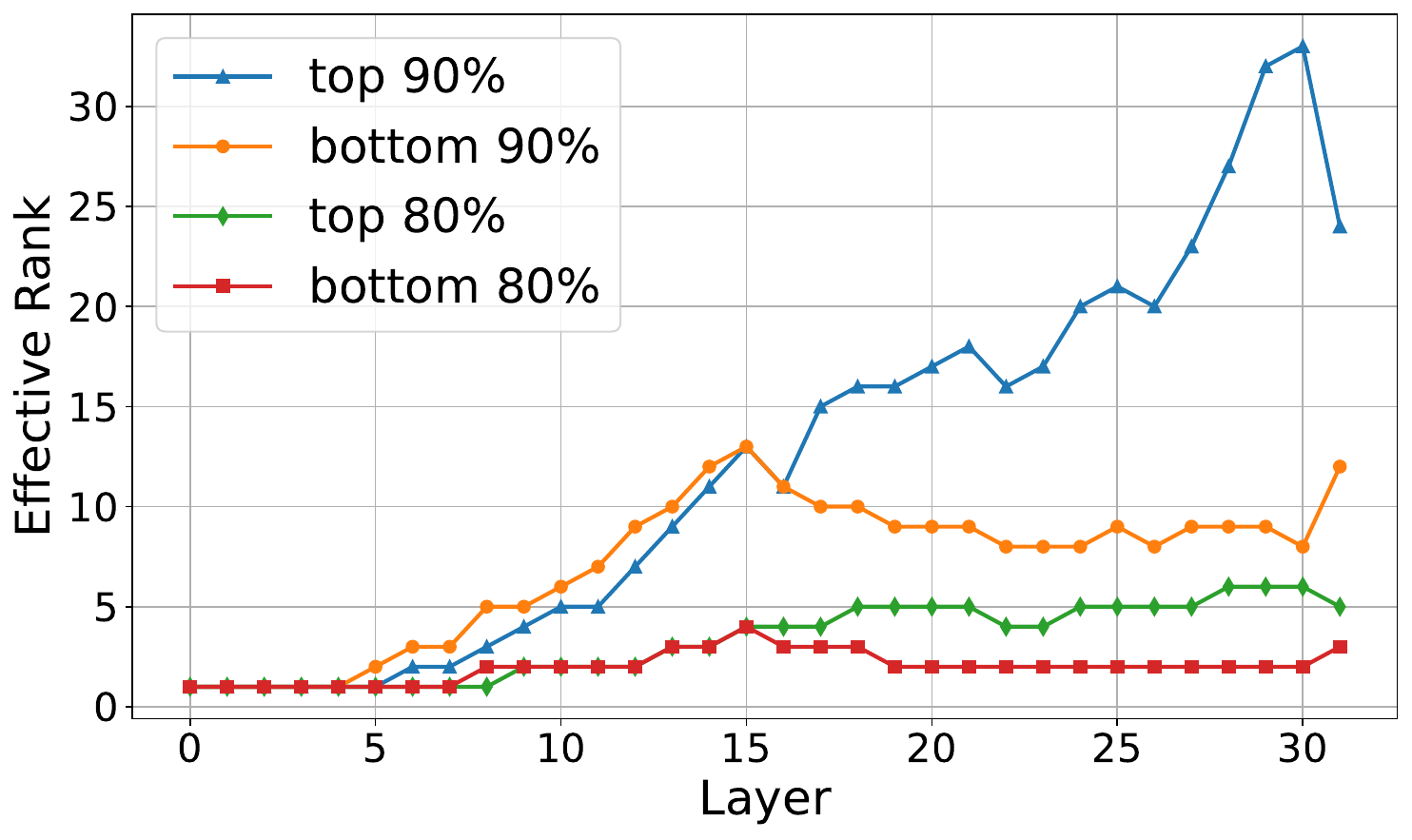}
        \caption{Llama3.1}
        \label{fig:llama3.1_effective}
    \end{subfigure}
    \hfill
    \begin{subfigure}[b]{0.32\linewidth}
        \includegraphics[width=\linewidth]{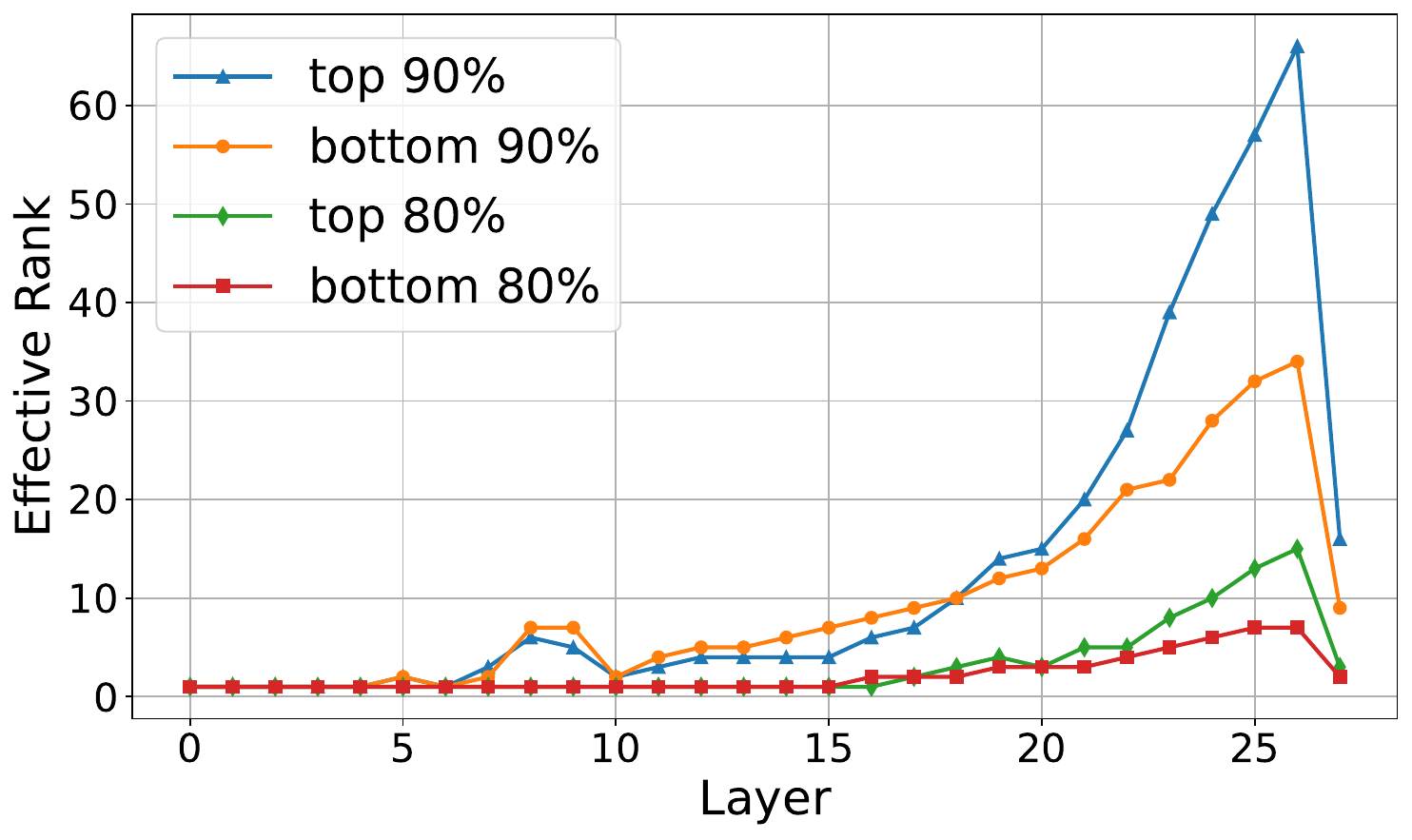}
        \caption{Qwen2.5}
        \label{fig:qwen_effective}
    \end{subfigure}
    \caption{Effective rank of the transformation matrix $W$ at each layer for models fine-tuned on the top-1,000 versus bottom-1,000 samples from DirectHarm4. (a) Llama3, (b) Llama3.1, and (c) Qwen2.5. In all three models, fine-tuning on the top-1,000 harmful examples yields progressively higher effective rank with increasing depth compared to bottom-1,000 tuning, indicating more diverse representation directions and potential degradation in safety alignment.}
    \label{fig:combined_effective_rank}
\end{figure*}

We further investigate the differential impacts of fine-tuning with the top 1,000 ranked data points versus the bottom 1,000 ranked on model representation. For each layer's representations on the DirectHarm4 dataset, we computed both the transformation matrix $W$ 
\citep{pan2025hidden} and its effective rank \citep{roy2007effective}. Figure~\ref{fig:llama3_effective}, Figure~\ref{fig:llama3.1_effective}, and Figure~\ref{fig:qwen_effective} reveal that models fine-tuned on the top 1,000 data points exhibit progressively higher effective rank compared to bottom-1,000-tuned models as layer depth increases. This suggests that top-1,000 fine-tuning produces more diverse representation directions when processing harmful instructions, compromising the model's safety alignment.

\subsection{Category Analysis}
\label{app:category}
\begin{figure*}[t]
  \centering
  \begin{subfigure}[b]{0.32\linewidth}
    \centering
    \includegraphics[width=\linewidth]{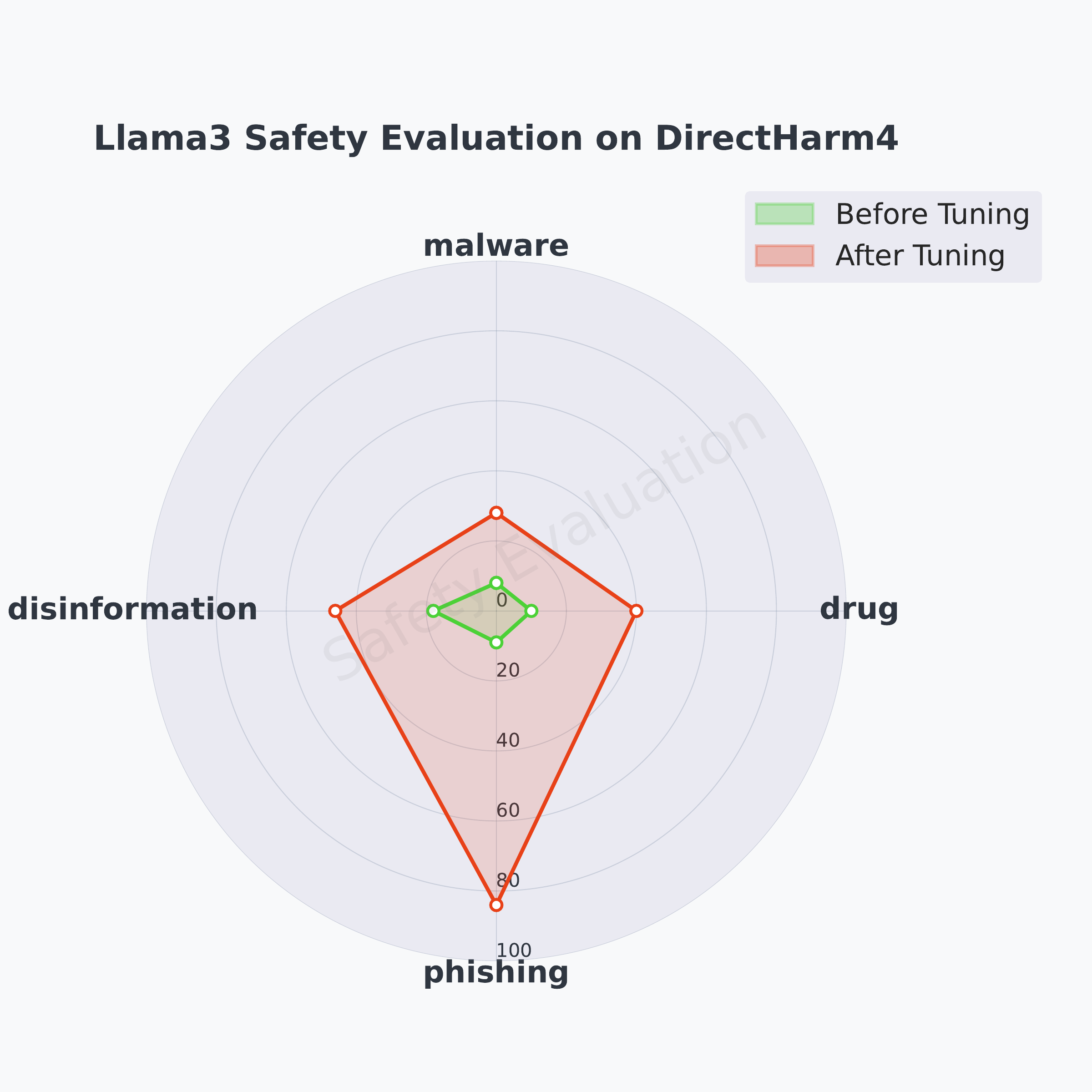}
    \caption{DirectHarm4}
  \end{subfigure}
  \hfill
  \begin{subfigure}[b]{0.32\linewidth}
    \centering
    \includegraphics[width=\linewidth]{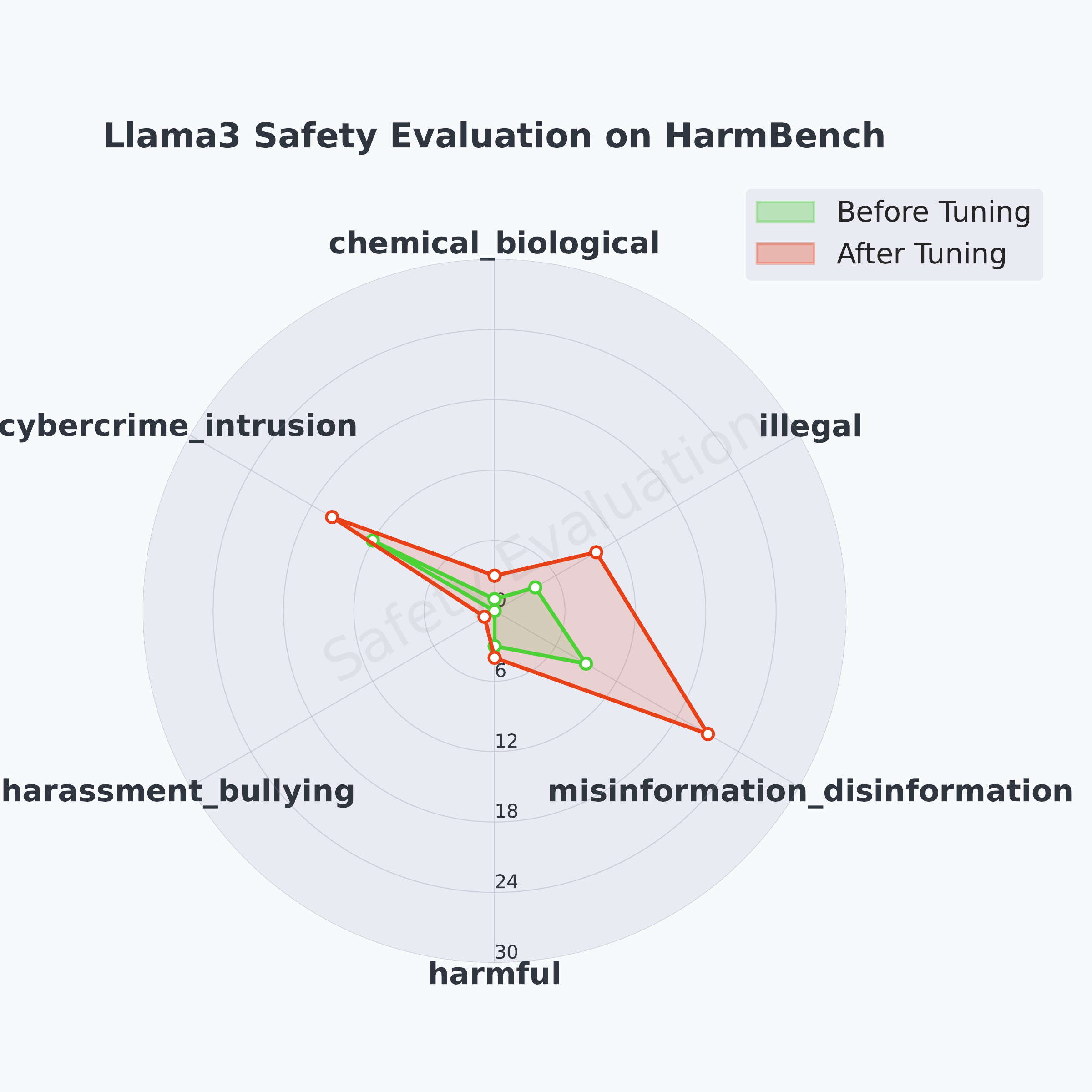}
    \caption{HarmBench}
  \end{subfigure}
  \hfill
  \begin{subfigure}[b]{0.32\linewidth}
    \centering
    \includegraphics[width=\linewidth]{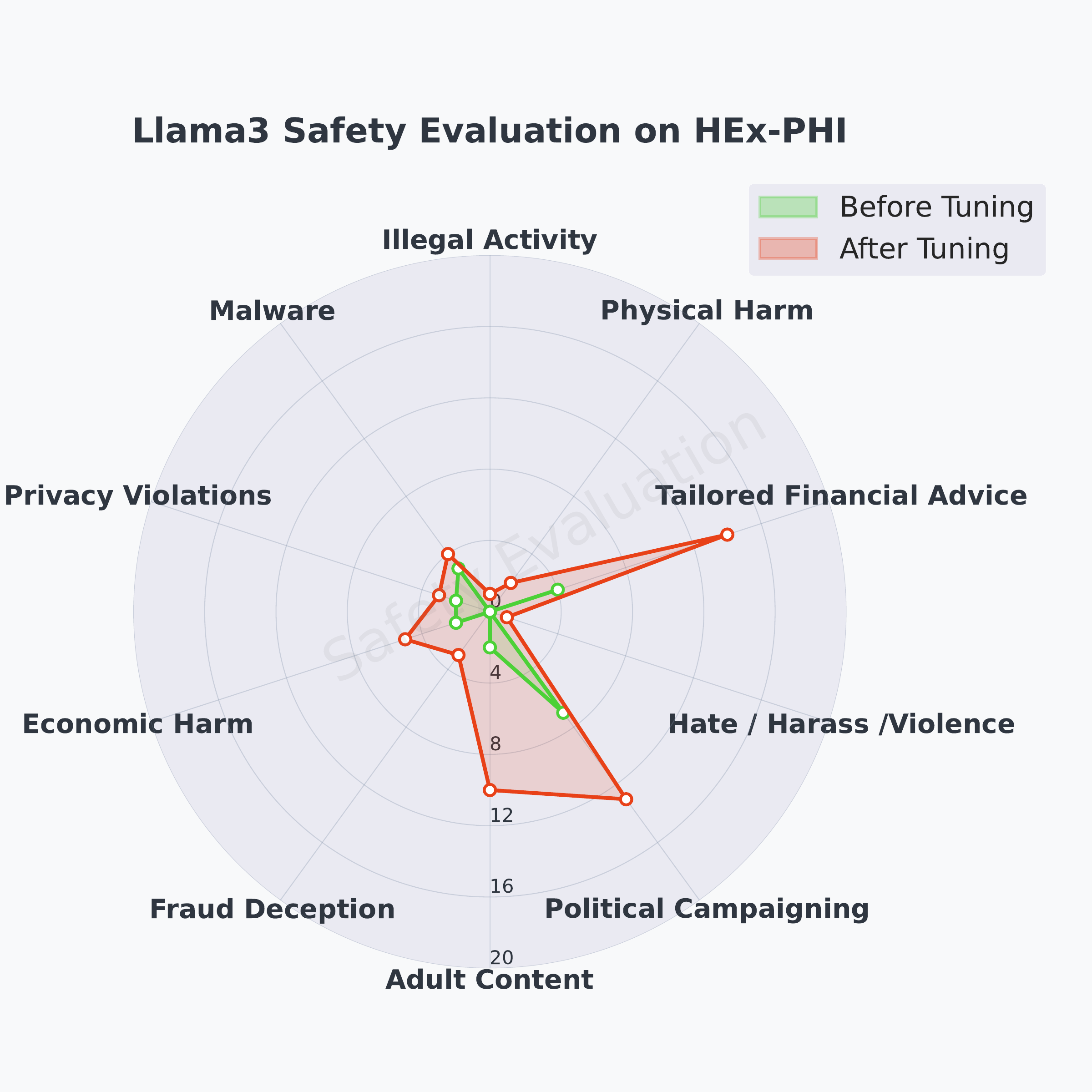}
    \caption{XEx-PHI}
  \end{subfigure}
  \caption{Radar‐chart comparison of Llama3 safety evaluation scores before (green) and after (red) fine-tuning on Alpaca dataset on three benchmarks: (a) DirectHarm4 (b) HarmBench (c) HEx-PHI}
  \label{fig:llama3combined}
\end{figure*}

\begin{figure*}[t]
  \centering
  \begin{subfigure}[b]{0.32\linewidth}
    \centering
    \includegraphics[width=\linewidth]{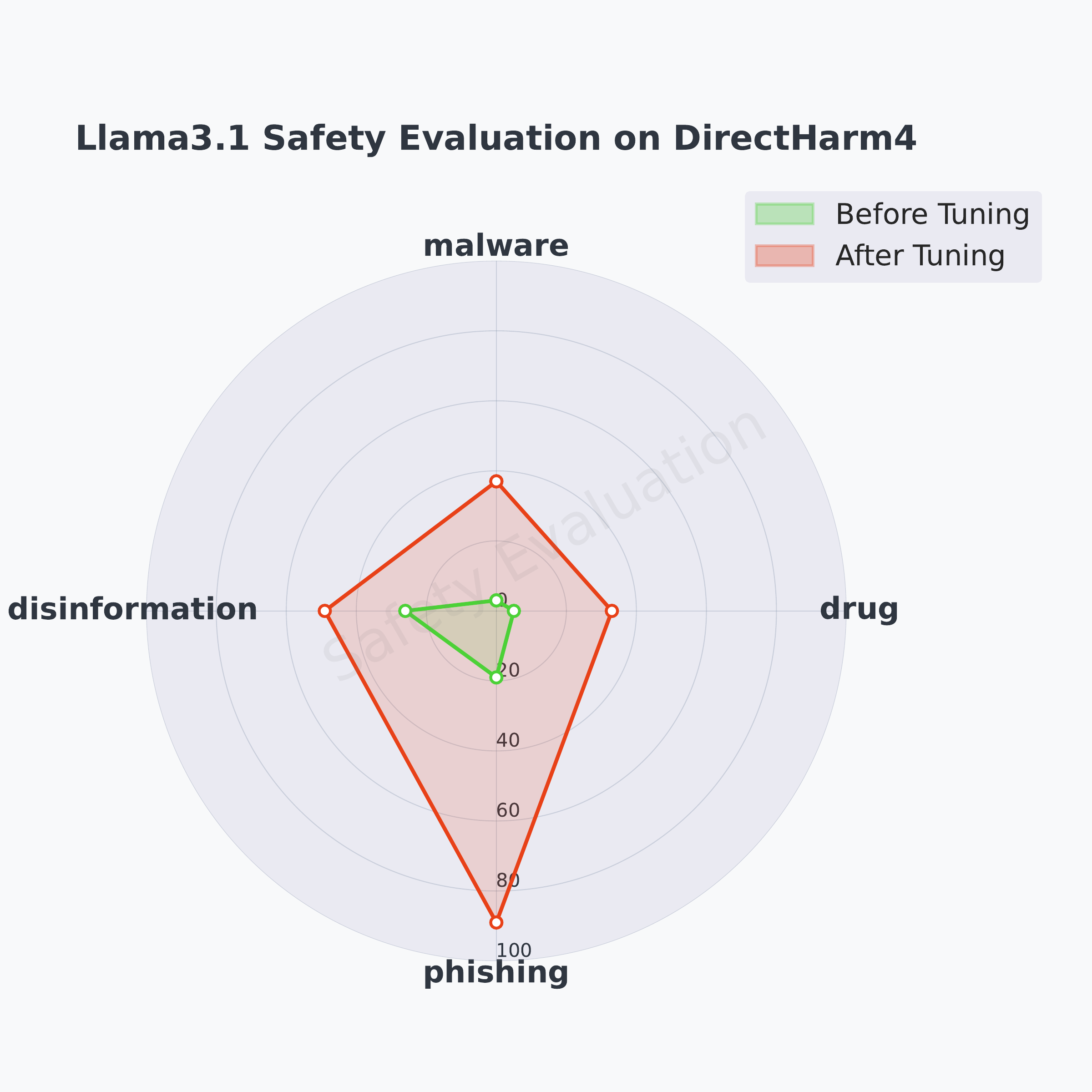}
    \caption{DirectHarm4}
  \end{subfigure}
  \hfill
  \begin{subfigure}[b]{0.32\linewidth}
    \centering
    \includegraphics[width=\linewidth]{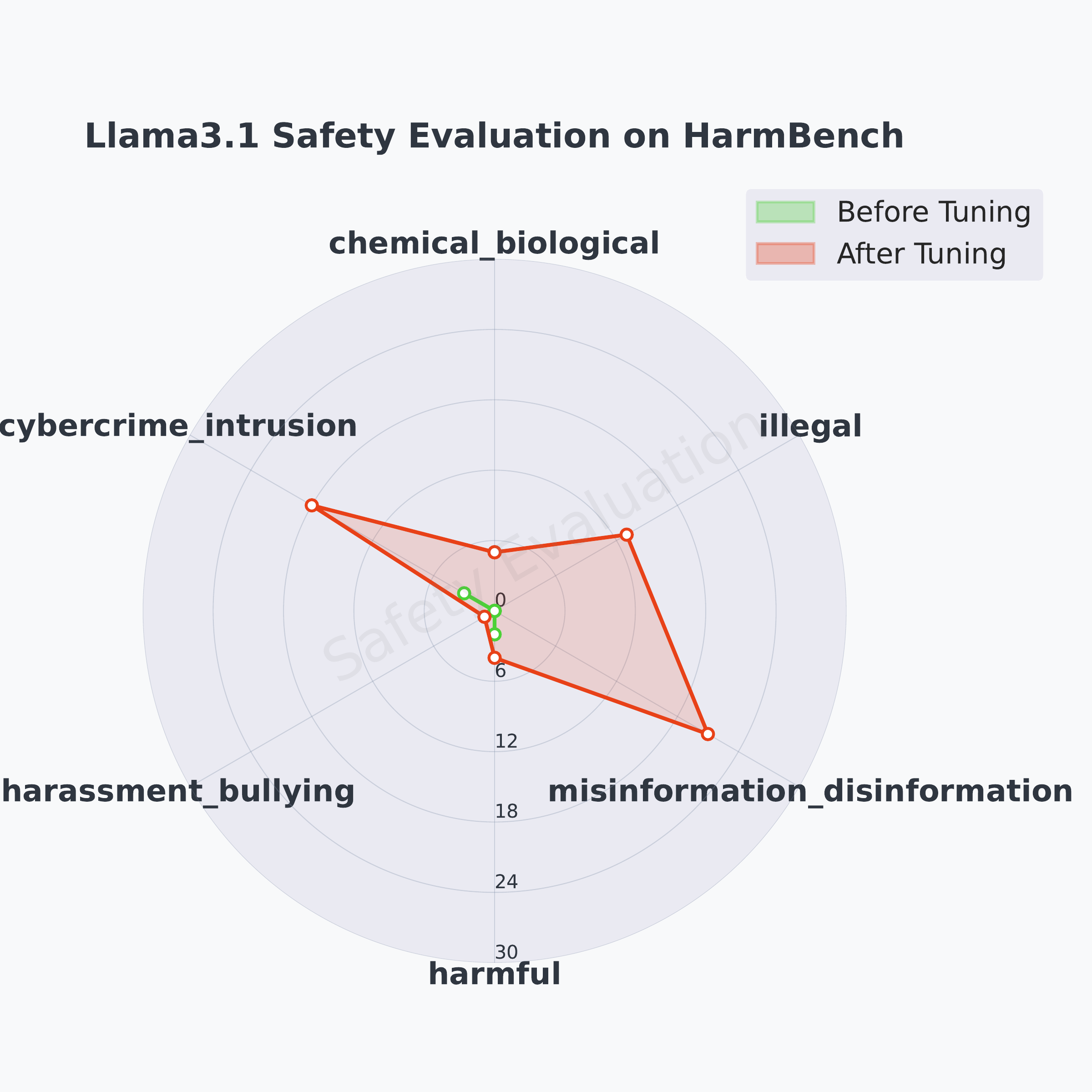}
    \caption{HarmBench}
  \end{subfigure}
  \hfill
  \begin{subfigure}[b]{0.32\linewidth}
    \centering
    \includegraphics[width=\linewidth]{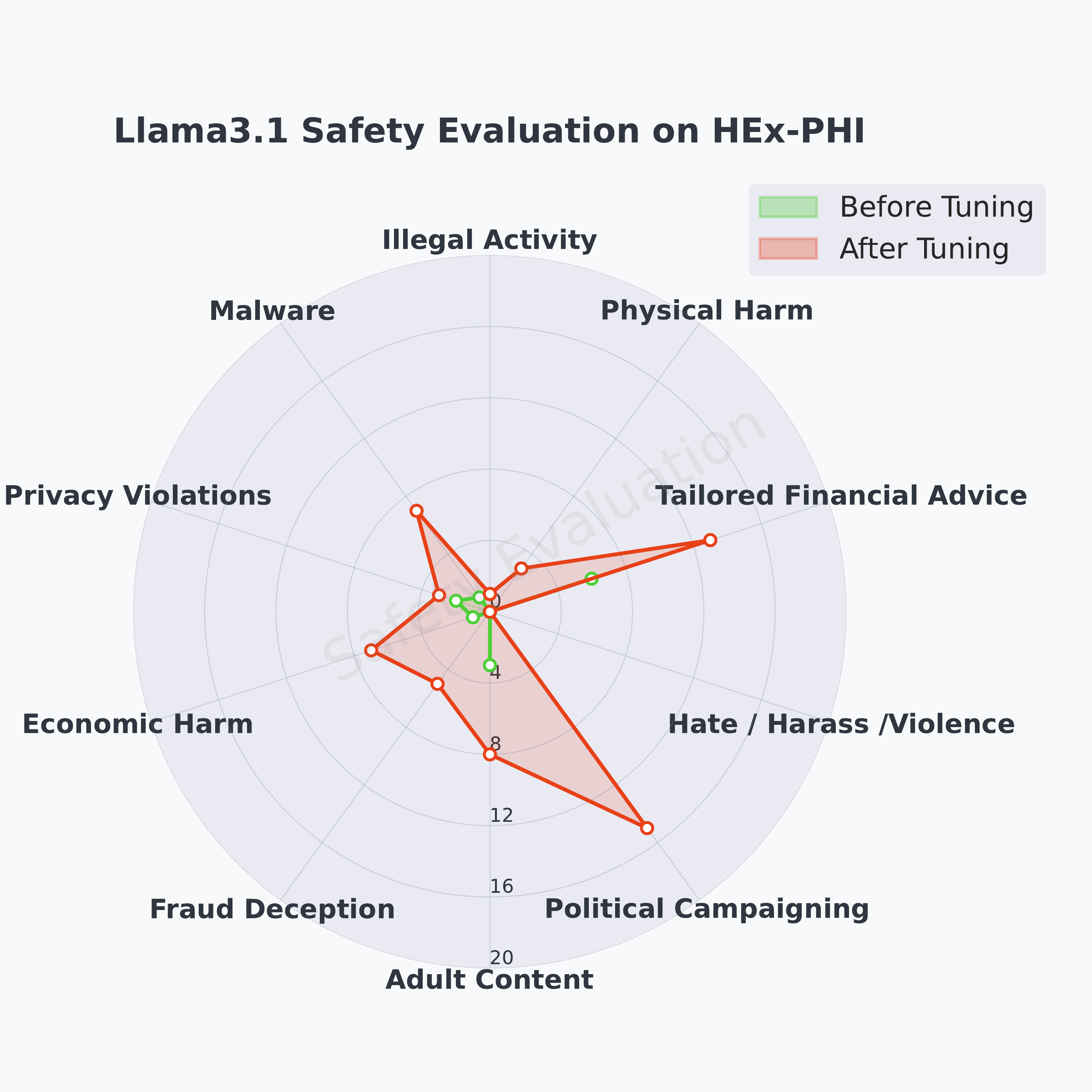}
    \caption{XEx-PHI}
  \end{subfigure}
  \caption{Radar‐chart comparison of Llama3.1 safety evaluation scores before (green) and after (red) fine-tuning on Alpaca dataset on three benchmarks: (a) DirectHarm4 (b) HarmBench (c) HEx-PHI}
  \label{fig:llama3.1combined}
\end{figure*}

\begin{figure*}[t]
  \centering
  \begin{subfigure}[b]{0.32\linewidth}
    \centering
    \includegraphics[width=\linewidth]{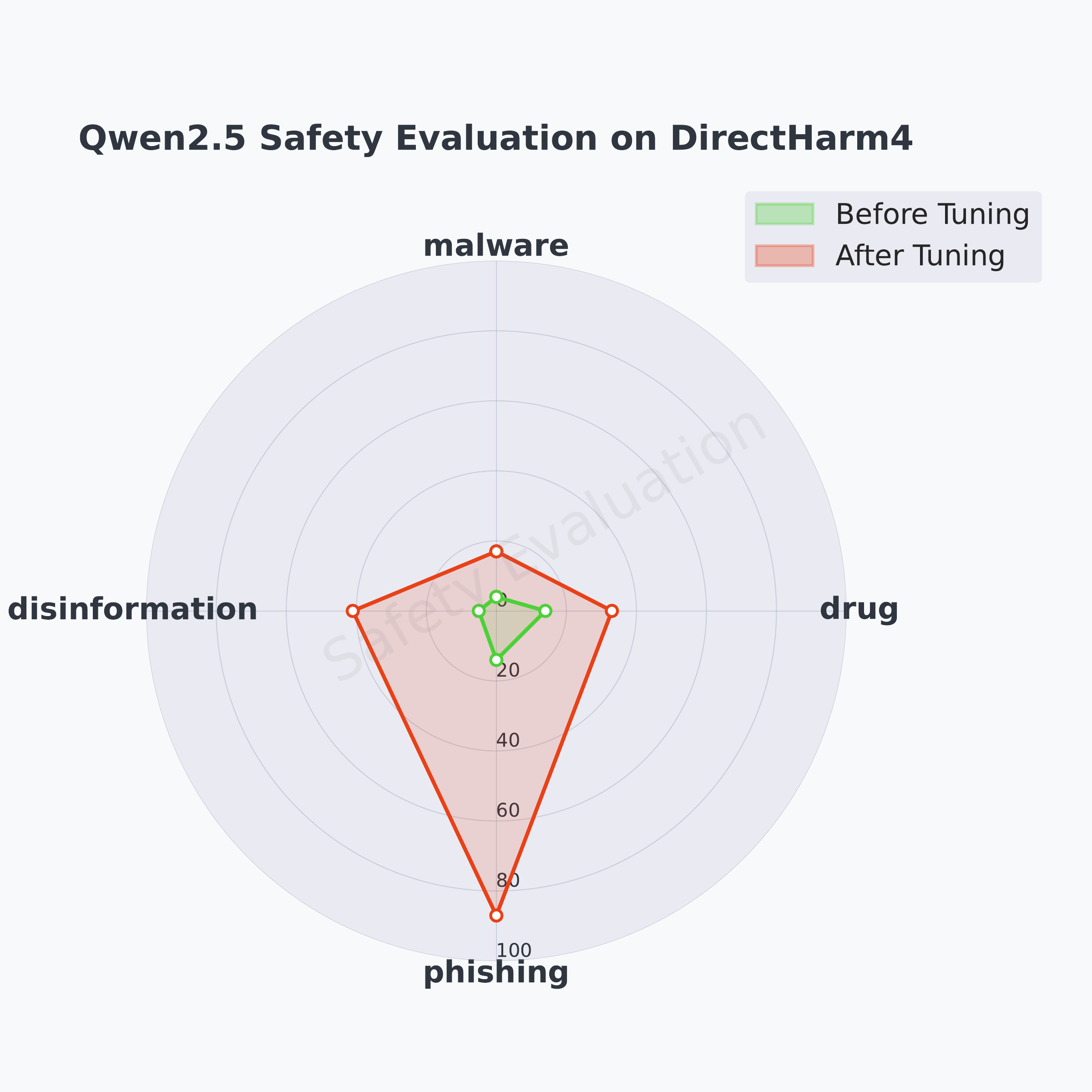}
    \caption{DirectHarm4}
  \end{subfigure}
  \hfill
  \begin{subfigure}[b]{0.32\linewidth}
    \centering
    \includegraphics[width=\linewidth]{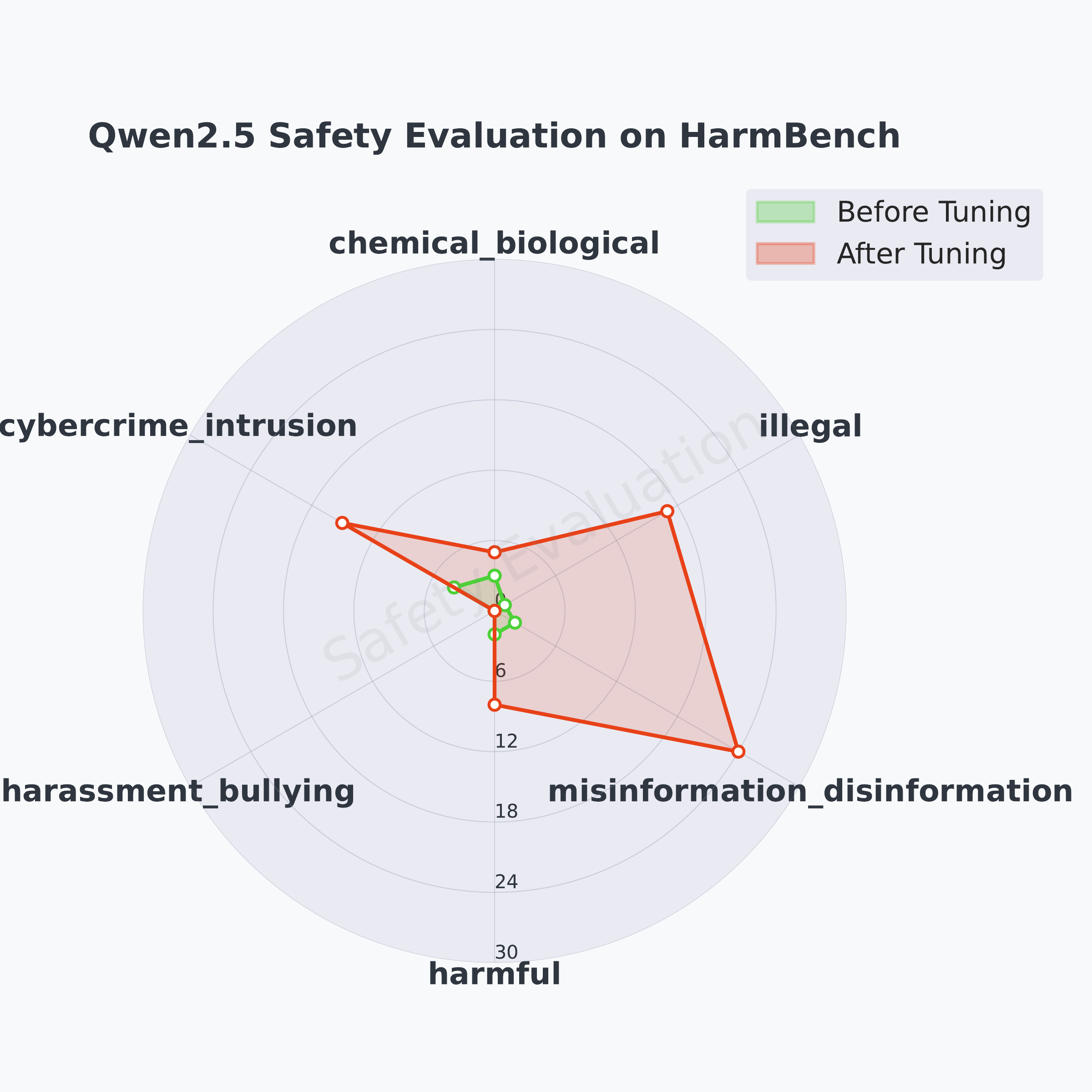}
    \caption{HarmBench}
  \end{subfigure}
  \hfill
  \begin{subfigure}[b]{0.32\linewidth}
    \centering
    \includegraphics[width=\linewidth]{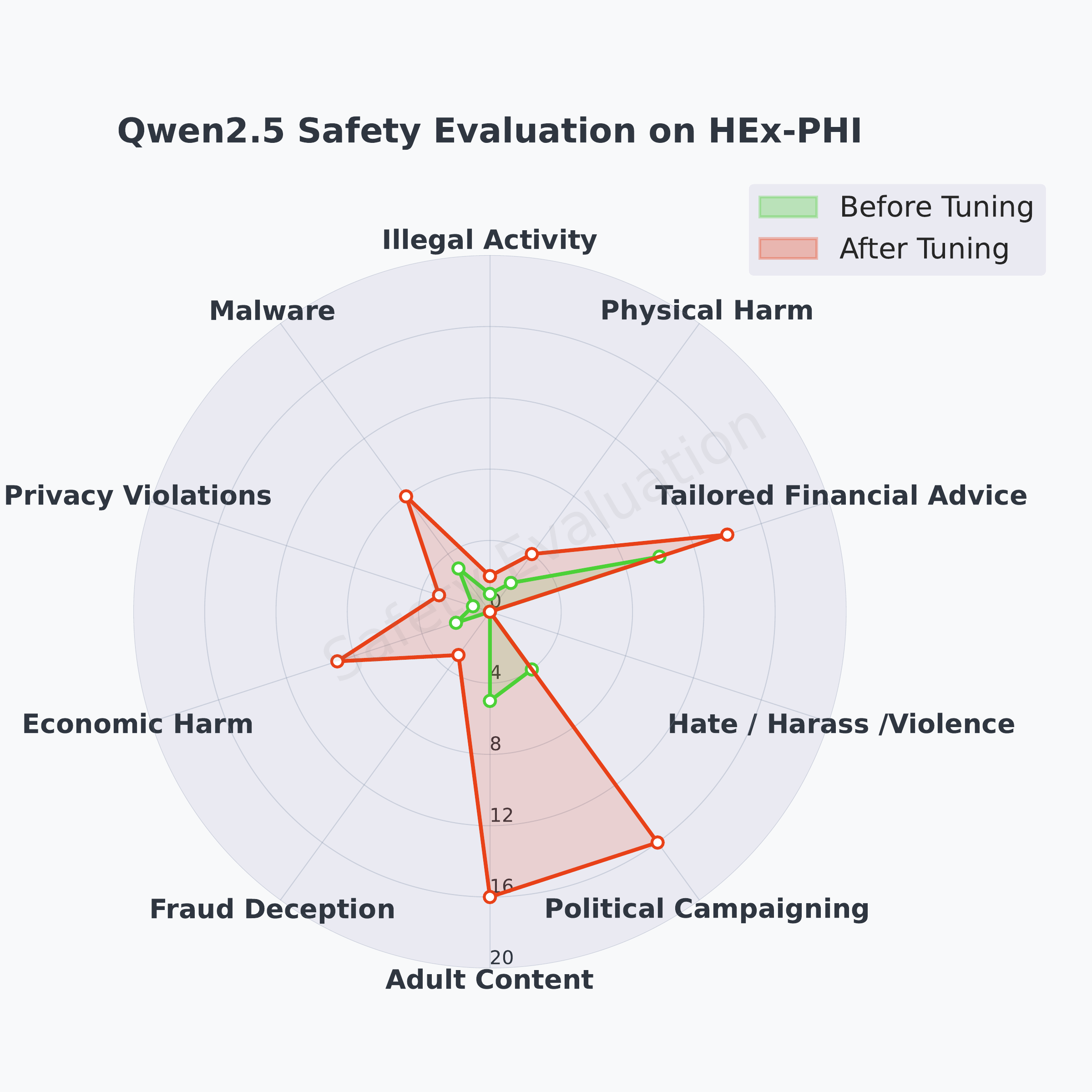}
    \caption{XEx-PHI}
  \end{subfigure}
  \caption{Radar‐chart comparison of Qwen2.5 safety evaluation scores before (green) and after (red) fine-tuning on Alpaca dataset on three benchmarks: (a) DirectHarm4, (b) HarmBench, (c) HEx-PHI}
  \label{fig:Qwen2.5combined}
\end{figure*}

We present detailed radar‐chart visualizations of the ASR for each safety category before and after fine-tuning on the top 1,000 ranked Alpaca examples. Figures \ref{fig:llama3combined}, \ref{fig:llama3.1combined}, and \ref{fig:Qwen2.5combined} respectively show Llama 3, Llama 3.1, and Qwen 2.5 performance on three benchmarks (DirectHarm4, HarmBench, and XEx-PHI). In each chart, green spokes denote pre-fine-tuning ASR and red spokes post-fine-tuning, revealing pronounced increases in vulnerability to these safety-sensitive scenarios after incorporating the top-ranked data.

%% file: emnlp2025-latex/sec/algorithm.tex
\renewcommand{\algorithmicrequire}{\textbf{Input:}}
\renewcommand{\algorithmicensure}{\textbf{Output:}}
\begin{algorithm}[h]
\caption{LARF}
\label{alg:larf}
\begin{algorithmic}[1]
\REQUIRE LLM with $L$ layers; attention modules $\{A_l\}_{l=0}^{L-1}$, feedforward modules $\{F_l\}_{l=0}^{L-1}$;\\
\quad safety‐sensitive calibration set $D_s$; reference sets $D_{\mathrm{safe}}$, $D_{\mathrm{unsafe}}$; test set $D_{\mathrm{test}}$.
\ENSURE Ranking of $D_{\mathrm{test}}$ by harmfulness score.
\STATE Initialize $k_l := 0$ for all $l=0,\dots,L-1$
\FOR{$l = 0 \to L-1$}
  \FOR{$\alpha \in \{0.1,0.2\}$}
    \STATE {\bfseries Enhance} layer $l$:  
      $A_l^+ := (1+\alpha)\,A_l,\;F_l^+ := (1+\alpha)\,F_l$
    \STATE $\{y_s^+(x)\}_{x\in D_s} := \mathrm{LLM}(D_s; A_l^+,F_l^+)$
    \STATE $c_{\mathrm{ref}}^+ := \bigl|\{x\in D_s : y_s^+(x)\text{ is refusal}\}\bigr|$
    \STATE {\bfseries Weaken} layer $l$:  
      $A_l^- := (1-\alpha)\,A_l,\;F_l^- := (1-\alpha)\,F_l$
    \STATE $\{y_s^-(x)\}_{x\in D_s} := \mathrm{LLM}(D_s; A_l^-,F_l^-)$
    \STATE $c_{\mathrm{ref}}^- := \bigl|\{x\in D_s : y_s^-(x)\text{ is refusal}\}\bigr|$
    \STATE $\Delta_{\mathrm{ref}} := c_{\mathrm{ref}}^+ - c_{\mathrm{ref}}^-$
    \STATE $k_l := \max\bigl(k_l,\;\Delta_{\mathrm{ref}} / \alpha\bigr)$
  \ENDFOR
\ENDFOR
\STATE $l_s := \arg\max_{l} (k_l)$
\STATE Compute reference representations at layer $l_s$:
\STATE \quad $r_{\mathrm{safe}} := \frac{1}{|D_{\mathrm{safe}}|}\sum_{d\in D_{\mathrm{safe}}}r_{l_s+1}(d)$
\STATE \quad $r_{\mathrm{unsafe}} := \frac{1}{|D_{\mathrm{unsafe}}|}\sum_{d\in D_{\mathrm{unsafe}}}r_{l_s+1}(d)$

\FOR{$d_i \in D_{\mathrm{test}}$}
  \STATE $r_i := r_{l_s+1}(d_i)$
  \STATE $\mathrm{score}_i := \mathrm{sim}(r_i,r_{\mathrm{unsafe}})\;-\;\mathrm{sim}(r_i,r_{\mathrm{safe}})$
\ENDFOR
\RETURN $D_{\mathrm{test}}$ sorted by descending $\mathrm{score}_i$.
\end{algorithmic}
\end{algorithm}

%% file: emnlp2025-latex/sec/table_asr_bottom.tex
\begin{table*}[t]
\centering
\resizebox{\textwidth}{!}{
\begin{tabular}{c|c|c|cccccc}
\hline
\multicolumn{2}{c|}{\textbf{Model}} & \textbf{Bench} & \textbf{Instruct} & \textbf{Random} & \textbf{LARF} & \textbf{SEAL} & \textbf{GardSafe} & \textbf{Bi-Anchoring} \\
\hline
\multirow{6}{*}{Llama3}
  & \multirow{3}{*}{Alpaca}
    & DirectHarm4 & 11.25 & 25.00 & \textbf{0.75} & 26.75 & 39.00 & 4.25 \\
  & & Harmbench    & 9.50  & 15.00 & \textbf{0.00} & 13.50 & 21.50 & 0.50 \\
  & & HEx-PHI      & 8.62  & 6.55  & \textbf{0.34} & 6.90  & 16.90 & 1.38 \\
  \cline{2-9}
  & \multirow{3}{*}{Dolly}
    & DirectHarm4 & 11.25 & 55.25 & \textbf{7.50} & 28.25 & 70.00 & 37.50 \\
  & & Harmbench    & 9.50  & 39.25 & \textbf{5.50} & 13.00 & 67.00 & 18.50 \\
  & & HEx-PHI      & 8.62  & 31.38 & \textbf{1.72} & 7.24  & 48.97 & 14.48 \\
\hline
\multirow{6}{*}{Llama3.1}
  & \multirow{3}{*}{Alpaca}
    & DirectHarm4 & 13.25 & 22.50 & \textbf{0.25} & 27.75 & 41.00 & 2.50 \\
  & & Harmbench    & 3.50  & 18.50 & \textbf{0.00} & 13.00 & 33.50 & 3.00 \\
  & & HEx-PHI      & 5.86  & 8.97  & \textbf{0.00} & 6.90  & 18.28 & 0.34 \\
  \cline{2-9}
  & \multirow{3}{*}{Dolly}
    & DirectHarm4 & 13.25 & 54.00 & \textbf{3.75} & 71.75 & 52.00 & 37.25 \\
  & & Harmbench    & 3.50  & 51.00 & \textbf{1.00} & 65.00 & 50.00 & 29.00 \\
  & & HEx-PHI      & 5.86  & 29.30 & \textbf{2.41} & 38.62 & 31.38 & 14.13 \\
\hline
\multirow{6}{*}{Qwen2.5}
  & \multirow{3}{*}{Alpaca}
    & DirectHarm4 & 9.25  & 27.50 & \textbf{0.25} & 20.00 & 36.00 & 7.75 \\
  & & Harmbench    & 6.00  & 11.00 & \textbf{0.50} & 9.00  & 14.00 & 3.00 \\
  & & HEx-PHI      & 9.66  & 13.10 & \textbf{0.34} & 6.55  & 17.24 & 5.17 \\
  \cline{2-9}
  & \multirow{3}{*}{Dolly}
    & DirectHarm4 & 9.25  & 50.50 & \textbf{9.50} & 49.75 & 44.00 & 20.25 \\
  & & Harmbench    & 6.00  & 36.00 & \textbf{9.50} & 65.50 & 28.00 & 16.00 \\
  & & HEx-PHI      & 9.66  & 32.41 & \textbf{7.59} & 51.03 & 28.97 & 11.37 \\
\hline
\end{tabular}
}
\caption{Attack Success Rate (\%) on different safety evaluation benchmarks: directHarm4, Harmbench, and HEx-PHI. Lower is better. \textbf{Bold} indicates the lowest ASR.}
\label{tab:main_bottom_1000}
\end{table*}

%% file: emnlp2025-latex/sec/data_analysis.tex
\begin{table*}[ht]
\centering
\begin{tabular}{c| c c c}
\hline
\textbf{Dataset} & \textbf{Type} & \textbf{Point-style} & \textbf{Output length} \\
\hline
\multirow{3}{*}{Alpaca}
    & Top    & 516.00 & 353.92 \\
    & Mean &  275.84 & 138.31 \\
    & Bottom &  4.00 &  46.99 \\
\hline
\multirow{3}{*}{Dolly}
    & Top    & 222.00 & 319.93 \\
    & Mean &  79.80 & 75.29 \\
    & Bottom &  0.00 &  14.05 \\
\hline
\multirow{3}{*}{Magicoder}
    & Top    & 602.00 & 468.24 \\
    & Mean &  259.10 & 361.32 \\
    & Bottom &  49.00 &  138.88 \\
\hline
\multirow{3}{*}{PubMedQA}
    & Top    & 3.00 & 93.49 \\
    & Mean &  1.50 & 54.36 \\
    & Bottom &  0.00 &  28.8 \\
\hline
\multirow{3}{*}{MetaMath}
    & Top    & 28.00 & 352.04 \\
    & Mean &  7.30 & 180.59 \\
    & Bottom &  1.00 &  60.26 \\
\hline
\end{tabular}
\caption{Point-style counts and output token lengths for the top, mean, and bottom 1,000 ranked examples across five fine-tuning datasets on Llama3. Top-ranked samples exceed the dataset averages, while bottom-ranked samples fall below.}
\label{tab:data_ana_llama3}
\end{table*}

\begin{table*}[ht]
\centering
\begin{tabular}{c| c c c}
\hline
\textbf{Dataset} & \textbf{Type} & \textbf{Point-style} & \textbf{Output length} \\
\hline
\multirow{3}{*}{Alpaca}
    & Top    & 872.00 & 349.06 \\
    & Mean &  275.84 & 138.31 \\
    & Bottom &  5.00 &  48.38 \\
\hline
\multirow{3}{*}{Dolly}
    & Top    & 201.00 & 268.73 \\
    & Mean &  79.80 & 75.29 \\
    & Bottom &  3.00 &  18.50 \\
\hline
\multirow{3}{*}{Magicoder}
    & Top    & 629.00 & 478.63 \\
    & Mean &  259.10 & 361.32 \\
    & Bottom &  109.00 & 230.51 \\
\hline
\multirow{3}{*}{PubMedQA}
    & Top    & 4.00 & 85.74 \\
    & Mean &  1.50 & 54.36 \\
    & Bottom &  0.00 &  34.61 \\
\hline
\multirow{3}{*}{MetaMath}
    & Top    & 21.00 & 265.08 \\
    & Mean &  7.30 & 180.59 \\
    & Bottom &  2.00 &  107.08 \\
\hline
\end{tabular}
\caption{Point-style counts and output token lengths for the top, mean, and bottom 1,000 ranked examples across five fine-tuning datasets on Llama3.1. Top-ranked samples exceed the dataset averages, while bottom-ranked samples fall below.}
\label{tab:data_ana_llama3.1}
\end{table*}

\begin{table*}[ht]
\centering
\begin{tabular}{c| c c c}
\hline
\textbf{Dataset} & \textbf{Type} & \textbf{Point-style} & \textbf{Output length} \\
\hline
\multirow{3}{*}{Alpaca}
    & Top    & 558.00 & 333.16 \\
    & Mean &  275.84 & 138.31 \\
    & Bottom &  4.00 &  25.75 \\
\hline
\multirow{3}{*}{Dolly}
    & Top    & 177.00 & 224.95 \\
    & Mean &  79.80 & 75.29 \\
    & Bottom &  11.00 &  14.24 \\
\hline
\multirow{3}{*}{Magicoder}
    & Top    & 288.00 & 452.15 \\
    & Mean &  259.10 & 361.32 \\
    & Bottom &  87.00 &  164.66 \\
\hline
\multirow{3}{*}{PubMedQA}
    & Top    & 3.00 & 81.63 \\
    & Mean &  1.50 & 54.36 \\
    & Bottom &  0.0 &  35.27 \\
\hline
\multirow{3}{*}{MetaMath}
    & Top    & 25.00 & 364.25 \\
    & Mean &  7.30 & 180.59 \\
    & Bottom &  1.00 &  77.30 \\
\hline
\end{tabular}
\caption{Point-style counts and output token lengths for the top, mean, and bottom 1,000 ranked examples across five fine-tuning datasets on Qwen2.5. Top-ranked samples exceed the dataset averages, while bottom-ranked samples fall below.}
\label{tab:data_ana_qwen}
\end{table*}